\documentclass[a4paper,10pt]{article}
\usepackage{jheppub}
\usepackage{graphicx}
\usepackage{graphics}
\usepackage[mathscr]{eucal}
\usepackage{amssymb}
\usepackage{amsbsy}
\usepackage{pifont}
\newcommand{\xmark}{\ding{55}}%
\usepackage{amsmath}
\usepackage{xspace}
\usepackage{listings}
\usepackage{ulem}
\usepackage{slashed}
\usepackage{verbatim}
\usepackage{axodraw4j}
\usepackage{caption}
\usepackage{subcaption}
\usepackage{multirow}
\usepackage{xcolor}
\usepackage{hyperref}
\usepackage[capitalise]{cleveref}
\usepackage[title,toc]{appendix}
\usepackage{cancel}
\usepackage{soul}
\usepackage{natbib}
\usepackage{mciteplus}
\usepackage{lineno}
\usepackage{placeins}
\usepackage{comment}

\usepackage[shortlabels]{enumitem}

\hypersetup{
colorlinks=true,
linkcolor=blue,
citecolor=blue,
urlcolor=blue
}

\definecolor{ultramarine}{rgb}{0.5, 0.125, 0.376}
\definecolor{mycolor}{rgb}{0.9, 0.5, 0.}
\definecolor{Anacolor}{rgb}{0.8, 0.5, 0.7}

\input{TII-decays-NEW-V4-defs.sty}

\title{Type-II Seesaw Higgs triplet productions and decays at the LHC}

\author[a]{Otilia A. Ducu,} 
\author[a]{Ana E. Dumitriu,} 
\author[a]{Adam Jinaru,}
\author[]{Romain Kukla,}
\author[b]{Emmanuel~Monnier,}
\author[c]{Gilbert Moultaka,} 
\author[a]{Alexandra Tudorache,}
\author[d]{Hanlin Xu}

\affiliation[a]{IFIN-HH Bucharest; Romania}
\affiliation[b]{CPPM, Aix-Marseille Universit\'e, CNRS/IN2P3, Marseille; France}
\affiliation[c]{Laboratoire Univers \& Particules de Montpellier (LUPM), Universit\'e de Montpellier, CNRS, Montpellier; France}
\affiliation[d]{Department of Modern Physics and State Key Laboratory of Particle Detection and Electronics,
University of Science and Technology of China, Hefei; China}

\emailAdd{oducu@cern.ch}
\emailAdd{ana.dumitriu@cern.ch}
\emailAdd{adam.jinaru@cern.ch}
\emailAdd{monnier@in2p3.fr}
\emailAdd{gilbert.moultaka@umontpellier.fr}
\emailAdd{atudorac@cern.ch}
\emailAdd{hanlin.xu@cern.ch}
\abstract{The Type-II Seesaw Model provides an attractive scenario to account for Majorana-neutrino masses. Its extended Higgs sector, if sufficiently light, can have a rich and distinctive phenomenology at the LHC while yielding automatically an essentially  Standard-Model-Higgs-like state. Several phenomenological studies have been devoted to the scalar sector of this model, as well as experimental searches focusing mostly on the (doubly-)charged states. In this paper we present an exhaustive study of the main production and decay channels of all the non-standard scalar states originating from the $\rm SU(2)_L$ doublet and a complex triplet of the model. We stick to scenarios where lepton-number-violating decays are suppressed, for which present experimental limits are still weak, highlighting theoretical parameter sensitivities that were not previously emphasized in the literature and the uncertainties they can induce for the experimental searches at the LHC. A comprehensive classification of the various cascade decays and corresponding Standard Model particle multiplicities is provided. As an illustration, a detailed prospective search study at the LHC with an ATLAS-like detector is carried out on some benchmark points, for charged, doubly-charged, and, for the first time, neutral state productions. 
}

\preprint{}

\begin{document}
\thedate
%\linenumbers
\maketitle
\tableofcontents
% ==============================================================================

%\section*{General comments} 

%%%%%%%%%%%%%%%%%%%%%%%%%%%%%%%%%%%%%%%%%%%%%%%%
%%%%%%%%%%%%%%%%%%%%%%%%%%%%%%%%%%%%%%%%%%%%%%%%
%%%%%%%%%%%%%%%%%%%%%%%%%%%%%%%%%%%%%%%%%%%%%%%%
\section{Introduction} 
\label{section1}

The discovery of a Standard-Model-like Higgs boson
at the LHC~\cite{Aad:2012tfa,Chatrchyan:2012ufa}
completed the last long-sought missing piece of the Standard Model. 
Nonetheless, the absence so far of clear and direct experimental indications at the TeV scale for new physics beyond the Standard Model, disappointed hopes based on aesthetic naturalness and mass hierarchy arguments for the most motivated scenarios such as supersymmetry, Higgs compositeness or models with extra space dimensions. While this might be a hint for the irrelevance of such arguments, the need for new physics is still motivated by several other shortcomings of the Standard Model. As a consequence, efforts to search for new particles in a wider context than the aforementioned scenarios got a new boost, including extensions of the Higgs sector that could hopefully answer some of the puzzles still remaining in particle physics and cosmology. 

In fathoming these problems, one interesting scenario addressing the origin of neutrino masses  is the Type-II Seesaw Model~\cite{Konetschny:1977bn, Cheng:1980qt, Lazarides:1980nt, 
Schechter:1980gr, Mohapatra:1980yp}. The addition of an $\rm SU(2)_L$ complex triplet to the Higgs sector allows a 
lepton-number-violating Yukawa operator that can generate (Majorana) neutrino masses  dynamically through spontaneous electroweak-symmetry breaking.
Strictly speaking, a seesaw mechanism operates when the mass scale of the new scalars is much higher than the electroweak (EW) scale, perhaps the GUT scale, where the size of the neutrino masses is a consequence of a very tiny vacuum expectation value (VEV) in the new scalars sector, thus keeping the related Yukawa coupling of order one.
This puts such scalars out of reach at the colliders. However, giving up this aesthetic feature, one can still consider scenarios with EW scale masses~\cite{deGouvea:2006gz,Perez:2008ha}, in which case rich and exotic signatures in the scalar sector can be searched for at the colliders.
Phenomenologically distinctive features of this model are due not only to the presence of two doubly-charged scalars (\Hpp), but also to that of two singly-charged (\Hp), two neutral CP-even ($h^0,H^0)$ and one CP-odd ($A^0$) scalars. Indeed, the latter despite being reminiscent of the scalar content of two-Higgs-doublet models (2HDM), have specific production and decay modes depending on the amount of mixing between the doublet and the triplet, that can make their experimental search strategies somewhat involved. Another distinctive property of the model is that one of the two CP-even scalars comes out naturally SM-like in most of the parameter space, irrespective of the mass spectrum of the other scalars. This welcome aesthetic feature, contrasting with doublet or singlet extensions where alignment or decoupling limits should be enforced, is rarely emphasized.

Several phenomenological studies addressed the search for a relatively light scalar sector of this model
\cite{Huitu:1996su,Chakrabarti:1998qy,Chun:2003ej,Muhlleitner:2003me,Akeroyd:2005gt,Dey:2008jm,FileviezPerez:2008jbu,delAguila:2008cj,Akeroyd:2009hb,Akeroyd:2010je,Akeroyd:2011zza,Arhrib:2011uy,Melfo:2011nx,Aoki:2011pz,Arhrib:2011vc,Akeroyd:2012nd,Chiang:2012dk,Chun:2012zu,Akeroyd:2012ms,Chun:2012jw,BhupalDev:2013xol,Englert:2013wga,Kanemura:2013vxa,Chun:2013vma,Kang:2014lwn,kang:2014jia,Kanemura:2014goa,Arhrib:2014nya,Han:2015hba,Han:2015sca,Das:2016bir,Mitra:2016wpr,Babu:2016rcr,BhupalDev:2018tox,Du:2018eaw,Antusch:2018svb,Li:2018jns,Ferreira:2019qpf,Anisha:2021fzf,Banerjee:2024jwn,Bolton:2024thn,Ashanujjaman:2023tlj,Giarnetti:2023dcr,Mandal:2022zmy,Chiang:2021lsx}, some of which focusing mainly on the exotic doubly-charged scalar; see also Refs.~\cite{Primulando:2019evb,Ashanujjaman:2021txz} for comprehensive reviews.

On the experimental side, searches for doubly-charged scalar bosons have been ongoing at various colliders and with different model assumptions: single or pair production followed by decays to a pair of leptons or to a pair of $W^\pm$ bosons with same electric charge (same-charge). Limits on single production of doubly-charged Higgs, akin to models like the Georgi-Machacek \cite{Georgi:1985nv,Chanowitz:1985ug}, were derived both from same-charge lepton~\cite{HERA:2006,CDF:2011} and same-charge $W^\pm$~\cite{CMS:2015,CMS:2018,CMS:2021wlt,ATLAS:2023dbw} final-state searches.
As for pair production with same-charge leptons final state, lower mass limits started off from 98.5~GeV~\cite{OPAL} to reach the present most stringent limits of 900--1080~GeV~\cite{ATLAS:2023pairmultilep}; although often presented within specific models such as the Zee-Babu neutrino mass model~\cite{Zee:1985id,Babu:1988ki,Nebot:2007bc}, or the Left-Right Symmetric Model~\cite{Pati:1974yy,Mohapatra:1974hk,Senjanovic:1975rk,BhupalDev:2016nfr,Borah:2016hqn}, these limits apply to the Type-II Seesaw Model as well when the decay to leptons is dominant~\cite{CMS:2012,ATLAS:2018multilepton,ATLAS:2023pairmultilep}.   

ATLAS analyses~\cite{ATLAS:2019pair, 
ATLAS:2021pairbosons} focused on pair production of doubly-charged scalars decaying to same-charge $W^\pm$ bosons, \cite{Kanemura:2013vxa,Kanemura:2014goa}, as well as on associated production with a singly-charged scalar decaying to gauge bosons.
The obtained exclusion limits on the mass range from 350~GeV for the first,   to 250~GeV for the second, at 95\%~CL.

The searches for singly-charged Higgs boson were conducted in two mass regimes: light \Hp masses ~\cite{ATLAS:2014otc,CMS:2015lsf,ATLAS:2018gfm,CMS:2019bfg,CMS:2015yvc,ATLAS:2013uxj,CMS:2018dzl, ATLAS:2023bzb, CMS:2019idx,ATLAS:2021xhq}, for which the \Hp mass is smaller than the top mass, and heavy \Hp masses~\cite{ATLAS:2015nkq, ATLAS:2018ntn, CMS:2019rlz, CMS:2020imj,ATLAS:2014otc,CMS:2015lsf,ATLAS:2016avi,ATLAS:2015edr,CMS:2017fgp,CMS:2021wlt,ATLAS:2020zzb,CMS:2022jqc}. 
Model independent upper limits on the BR, or on the $\sigma \times\mathrm{BR}$ product, were imposed and then equivalent exclusion regions were extracted in the context of MSSM or Type-II 2HDM models. The searches for neutral Higgs bosons followed similar lines, considering the same models, by LEP~\cite{ALEPH:2006tnd} and Tevatron~\cite{CDF:2009esh,D0:2010etq,D0:2011yor,CDF:2011jcu} in the beginning, through ATLAS~\cite{ATLAS:2019tpq,ATLAS:2012ube,ATLAS:2019odt, ATLAS:2014vhc, ATLAS:2016ivh, ATLAS:2017eiz, ATLAS:2020zms} and CMS~\cite{CMS:2013baf,CMS:2015grx, CMS:2018hir,CMS:2015ooa,CMS:2019mij, CMS:2011lzj,CMS:2012bkm,CMS:2014ccx,CMS:2018rmh}, ending with the most recent  ATLAS~\cite{ATLAS:2023szc} and CMS~\cite{CMS:2024mtn} results, which put tight constraints on the non-degenerate $m_{A^0}$ and $m_{H^0}$ plane. 

Most of the above limits, in particular the strong ones coming from single production and/or decays to same-charge leptons, although dubbed 'model-independent', are in fact either irrelevant or can be avoided
in the case of the Type-II Seesaw Model.
For one thing, the scalars of this model are mainly produced in pairs or in association, dominantly through Drell-Yan (DY) processes. 
Single production modes are suppressed by the small triplet VEV with respect to the EW scale, in contrast with the Georgi-Machacek model. For another, this same VEV can still be chosen large enough to suppress decays to same-charge leptons as compared to same-charge $W^\pm$-bosons.

In the present paper we consider exclusively this last scenario for which mass exclusion limits are still moderate~\cite{ATLAS:2019pair, ATLAS:2021pairbosons}. The aim is to provide an exhaustive study of the phenomenology of the Type-II-Seesaw-Model scalar sector, and possible repercussions on searches at the LHC, when direct lepton-number-violating (LNV) decays to leptons are suppressed with respect to decays to other SM particles. 
To ensure this suppression we fix throughout the study the triplet VEV to a relatively large benchmark value (but still compatible with the experimental value of the SM $\rho$-parameter \cite{ParticleDataGroup:2024cfk}). This will allow to highlight an important sensitivity of the various decay branching ratios (BRs) to another independent parameter, namely the mixing angle that controls the closeness of one of the two CP-even scalars of the model to a SM-like Higgs. This physically important parameter is often approximated in the literature by fixing it to a value proportional to the ratio of the triplet to doublet VEVs,  
thus masking its proper sensitivity when the VEVs are fixed. In particular, it would leave unnoticed that the panorama of relative contributions of the different decay channels do not depend only on the mass of the decaying scalar, as shown for instance in \cite{Primulando:2019evb}, but also strongly on the above mentioned mixing angle, even for fixed scalar masses. The outcome of this sensitivity, convoluted with the effects of unknown mass splittings and the triplet VEV, will be an irreducible uncertainty that mitigates exclusion limits from a given signal model. This calls for a global search strategy in order to explore the most promising production and decay modes for the (HL-)LHC. As an illustration we present prospect studies relying on a few benchmark scenarios for the production and decays of (doubly-)charged or neutral scalars through a detailed cutflow analysis at the LHC assuming an ATLAS-like detector. 
While the (doubly-)charged analysis is guided by the previous  ATLAS studies, the one for the neutral states, though preliminary, is performed here for the first time, exploring the sensitivity at the end of LHC Run-3 and HL-LHC.   

The rest of the paper is organized as follows: In \cref{sec:TIISM} we recall the main ingredients of the model, define a convenient parameterization that avoids some technical pitfalls, then highlight the generic features of the mass spectrum. \Cref{sec:LHCpheno} is devoted to an overview of the production cross-sections and a detailed discussion of the decay branching ratios. 
A synopsis of all possible SM intermediate and final states, multiplicities and cascade decays is given in \cref{sec:allFinalStates}. An exploratory prospective analysis for the experimental search at the LHC is carried out in \cref{sec:benchmark-exp}, for three different benchmark model points and specific decay modes.  
\cref{sec:further_discussions} discusses sensitivities to modifications of the benchmark points. We conclude in \cref{sec:conclusion} and provide  further  material in the appendices.

%%%%%%%%%%%%%%%%%%%%%%%%%%%%%%%%%%%%%%%%%%%%%%%%
%%%%%%%%%%%%%%%%%%%%%%%%%%%%%%%%%%%%%%%%%%%%%%%%
%%%%%%%%%%%%%%%%%%%%%%%%%%%%%%%%%%%%%%%%%%%%%%%%
\FloatBarrier
\section{The scalar sector of the Type-II Seesaw Model \label{sec:TIISM} }

%%%%%%%%%%%%%%%%%%%%%%%%%%%%%%%%%%%%%%%%%%%%%%%%
\subsection{Ingredients \label{sec:theModel}}
On top of the $\rm SU(2)_L$-doublet scalar field $H$ of the SM with hypercharge 
$Y_H=1$, one assumes a  complex scalar triplet $\Delta$ with hypercharge $Y_\Delta=2$ (in the $Q = T_3 + \frac{Y}{2}$ convention).  
The ingredients of the model have been described in various works, however often with varying notations. We refer the reader
to Refs.~\cite{Perez:2008ha,Arhrib:2011uy} of which we follow the notations. 
The couplings of the new scalar components to the gauge bosons are uniquely fixed
by the gauge invariant kinetic terms. Their self-couplings and couplings to the doublet components are given by the most general renormalizable potential $V(H, \Delta)$ that takes the form:
%${\mathcal{L}}_{Y}$ is the needed Yukawa interaction to generate neutrinos masses (Appendix $\mathbf{A}$.3).
\begin{equation}
\begin{aligned}
V(H, \Delta) =& - \! m_H^2{H^\dagger{H}}+\frac{\lambda}{4}(H^\dagger{H})^2+M_\Delta^2Tr(\Delta^{\dagger}{\Delta})
+[\mu(H^T{i}\sigma^2\Delta^{\dagger}H)+{\rm h.c.}]  \\
&+ \! \lambda_1(H^\dagger{H})Tr(\Delta^{\dagger}{\Delta})+\lambda_2(Tr\Delta^{\dagger}{\Delta})^2
+\lambda_3Tr(\Delta^{\dagger}{\Delta})^2 +\lambda_4{H^\dagger\Delta\Delta^{\dagger}H}
%+\lambda_5{H^\dagger\Delta^{\dagger}\Delta{H}}
\label{eq:Vpot}
\end{aligned}
\end{equation}
Dirac-type Yukawa couplings to quarks or leptons involving $\Delta$ are forbidden by the gauge symmetries. The only allowed coupling is of the Majorana type, lepton-number violating and of the form
$``-  L^T Y_{\nu} \Delta L  + {\rm h.c.}"$ where $L$ denotes a lepton doublet and $Y_{\nu}$ the Yukawa coupling matrix in flavor space. 
After spontaneous electroweak symmetry breaking, dictated by the structure of $V(H, \Delta)$, the electrically neutral components 
of $H$ and $\Delta$ acquire VEVs denoted respectively $v_d$ and $v_t$, and the $10$ components of the scalar sector reorganize into
%\st{6 new scalars: a further CP-even boson $H^0$, here taken as heavier than the SM-like Higgs boson, a CP-odd scalar $A^0$, singly-charged \Hp and a doubly-charged $H^{\pm\pm}$ scalars.}
$7$ physical massive states, $h^0, H^0, A^0, H^{\pm}, H^{\pm \pm}$ and three goldstone bosons, $G^0, G^\pm$.

We recall here 
the content of these states and their masses in terms of the multiplet component states, the mixing angles, the couplings in the potential and the VEVs of the $\rm SU(2)_L$ doublet and triplet fields \cite{Arhrib:2011uy}: 
\begin{align}
&h^0 = \cos\alpha \, h + \sin\alpha \, \xi^0, 
~~~H^0 = - \sin\alpha \, h + \cos\alpha \, \xi^0, \label{eq:CP-even}\\
%\vspace{.2cm}
&A^0 = - \sin \beta \, Z_1 + \cos \beta \, Z_2, 
~~~G^0 =  \cos \beta \,  Z_1 + \sin \beta \, Z_2, \label{eq:CP-odd} \\
%\vspace{.2cm}
&G^\pm = \cos\beta^{'} \phi^{\pm}+ \sin\beta^{'} \delta^{\pm},
~~~H^\pm = -\sin\beta^{'} \phi^{\pm}+ \cos\beta^{'}
\delta^{\pm},  \label{eq:charged}
\end{align}

\begin{align}
m_{H^{\pm\pm}}^2&=\frac{\sqrt{2}\mu{v_d^2}- \lambda_4v_d^2v_t-2\lambda_3v_t^3}{2v_t},  \label{eq:mHpmpm}  \\
m_{H^\pm}^2&=\frac{(v_d^2+2 v_t^2)[2\sqrt{2}\mu- \lambda_4 v_t]}{4v_t}, \ m_{G^\pm}=0, \label{eq:mHpm} \\
m_{A^0}^2 &= \frac{\mu(v_d^2+4v_t^2)}{\sqrt{2}v_t},  \ m_{G^0}=0, \label{eq:mA0}\\
m_{h^0, H^0}^2 &=\frac{1}{2}[A+C \mp \sqrt{(A-C)^2+4B^2}], \label{eq:mh0}
\end{align}

where
\begin{align}
A = \frac{\lambda}{2}{v_d^2}, 
  B &=v_d[-\sqrt{2}\mu+(\lambda_1+\lambda_4)v_t], \ 
  C = \frac{\sqrt{2}\mu{v_d^2}+4(\lambda_2+\lambda_3)v_t^3}{2v_t},
  \label{eq:ABC}
\end{align}

and
\begin{align}
\sin\alpha = -\frac{sgn[B] {\epsilon_{\alpha}}}{\sqrt{2}}\left(1 + \frac{(A - C)}{\sqrt{(A - C)^2 + 4 B^2}}\right)^{\!\!\frac12},  \cos\alpha=    \frac{\epsilon_{\alpha}}{\sqrt{2}}\left(1 - \frac{(A - C)}{\sqrt{ (A - C)^2 + 4 B^2  }}\right)^{\!\!\frac12} \label{eq:sinalpha}
\end{align}
\begin{align}
\sin\beta &= \epsilon_{\beta} \frac{2 v_t}{\sqrt{v_d^2 + 4 v_t^2}}, \
\cos\beta  = \epsilon_{\beta} \frac{v_d}{\sqrt{v_d^2 + 4 v_t^2}},
\\
\sin\beta^{'}&= \epsilon_{\beta^{'}} \frac{\sqrt{2}v_t}{\sqrt{v_d^2+2v_t^2}}, \
\cos\beta^{'}= \epsilon_{\beta^{'}} \frac{v_d}{\sqrt{v_d^2+2v_t^2}}.
\end{align}

Here $\epsilon_{\alpha}$, $\epsilon_{\beta}$ and $\epsilon_{\beta^{'}}$ denote arbitrary signs, and the relative sign between \sina and $\cos\alpha$
is given by the sign of $-B$. The doublet and triplet VEVs, $v_d$ and $v_t$, 
fix unambiguously (modulo $\pi$)  the mixing angles $\beta$ and $\beta'$.
The fields $h$($\xi^0$) and $Z_1$($Z_2$) appearing respectively in \cref{eq:CP-even,eq:CP-odd} denote the real and imaginary parts of the neutral component of the $\rm SU(2)_L$ doublet(triplet) after the VEV shifts. $\phi^{\pm}$ and $\delta^\pm$ appearing in \cref{eq:charged} are the singly-charged components of the doublet and triplet respectively. 
The Higgs mechanism leads to the theoretical tree-level $Z$ and $W^\pm$ boson masses,
\begin{align}
M_Z^2&=
%\frac{(g^2+{g'}^2)(v_d^2+4v_t^2)}{4} 
     \frac{g^2(v_d^2+4v_t^2)}{4\cos^2\theta_W} ,\label{eq:mZ}\\
M_W^2&=\frac{g^2(v_d^2+2v_t^2)}{4}  ,\label{eq:mW}
\end{align}
where $g$ is the $\rm SU(2)_L$ gauge coupling and $\theta_W$ the weak angle, and the tree-level $\rho$-parameter
\begin{align}
\rho&=\frac{v_d^2+2v_t^2}{v_d^2+4v_t^2}. \label{eq:rho_param}
\end{align}
Even though \cref{eq:rho_param} implies $\rho <1$, which would be in tension with the central value of the $\rho$-parameter from global fits,
taking  $\sqrt{v_d^2+ 2 v_t^2}=246\rm~GeV$ and $v_t\ll v_d$ ensures consistency of this parameter and of the gauge boson masses with the experimental values 
\cite{ParticleDataGroup:2024cfk}.

As seen from \cref{eq:CP-even}, the lighter ($h^0$) and heavier ($H^0$) CP-even states will have an $\rm SU(2)_L$-doublet component, $h$, controlled by $\cos \alpha$ and $\sin\alpha$ respectively. Thus a very small $\sin\alpha$ ($\cos \alpha$) will mean the lighter (heavier) state behaves as a SM Higgs.
Similarly, $A^0$ and \Hp will have  very small mixings,
respectively $\sin\beta$ and $\sin\beta'$, with the $\rm SU(2)_L$-doublet components, therefore remaining almost purely triplet states while $H^{\pm\pm}$ is obviously pure triplet. Together with
their lepton number violating couplings and their couplings to the electroweak gauge bosons, this fully determines their couplings to the SM particle content. In particular, $H^0, A^0$ and \Hp  will have a pattern of couplings rather different from other models comprising these particles, such as the typical 2HDM.

%%%%%%%%%%%%%%%%%%%%%%%%%%%%%%%%%%%%%%%%%%%%%%%%
\FloatBarrier
\subsection{Optimizing the model parameterization}
\label{sec:parameters}

The previous model description shows that a straightforward approach to the phenmenological study would be to take as input the  mass-parameter $\mu$, the doublet and triplet VEVs, $v_d$ and $v_t$, and the complete set of couplings
$\lambda, \lambda_{i=1,...4}$, deducing from them the physical states, the mass spectrum and couplings among the scalars as well as the doublet-triplet-mixing angles. The latter mixing then fixes the couplings to the gauge bosons induced by the gauge couplings, and to the quarks and leptons induced by the Dirac-type Yukawa couplings.

This approach is however not very convenient. The physical masses are not directly tractable, preventing a simple procedure of focusing on a required mass scale and mass splittings (rather than an extraction from long scans). In particular, even though a CP-even state with SM-like properties comes out naturally in the model, targeting the right mass around 125~GeV, as well as the favored scenario of it being the lighter CP-even mass, i.e. $|\sin\alpha| \ll 1$ (see \cref{eq:CP-even}), is not automatic. Apart from a very tiny region in $\mu$ where the mixing can vary a lot, there is a clear separation between two scenarios where either the lighter $h$ or the heavier $H^{0}$ is the SM-like Higgs boson.

An ideal approach would have been to take all five scalar masses together with \sina as input, then to infer $\mu$ and the five $\lambda's$ using the inverted relations given in Ref.~\cite{Arhrib:2011uy}. However, here one hits another problem: a huge sensitivity of $\lambda_2$ and $\lambda_3$ to minute variations of the physical masses. This effect is due to the presence of $({m}/{v_t})^{2}$ factors in these two couplings, where $m$ is a typical scalar mass and $v_t\ll m$. Such a sensitivity is not just a side technical issue to be avoided. It can be seen as a real theoretical problem in the model when it comes to interpreting the experimental limits. 
Taking for illustration $v_t \simeq .1$, 
$\mu \sim$~.1~-~.7~GeV and a set of couplings satisfying $|\lambda_i| \lesssim 1$,  leading to scalar masses in the range 200~-~500~GeV, then using the inverted relations to retrieve the $\lambda$'s from the masses, one finds that 
a variation as small as $\frac{\delta m}{m} \sim 10^{-6}$ from the initial mass values can induce up to a factor 200 enhancement on $\lambda_2$ and $\lambda_3$! This is a genuine effect, not attributable to some numerical precision loss. It means that one can easily be pushed into a non-perturbative regime in the scalar sector, even within any foreseeable experimental resolution on masses. It follows that one should be careful in interpreting experimental limits on masses based on the evaluation of production cross-sections in the perturbative regime. As well-known (see also \cref{sec:prodModes}), pair or associated productions proceed mainly through DY processes controlled by the gauge couplings. If $\lambda_2$ and $\lambda_3$ become very large then contributions to the DY from s-channel exchange of scalars, even with suppressed couplings to quarks, would dominate, giving a wrong tree-level result for the cross-section. In practice, this would typically happen if for instance the $\lambda$'s are computed internally from input masses implemented say in  UFO modules for an automatic calculation. 

To avoid such ambiguities it thus appears mandatory  to define a strategy where $\lambda_2$ and $\lambda_3$ are input.\footnote{The same approach was adopted  in Ref.~\cite{Primulando:2019evb}, although there the justification for this choice was more for convenience, with no mention of the perturbativity issue.} However to keep the benefit of having benchmark masses as input we choose the following hybrid set of parameters: $m_{h^0}, m_{H^\pm}, m_{H^{\pm \pm}}, \sin\alpha, \lambda_2, \lambda_3,v_d$ and $v_t$. An overview of the technical  implementation of this strategy is given in \cref{app:paramstrategy}.
Such a strategy allows to control directly the masses of the benchmark (doubly-)charged scalar states,
fix once and for all one CP-even state mass to the experimental SM-like Higgs mass value, and study directly the sensitivity to the $\sin\alpha$ mixing. The latter will be shown to have in some cases important effects on the branching ratios of the scalar states decays, and thus on the experimental search strategy,  
as discussed at length in \cref{sec:decayModes}.

In practice we explore the parameter space by benchmarking our scan on the doubly-charged Higgs boson mass,
with a first dimension being the mass splitting with the singly-charged Higgs boson.
Theoretical constraints apply on the model parameters,
in particular on the $\lambda$'s in order to ensure unitarity and boundedness from below of the potential~\cite{Arhrib:2011uy,Bonilla:2015eha,Moultaka:2020dmb}.
 The couplings $\lambda, \lambda_1$ and $\lambda_4$ are uniquely determined from the above set of input. Then, together with $\lambda_2$ and $\lambda_3$, one can check for the consistency of these couplings with respect to the  theoretical constraints.  
Other phenomenological constraints should be taken into account. The scalar triplet gives contributions to the so-called oblique parameters, S, T and U \cite{Peskin:1991sw} and their extensions \cite{Burgess:1993mg,Lavoura1994},   
that enter the SM precision observables. As for any new scalar multiplets coupling to the SM gauge bosons, the contributions to the oblique parameters will depend in our case on the triplet-like Higgs masses \mHpp, \mHp and $m_{H^0}$. As first noted in Ref.~\cite{Chun:2012jw}, the electroweak precision data recast on S, T and U parameters translate into a constraint on the mass splitting $\left | \Delta m_{H^{\pm\pm}, H^\pm}\right |\lesssim 40$~GeV, where   
%\begin{align}
$\Delta m_{H^{\pm\pm}, H^\pm}  = m_{H^{\pm\pm}} - m_{H^\pm}$
%\end{align}, 
(see Refs.~\cite{Cheng:2022hbo} and~\cite{Primulando:2019evb,Ashanujjaman:2022tdn} for more recent updates).
Some analyses~\cite{Das:2016bir,Primulando:2019evb}  consider a wider $\Delta m_{H^{\pm\pm}, H^\pm}$  interval [-50,50], or even put tighter constraints for $v_t$ values similar to the one we consider, $\Delta m_{H^{\pm\pm}, H^\pm} \in [-30,40]$. We choose to be even more conservative, taking the mass splitting in the interval  [-20,20],
\begin{align}
\left|\Delta m_{H^{\pm\pm}, H^\pm}\right| & \lesssim 20 \, \rm{GeV} \label{eq:Deltam_20} .
\end{align}

It has been shown that the magnitude of the triplet VEV $v_{t}$ enhances or suppresses the couplings, thus impacting the available final states, in particular for doubly-charged and singly-charged Higgs bosons~\cite{Mitra:2016wpr,Melfo:2011nx,Ashanujjaman:2021txz}.
Illustration with a given choice of masses identified three ranges:
small $(v_t \lesssim 10^{-6}\rm GeV)$, intermediate $(10^{-6}\rm GeV \lesssim v_t \lesssim 10^{-2}\rm GeV) $ and large  $(v_t \gtrsim 10^{-2}\rm GeV)$.
In the present paper we focus on the large $v_{t}$ scenario taking a benchmark value of $v_t=0.1$~GeV.\footnote{With such a value of $v_t$ way larger than the neutrino mass scale, the Majorana-type Yukawa coupling matrix will have to have negligibly small entries in flavor space to account for tiny neutrino masses. The LNV decays, $H^{\pm \pm} \to \ell^\pm \ell^\pm$, $H^\pm \to \ell^\pm \nu_\ell$ and $H^0,A^0 \to \nu \nu$ will thus have very small branching fractions and will be ignored throughout the present study. See also the beginning of \cref{sec:Hpp}. For the same reason, lepton-flavor-violating processes, see e.g. Ref.~\cite{Banerjee:2024jwn}, will be suppressed too, and do not imply significant constraints on the scalar masses in our analysis.} Finally, further phenomenological constraints on the model parameters can come from the experimental measurements of the SM Higgs rare decays to $\gamma\gamma$ \cite{CMS:2018piu,ATLAS:2018hxb} and $\gamma Z$ \cite{ATLAS:2023yqk}. Indeed, even if, as we will see in the next section, $h^0$ has essentially SM-like Higgs properties, this is true only at the tree-level in the model. The loop-induced $h^0 \to \gamma\gamma$ and $h^0 \to \gamma Z$ decays depend on the (doubly-)charged scalar masses as well as on 
$\lambda_1$~\cite{Arhrib:2011vc,Akeroyd:2012ms}.

%%%%%%%%%%%%%%%%%%%%%%%%%%%%%%%%%%%%%%%%%%%%%%%%
\FloatBarrier
\subsection{Mass spectrum evolution and   \texorpdfstring{$\sin \alpha$}{}
\label{sec:mass_spectrum}}

\begin{figure}[htb!]
\begin{center}   
    {\includegraphics[width=0.48\textwidth]{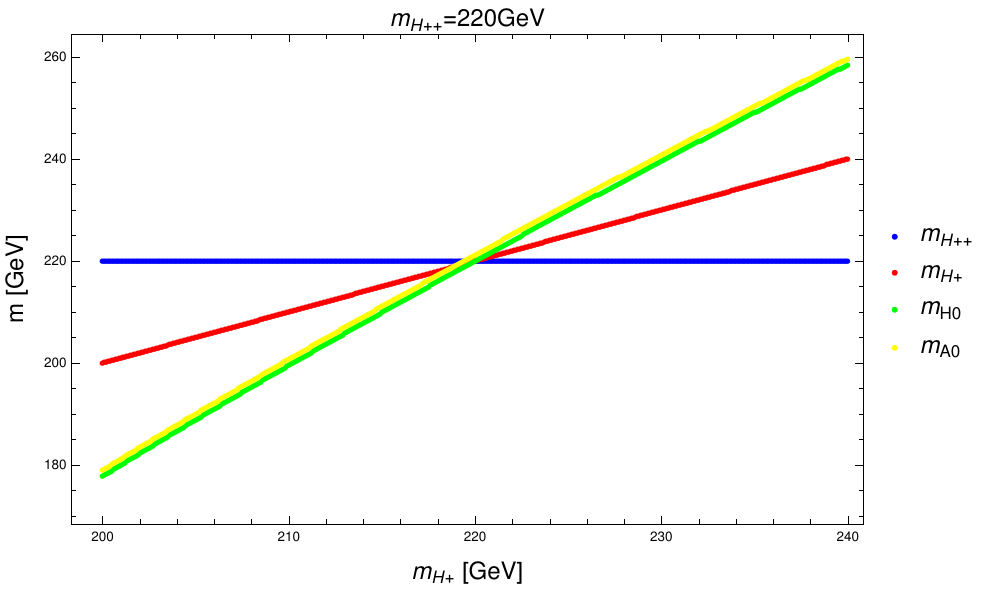}}
    {\includegraphics[width=0.48\textwidth]{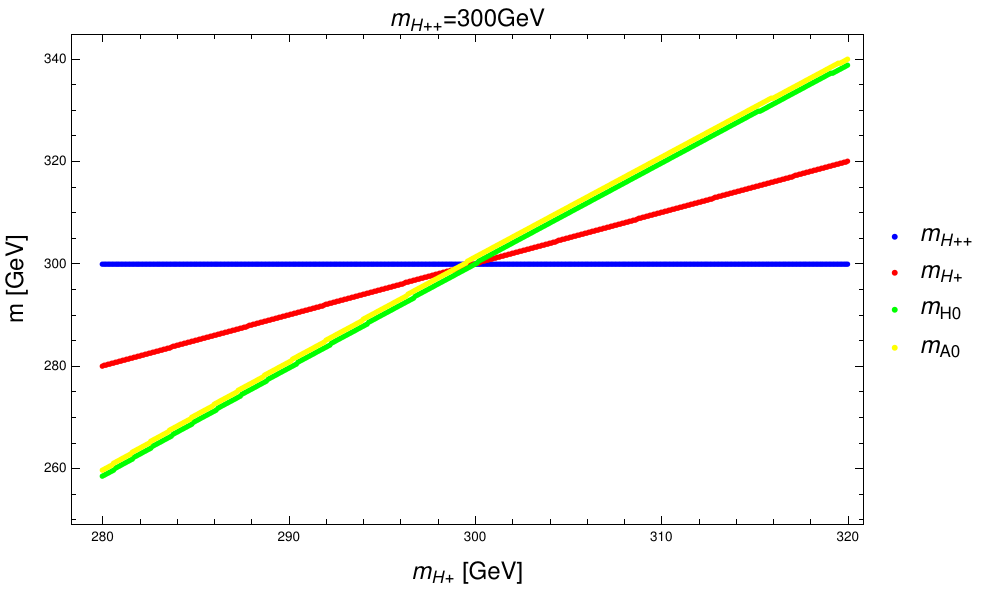}}
    {\includegraphics[width=0.48\textwidth]{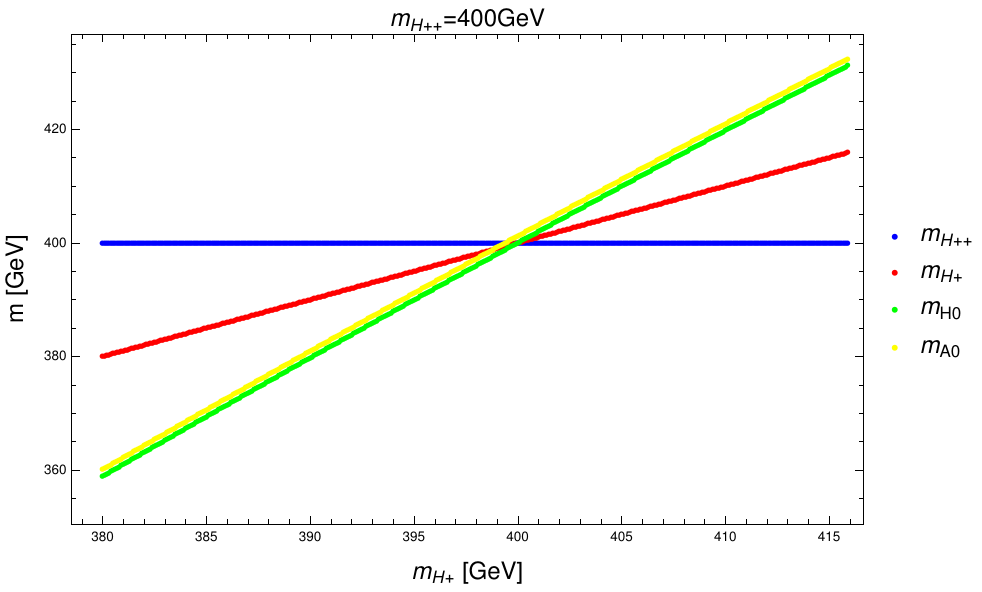}}   
\end{center}
\caption{\label{fig:spectrum-evol} 
Mass spectrum evolution as a function of the  \Hp mass in the range 
$\left| m_{\rm H^{\pm \pm}} - m_{\rm H^{\pm \pm} }\right| \leq 20 \rm~GeV$, for three fixed values of the \Hpp mass, $220$~GeV,  $300$~GeV and $400$~GeV, with $\lambda_2=\lambda_3=0.1, v_t=0.1$~GeV, $\sin \! \alpha =8.1\times 10^{-4}$ and the SM-like Higgs mass fixed to $m_{h^0}=125$~GeV. See text for further discussions.}
\end{figure}

We recall here the trend of the mass spectrum as well as that of \sina, fixing
$m_{h^0}=125$~GeV and $v_t=0.1$~GeV. From \cref{eq:mHpmpm,eq:mHpm,eq:mh0,eq:ABC} one finds the general trend of the splitting between the squared masses, $m_{H^{\pm\pm}}^2 - m_{H^{\pm}}^2 \simeq m_{H^{\pm}}^2 - m_{H^0}^2 \simeq - \frac{\lambda_4}{4}v_d^2$, \cite{Melfo:2011nx}, up to ${\cal O}(v_t^2)$ and ${\cal O}(\mu v_t)$ corrections. Similarly one has $m_{H^0}^2 \simeq m_{A^0}^2$, up to ${\cal O}(v_t^2)$ corrections. Although deviations from this general trend are tiny, it is important to keep track of the detailed configurations, since, as we will see in the subsequent sections, even small variations of the parameters can induce significant modifications to a given hypothetically dominant final state and thus to the experimental search strategies based on this final state signals. \cref{fig:spectrum-evol} shows the mass evolution of the scalar states for three different benchmark \Hpp masses, using the scan strategy described in the previous section. Changing the mass splitting between the charged Higgs bosons impacts linearly the other scalars, however with a similar scaling independently of the overall mass scale, as seen from the three plots of \cref{fig:spectrum-evol}. At the point of exact \Hpp-\Hp mass degeneracy, the four scalar states become essentially degenerate. This occurs for $\lambda_4 \simeq -4 v_t (\sqrt{2} \mu + \lambda_3 v_t)/v_d^2$. Then the mass hierarchy is flipped on either sides of this point, which obviously flips possible cascade decay channels of one state into another. 
Also the scalar mass range under consideration will correspond to a $\mu$ parameter in the range $[0.07,0.45]$. The two possible mass hierarchies illustrated in \cref{fig:spectrum-evol} will be exploited further in \cref{sec:allFinalStates}. 
%The SM Higgs boson mass is fixed by the value of $\lambda_{0}$ which is taken in a range so that $m_{h^{0}} \in [125.0, 125.12]$~GeV.

\begin{figure}[!htb]
\begin{center}
\begin{center}
    {\includegraphics[width=0.6\textwidth,keepaspectratio]{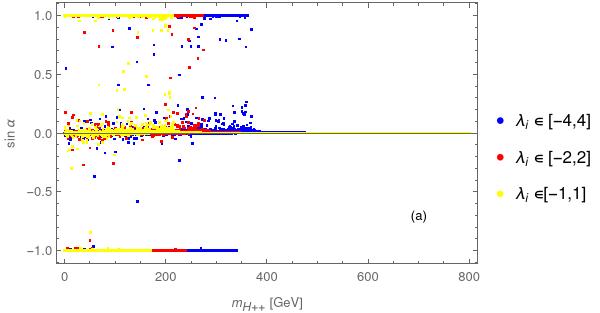}}
\end{center}
    {\includegraphics[width=0.4\textwidth,keepaspectratio]{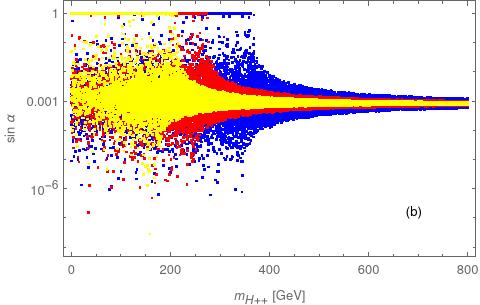}}
    {\includegraphics[width=0.405\textwidth,keepaspectratio]{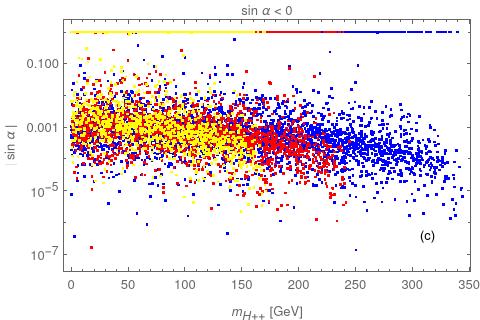}}
\end{center}
\caption{\label{fig:sinalpha} \sina scatter-plot as a function of \mHpp and a random scan on the $\lambda_i$'s over the indicated ranges, fixing $m_{h^0}$ or $m_{H^0}=125$~GeV, $v_d=246$~GeV and $v_t=0.1$~GeV; (a) the general trend; (b) and (c) zooming in respectively on positive and negative \sina (showing absolute values for the latter).
}
\end{figure}

The mixing angle $\alpha$ plays an important role since $\sin\alpha$ parameterizes the deviation of the lighter CP-even state $h^0$ from a SM-like Higgs, cf.~\cref{eq:CP-even}. A scatter plot of this parameter is in \cref{fig:sinalpha}, and shows the general trend. From \cref{fig:sinalpha}~(a), one can  clearly see that apart from very rare instances, one of the two CP-even states behaves like a SM Higgs, the lighter $h^0$ for $\sin\alpha \simeq 0$, the heavier $H^0$ for $\sin\alpha \simeq \pm 1$, cf. \cref{eq:CP-even}.  
Zooming in on the positive and negative \sina regions, \cref{fig:sinalpha}~(b) and~(c), shows that 
$|\sin\alpha|$ lies preferably around $10^{-3}$ even though it can be much smaller, or around $1$. 
Note that the $\sin\alpha \simeq 1$ and $\sin\alpha < 0$ configurations become unphysical for \mHpp $\gtrsim 380$~GeV due to the appearance of tachyonic states. Note also that \cref{fig:sinalpha} does not take into account unitarity and boundedness-from-below-of-the-potential constraints on the $\lambda_i$'s, nor the phenomenological upper bound on $|\Delta m_{H^{\pm\pm}, H^\pm}|$. We checked that these constraints, although reducing somewhat the scatter plots, do not modify the spread of allowed \sina values.  
In the scenario where $h^0$ is SM-like, positive \sina is often approximated by $2.46 \, v_t/v_d$ \cite{Primulando:2019evb} (of order $10^{-3}$ in our case, that is indeed the location of the funnel seen in \cref{fig:sinalpha}). However, the fact that \sina can still take values one or two orders of magnitude lower than or in excess of this nominal value, as evident from the scatter plots, is of prime importance. Indeed, as we will see in \cref{sec:decayModes}, the decay branching ratios are very sensitive to variations of \sina.

%%%%%%%%%%%%%%%%%%%%%%%%%%%%%%%%%%%%%%%%%%%%%%%%
%%%%%%%%%%%%%%%%%%%%%%%%%%%%%%%%%%%%%%%%%%%%%%%%
%%%%%%%%%%%%%%%%%%%%%%%%%%%%%%%%%%%%%%%%%%%%%%%%
\section{LHC phenomenology of the scalar sector \label{sec:LHCpheno}}

In the following we will rely on the scan strategy described in \cref{sec:parameters} for which the input parameters are $m_{h^0}, m_{H^{\pm \pm}}, \Delta m_{H^{\pm\pm} , H^\pm} , \sin\alpha, \lambda_2, \lambda_3,v_d$ and $v_t$. Among these, we fix
$m_{h^0}, v_d$ and $v_t$. Adopting the scenario where the lighter CP-even state is SM-like, we take $m_{h^0}=125$~GeV.\footnote{The scenario where $H^0$ is SM-like was investigated in Ref.~\cite{Arhrib:2014nya} and appears somewhat contrived in view of the LEP limits.}
The $v_t$ parameter should remain much smaller than the EW scale to satisfy the constraints on the $\rho$-parameter. We will fix it to $v_t=0.1$~GeV.
This value yields a tree-level $\rho$-parameter~\cref{eq:rho_param}, consistent at the $1.6\sigma$ level with the reported global fit result $\rho_0 = 1.00031\pm0.00019$~\cite{ParticleDataGroup:2024cfk}. As already pointed out, for such large-$v_t$ scenarios the LNV decays of \Hpp, for which the present experimental limits are the most stringent, become highly suppressed. We also fix $v_d$ to $\simeq 246$~GeV to keep $M_Z$ and $M_W$ in the right ballpark.

%%%%%%%%%%%%%%%%%%%%%%%%%%%%%%%%%%%%%%%%%%%%%%%%
\FloatBarrier
\subsection{Overview of the production channels}
\label{sec:prodModes}

Since $v_t$ remains very small compared to the EW scale, it follows that, apart from the SM-like $h^0$,  single \Hpp or \Hp or $H^0$  productions that would proceed via vector boson fusion (VBF), where the couplings are proportional to $v_t$, are suppressed as compared to pair or associated two-scalar production. This is in contrast 
with models protecting tree-level custodial symmetry with two $SU(2)_L$ triplets \cite{Georgi:1985nv,Chanowitz:1985ug} where the triplet VEVs can be very large---without conflicting with the experimental bounds on the
$\rho$ parameter---and the single scalar production sizable. Moreover, single production of $H^0$ or $A^0$ via gluon fusion (ggF) is also suppressed due to their very small doublet content---the only one that couples to fermions---cf. \cref{eq:CP-even,eq:CP-odd}.
Note also that, apart from the SM-like $h^0$, scalar productions in association with a gauge boson  are also suppressed by powers of $v_t/v_d$. These productions will not be considered further.
 
We list in \cref{table:prod_channels} the two-scalar production 
channels at the LHC.
%and comment briefly on the expected relative magnitudes of their cross-sections. 
The initial state subprocess can be either DY quark annihilation ($q\bar{q}^{{}^\text{\tiny (}{}'{}^\text{\tiny )}}$) or VBF or ggF. The leading production channels proceed via the $q\bar{q}^{{}^\text{\tiny (}{}'{}^\text{\tiny )}}$ subprocesses with intermediate s-channel $Z (\gamma)$ or $W^\pm$ exchange. They are controlled only by gauge couplings and masses of the two produced objects, and thus the least model-dependent. VBF channels can become relevant for heavier objects where longitudinal $W^\pm$'s or Z's would lead to some enhancement. These channels, requiring two forward jets, will also bring in more model-dependence through the sensitivity to unknown couplings in the purely scalar sector as well as masses of intermediate states.\footnote{Concurrently, the channels with two additional jets yield cross-sections that are smaller by a factor of 2 to 4 compared to the channels without, when considering the entire phase-space.} Thus, for a complete MC generation, when at the parton showering stage,  the process
with extra 2 jets can also be added, provided that a matching scheme that avoids double-counting the matrix-element and parton-shower jets is also implemented.
The ggF production of pairs of scalars  is subdominant for various reasons: the s-channel exchanges of  $H^0$ or $A^0$ are suppressed by a $\sin^2\alpha$ factor through their couplings to the top-quark loop as a result of their dominant triplet component, and further $v_t^2$, $\mu^2$ and $\mu\,v_t$ suppressions from the final state vertices when they exist. While single photon emission is forbidden at the ggF vertex \cite{Smolyakov:1982}, the off-shell $Z$-mediated s-channel exchange is suppressed either because the corresponding vertices are forbidden, e.g. for $H^0H^0$, $A^0A^0$ and $H^0h^0$, or due to the mass degeneracy of the pair of produced scalars, \Hpp\!\!\Hmm, \Hp\!\!\Hm and $H^0A^0$ \cite{delAguila:1990yw}, or simply due to the loop suppression as compared to the DY and VBF counterparts with a Z exchange, as for $A^0h^0$.
One is left with the ggF $h^0$-mediated productions. These could be comparable to the DY or VBF contributions due to the combined effects of the enhancing top quark Yukawa coupling and  absence of suppressions at the final state vertices such as for \Hpp\!\!\Hmm, \Hp\!\!\Hm, $H^0H^0$ and $A^0A^0$ productions~---~Note though that the final state $H^0h^0$ has $v_t^2, \mu^2$ and $\mu\,v_t$ suppressions, and the final state $A^0h^0$ is not produced from the $h^0$ exchange. At any rate, the $h^0$-mediated productions bring about a model dependence on the scalar couplings, $\lambda_1$ and $\lambda_4$, see \cref{eq:Vpot}. At the LHC energies and scalar masses under consideration ($\gtrsim 200$~GeV), the $h^0$-induced ggF contributions are expected to remain subdominant as long as these scalar  couplings  remain perturbatively small~\cite{Hessler:2014ssa}. In the present paper we are interested in the charged states productions for which the ggF is subdominant~\cite{Fuks:2019clu}, and the associated neutral states production for which the ggF is suppressed for both $Z$ and $h^0$ mediation, as argued above.

\begin{table}[t!]
\begin{center}
\renewcommand{\arraystretch}{1.5}
    \caption{\label{table:prod_channels} 
       The complete list of pair and associated production channels of all the scalars of the model. The relevant subprocess initial states are indicated, as well as the corresponding main dependence on couplings and intermediate states. Bracketed ggF indicates its subdominance. See text for further discussions.
    }
    \resizebox{0.9\textwidth}{!}{
    \begin{tabular}{c|c|c} \hline\hline
        \ \ initial state (pp) \ \  & sensitivity & ~~~~final state~~~~ \\ \hline
        $q \bar q, \gamma \gamma, Z \gamma, Z Z, W^\pm W^\mp$, $[ggF]$ & gauge couplings, $H^\pm$, $\lambda_1, h^0$   &  $H^{\pm \pm} H^{\mp \mp}$ \\ 
        $q \bar q', \gamma W^\pm, Z W^\pm$ & gauge couplings, $H^{\pm\pm}, H^{\mp}$ & $H^{\pm\pm} H^{\mp}$ \\ 
        $q \bar q, \gamma \gamma, Z \gamma, Z Z, W^\pm W^\mp$, $[ggF]$ & gauge couplings,  $H^\pm, A^0$,  $2 \lambda_1 + \lambda_4, h^0$ &  $H^{\pm} H^{\mp}$  \\ \hline
        $q \bar q', \gamma W^\pm, Z W^\pm$     & gauge couplings, $H^{\pm}, A^0$  &   $H^\pm H^0$ \\ 
        $q \bar q', \gamma W^\pm, Z W^\pm$     & gauge couplings, $H^{\pm}, A^0$  &  $H^\pm A^0$  \\ 
        $q \bar q', \gamma W^\pm, Z W^\pm$     & mixing- suppressed [gauge couplings, $H^{\pm}, A^0$]  &  $H^\pm h^0$  \\ 
        $W^\pm W^\pm$     & gauge couplings, $H^{\pm}$  &   $H^{\pm\pm} H^0$ \\ 
        $W^\pm W^\pm$     & gauge couplings, $H^{\pm}$  &  $H^{\pm\pm} A^0$  \\ 
        $W^\pm W^\pm$     & mixing-suppressed [gauge couplings, $H^{\pm}$]  &  $H^{\pm\pm} h^0$  \\ \hline
        $q \bar q, Z Z, W^\pm W^\mp$, $ggF$     & gauge couplings, $H^{\pm}, A^0$, $\lambda_1 + \lambda_4, h^0$  &  $H^0 H^0$  \\ 
        $q \bar q, W^\pm W^\mp$, $[ggF]$     &  gauge couplings, $H^{\pm}$ &  $H^0 A^0$  \\ 
        $q \bar q, W^\pm W^\mp$, $ggF$    &  gauge couplings, $H^{\pm}$, $h^0$  &  $H^0 h^0$  \\ 
        $q \bar q, Z Z, W^\pm W^\mp$, $ggF$     & gauge couplings, $H^{\pm}, H^0$, $\lambda_1 + \lambda_4, h^0$  &  $A^0 A^0$  \\ 
        $q \bar q, W^\pm W^\mp$, $[ggF]$     & mixing-suppressed [gauge couplings, $H^{\pm}$]  &  $A^0 h^0$  \\ \hline
         SM %\gilbertcom{SM di-Higgs to discuss with Elizabeth}   
         & SM  &  $h^0h^0$  \\ \hline\hline
     \end{tabular}
     }
\end{center}
\end{table}

In the following, the various numerical evaluations of production cross-sections are obtained with \texttt{MadGraph} version~\verb|MG5_aMC_v3.5.3| \cite{Alwall:2014hca,Frederix:2018nkq}, relying on the parton distribution functions set\newline 
\verb|NNPDF30_nlo_as_0118_hessian|~\cite{Carrazza_2015}, and using 
UFO modules of the Type-II Seesaw Model generated with FeynRules~\cite{Christensen:2008py,Degrande:2011ua,Alloul:2013bka}.\footnote{We thank Lorenzo Basso for providing these modules and for collaboration at a very early stage of the study.}

\begin{figure}[!ht]
    \begin{center}
          %{\includegraphics[width=1\textwidth,height=.55\textwidth]{figures/image_modified.png}}
          {\includegraphics[width=1\textwidth,height=.55\textwidth]{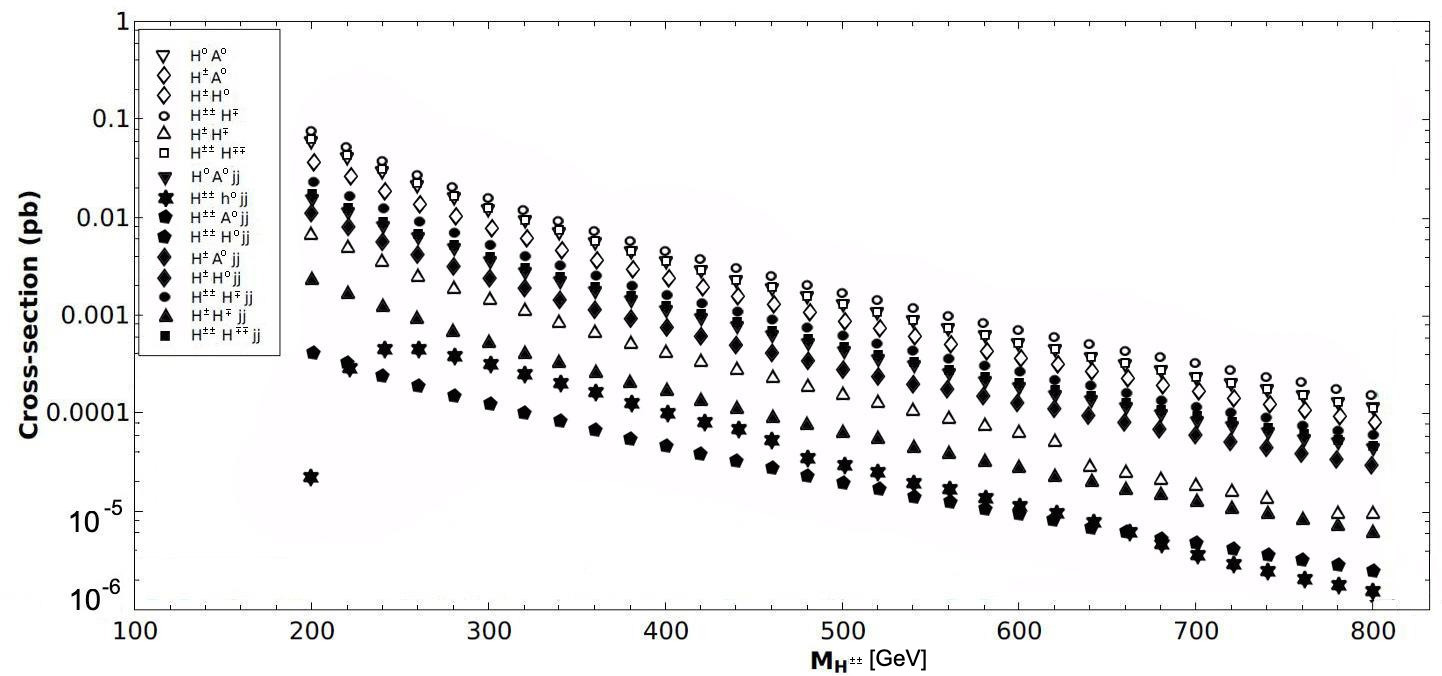}}
      \end{center}
\caption{\label{fig:prod_xs3} The leading cross-sections for pair and associated productions of the scalar bosons at 13~TeV as a function of the doubly-charged Higgs boson mass, for DY processes (open markers). Extra two jets for the same processes~---~the VBF production mode~---~are also shown (black markers). Cross-sections lower than 10$^{-6}$ pb are not shown, nor the $ggF$ contributions. The $h^0$ production cross-sections involving only SM particles as these are essentially the SM ones. See text for further discussions.} 
\end{figure}

\Cref{fig:prod_xs3} depicts the leading  $pp$ cross-sections, at center-of-mass energy $\sqrt{s}=13$~TeV, for pair and associated  productions as a function of \mHpp and assuming mass degenerate scalars. The results are obtained by convoluting leading-order matrix elements with next-to-leading-order parton distribution functions. Given the uncertainties related to the PDF choices as well as the dependencies on the factorization and renormalization scales, this level of approximation is sufficient for the purpose of assessing  a qualitative comparison of the electroweak-sector effects of the model. Further k-factors will be included in the analysis of \cref{sec:benchmark-exp}. (A more sophisticated treatment including higher-order QCD effects in the case of charged scalar productions can be found in \cite{Fuks:2019clu}.)

As expected, the DY productions
(the open markers in the upper part of \cref{fig:prod_xs3}) dominate largely over the VBF productions (the black markers) for each final state separately, but also for all final states globally with the exception of the DY-produced $\Hp\Hm$ final state (the open triangles on the plot). The suppression of the latter results from a destructive interference between the $\gamma^*$ and $Z^*$ s-channel contributions. We see from \cref{fig:prod_xs3} that the $\Hpp\Hmm$ production is dominated by the $\Hpp\Hm$ production over the whole range of masses, followed closely by the $H^0A^0$ final state, then 
by $H^\pm H^0$ and $H^{\pm} A^0$, the latter two being equal. 
This hierarchy can however slightly change in case of  mass splitting among the scalars. For relatively light \Hpp ($200-300$~GeV) varying $\Delta m_{H^{\pm\pm}, H^\pm}$ by $\pm~20$~GeV, as advocated in \cref{sec:parameters}, induces an increase (decrease) of $H^0A^0$ by 60 (50)\%, while $\Hpp\Hm$ varies only by 15-20\% both ways. 
This allows the former final state to dominate over the latter in sizable parts of the parameter space. For a given \Hpp mass scale, the effect on final states containing $H^0$ or $A^0$ is larger than that on $\Hpp\Hm$. This is expected from the generic mass hierarchy illustrated in \cref{fig:spectrum-evol}: when $\Delta m_{H^{\pm\pm}, H^\pm} >0$ one always has 
$m_{H^0}, m_{A^0} < m_{H^\pm}$  thus more phase space available for the neutral states, while 
$\Delta m_{H^{\pm\pm}, H^\pm} <0$ comes always with $m_{H^0}, m_{A^0} > m_{H^\pm}$, whence a decrease of phase space for the neutral states.
For heavier \Hpp ($\gtrsim 600$~GeV) the effect becomes less pronounced, see 
\cref{table:precent_prod_channels}. All in all, the production cross-sections for the three final states are comparable while that of $H^\pm H^0$ (and $H^{\pm} A^0$) remains somewhat below.
The DY results agree with those of Ref.~\cite{Ashanujjaman:2021txz}. Note that associated and pair production cross-sections for  $H^\pm h^0, H^0h^0, A^0h^0, H^0H^0, A^0A^0$, not shown in \cref{fig:prod_xs3}, are found to be below $10^{-6}$~pb for both the DY and VBF channels. Finally, bearing in mind PDF uncertainties, we compared for illustration the results of
\verb|NNPDF30_nlo_as_0118_hessian| to two other (leading-order) PDFs, \verb|NNPDF31_lo_as_0130| and \verb|NNPDF23_lo_as_0130_qed|, as well as when varying the renormalization and factorization scales, for the \Hpp\!\!\Hm associated production. The resulting cumulative effects of order $10 -15\%$ on the cross-section agree with typical expectations. The relative magnitudes of the various $q\bar{q}^{{}^\text{\tiny (}{}'{}^\text{\tiny )}}\!\!$-initiated final state cross-sections remain however unaffected.

\begin{table}[t!]
\begin{center}
    \caption{\label{table:precent_prod_channels} Relative variations of the DY production cross-sections when the \Hpp/\Hp mass splitting is varied in the $\pm 20$~GeV range, taking as reference points the cross-section values at degenerate masses.}
   \resizebox{0.7\textwidth}{!}{
    \begin{tabular}{|ccccc|} \hline
    \multicolumn{5}{|c|}{$\delta \sigma_{\rm DY}/\sigma_{\rm DY}$(\%) for $\Delta m_{H^{\pm\pm}, H\pm}=$ +20/-20~GeV} \\ \hline
    $m_{H^{\pm\pm}}$~(GeV) & 200 & 300 & 400 & 600 \\
    \hline
    $H^0 A^0$ & +66/-56\% & +50/-55\% &+50/-35\% & +30/-27\%\\
    $H^\pm H^0(A^0)$ & +53/-47\%& +35/-47\%& +40/-27\% & +23/-26\% \\
    \Hpp \Hm & +21/-19\%& +15/-19\%& +17/-10 \%& +8/-10\% \\ \hline
 \end{tabular}
     }
\end{center}
\end{table}

\subsection{Overview of the decay channels } 
\label{sec:decayModes}

In this section we examine in some detail the various decay channels taking into consideration all possible configurations, including (cascade) decays mediated by off-shell states. This leads to a very rich phenomenology. In addition to $h^0$, which behaves in the small \sina scenario essentially like the SM Higgs boson, the six other scalars have multiple decay modes. For each of them, we show  a panorama of the channels that open or close depending on their masses as dictated by the spectrum, cf. \cref{fig:spectrum-evol}. 

In particular, we identify a consequential sensitivity to the actual values of $\sin\alpha$, leading to significant variations of the branching fractions for some decays even when kinematically open. It should be stressed that this gives rise to a non-reducible uncertainty since the range of variation of \sina within which important effects occur, can be as small as
$[10^{-4},10^{-3}]$. Indeed, \sina parameterizes the deviation of $h^0$ from SM-likeness; but values as tiny as the ones considered here are too small to be experimentally probed in Higgs boson measurements at the LHC given the expected precision in the foreseeable future.

In the following we will illustrate the results taking three benchmark values for \mHpp and three values of \sina. We consider only tree-level results, thus ignoring possible loop induced rare decays,  and neglect the effects of running couplings or masses.

%%%%%%%%%%%%%%%%%%%%%%%%%%%%%%%%%%%%%%%%%%%%%%%%
\subsubsection{The \texorpdfstring{${H^{\pm\pm}}$}{} Higgs boson \label{sec:Hpp}}

This exotic scalar boson is characteristic of the model and has a simple decay pattern due to its unusual electric charge. It can decay to same-charge leptons through the lepton number violating operator that also generates Majorana-neutrino masses, to same-charge $W^\pm$ bosons through the kinetic term of the triplet multiplet after developing a VEV, to same-charge $W^{\pm *}$ and
$H^\pm$ in association, again induced by the kinetic term but irrespective of the VEV, and finally to same-charge $H^\pm$ pairs, induced by terms in the scalar potential.  For the considered $H^{\pm\pm}$ benchmark masses above $220$~GeV, the $\ell\ell$ and $W^{\pm}W^{\pm}$ decays are obviously on shell. However, the LNV decay width, $\Gamma(H^{\pm \pm} \rightarrow \ell^{\pm}\ell^{\pm})=m_{\nu_\ell}^2 \,m_{H^{\pm\pm}}/(16\, \pi \,v_t^2)$  (see e.g. \cite{Perez:2008ha})
being proportional to the square of a Majorana-neutrino mass will be highly suppressed with respect to the bosonic one, see \cref{eq:H++W+W+}, for the chosen value of the triplet VEV. 
For instance, taking $m_{\nu_\ell} \sim {\cal O}(1)$~eV and $m_{H^{\pm\pm}} \gtrsim 200$~GeV,
the two widths are comparable for $v_t \simeq 4.5 \times 10^{-4}$, while for our benchmark value of $v_t \simeq 0.1$ the $W^{\pm}W^{\pm}$ channel overshoots the $\ell\ell$ channel by ten orders of magnitude!
The two other decay channels, with at least one charged Higgs, will have one particle decaying 
off shell because of the chosen mass splitting requirement between the charged and doubly-charged Higgses, see \cref{eq:Deltam_20}, as explained at the end of \cref{sec:parameters}. 
Nonetheless, they also largely overwhelm the two-lepton channel.

\begin{figure}[!ht]
    \begin{center}
      {\includegraphics[width=0.49\textwidth]{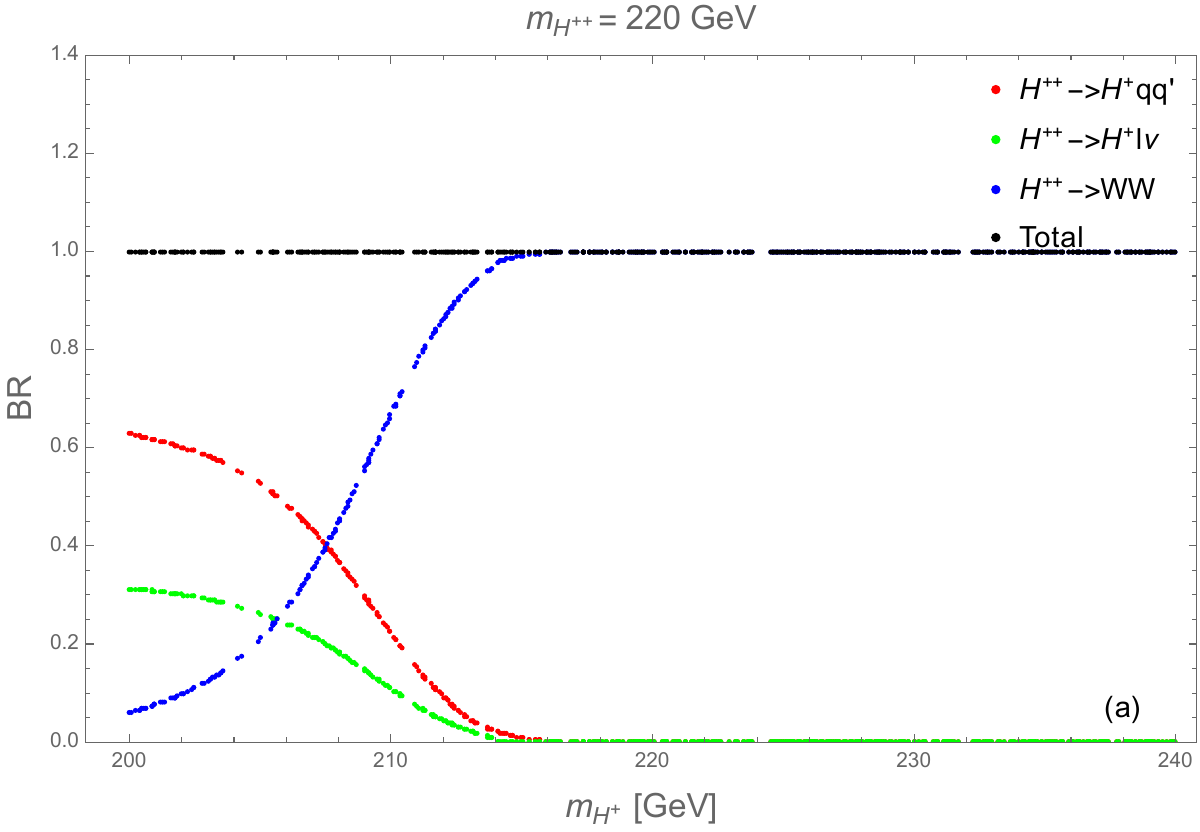}}
      {\includegraphics[width=0.49\textwidth]{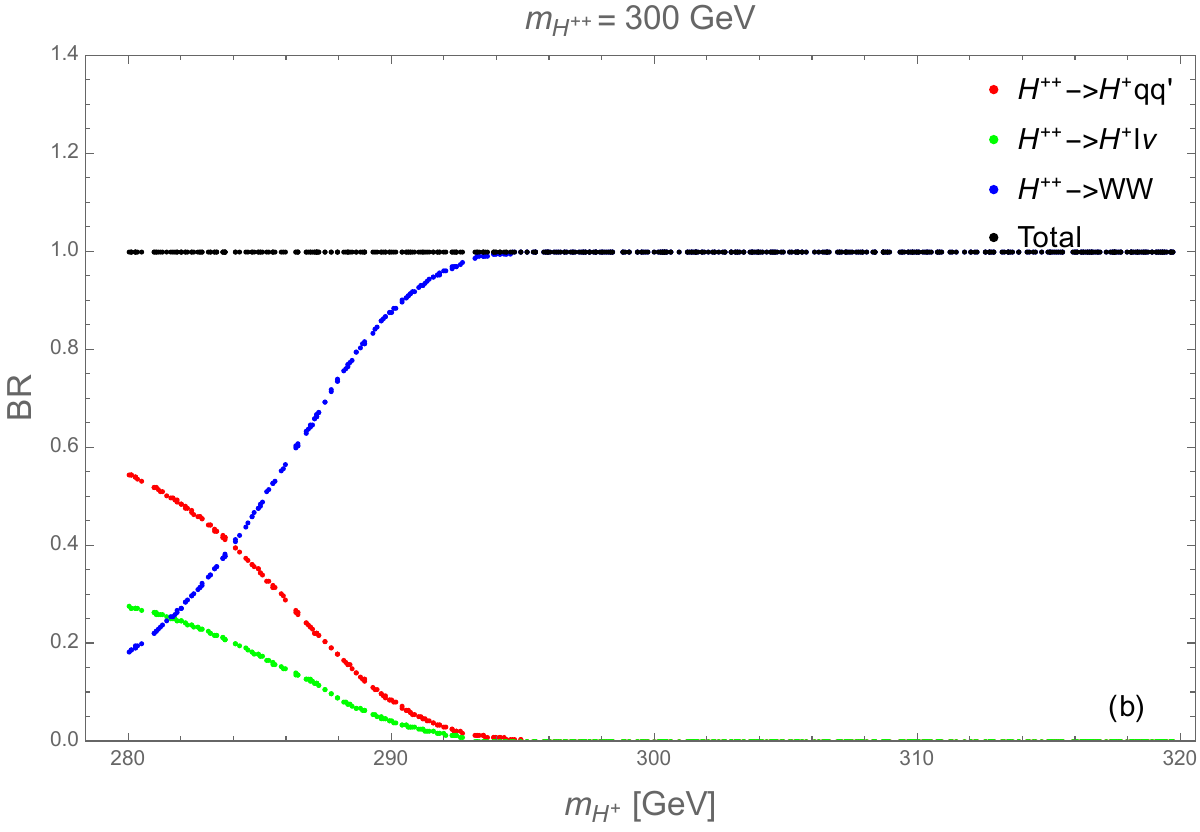}}   
      {\includegraphics[width=0.49\textwidth]{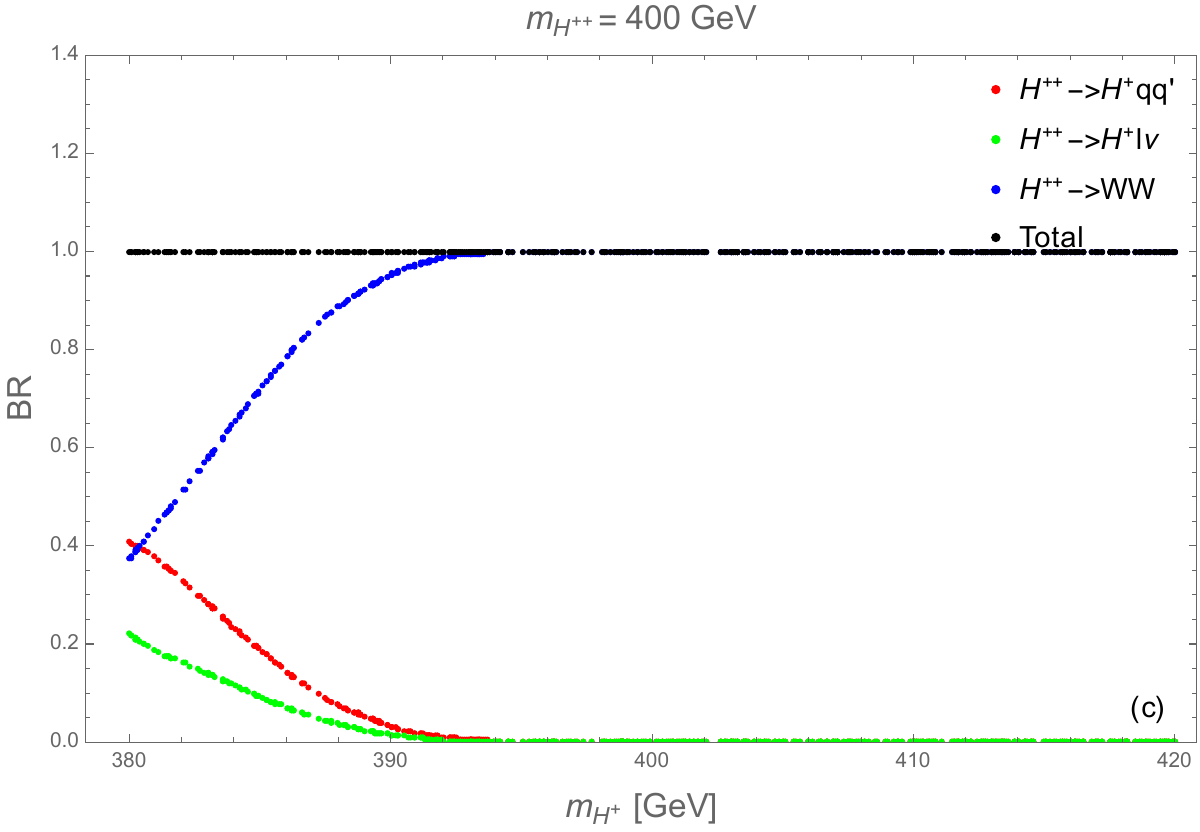}}
\end{center}
\caption{\label{fig:BR_Hpp}Decay branching ratio for the doubly-charged Higgs boson for a mass of 220~GeV, 300~GeV and 400~GeV ((a), (b), (c)) as a function of the singly-charged Higgs boson mass. 
All the decay widths are independent of \sina.
}
\end{figure}

As seen from \cref{fig:BR_Hpp}, the same-charge $W^\pm W^\pm$  on-shell decay dominates over most of the charged-Higgs mass range despite a $v_t^2$ suppression in the width, see \cref{eq:H++W+W+}. However, for a singly-charged Higgs on the lighter
side of the spectrum, three-body decays to an on-shell charged Higgs and two fermions, mediated by an off-shell boson, dominate. The latter come mainly from an off-shell $W^{\pm*}$. This results from a balance between the non-suppression of the coupling ($\cos^2\beta' \simeq 1$ in \cref{eq:H++W+H+}) and the suppression due to off-shellness. We also checked that approximate analytical expression for the off-shell $H^{\pm \pm} \to H^\pm W^{\pm *}$ decay, \cite{Aoki:2011pz}, reproduces very well our numerical evaluation using \texttt{MadGraph}.
We show separately the $H^{\pm \pm} \rightarrow H^\pm q \bar q'$ and $H^{\pm \pm} \rightarrow H^\pm \ell \nu$ contributions.  A factor of roughly 2 is seen between these two channels for the corresponding branching fractions, across all considered mass ranges.
 This is to be expected since, (neglecting effects from off-diagonal CKM and PNMS matrix entries) there are two contributions from the light quarks (times the color factor~3) and three contributions from the leptons, the $W ^\pm$ coupling being the same to all.
The decay channel $H^{\pm \pm} \rightarrow H^\pm H^{\pm *}$ with an off-shell $H^\pm$ is also present though significantly suppressed. This is due to a $v_t$ suppression, together with small values of $\mu$, in the coupling  $H^{\pm \pm} H^\mp H^{\mp }$ originating from the potential \cref{eq:Vpot}, as compared to the momentum-dependent coupling $H^{\pm \pm} H^\mp W^{\mp }$ originating from the kinetic terms; even for on-shell the relative suppression in the widths  is $\lesssim (v_t/m_{H^{\pm \pm}})^2$, cf. \cref{eq:H++W+H+,eq:H++H+H+}. 
Let us also note that, in contrast with what we will see for the other states, there is no sensitivity to \sina in the main \Hpp decays,  the coupling being either the gauge coupling, for the $H^{\pm \pm}  H^\mp W^\mp$ vertex, or  $v_t$ times the square of the gauge coupling, for the $H^{\pm \pm}  W^\mp W^\mp$ vertex.

The mass window within which the $H^{\pm \pm} \rightarrow H^{\pm}W^{\pm *}$ decay channel dominates narrows down with increasing masses of the doubly-charged Higgs boson. This trend can be seen in \cref{fig:BR_Hpp} for $m_{H^{\pm \pm}} = 220$, $300$ and $400$~GeV, and can be understood as resulting from the upper bound on the allowed mass splitting $|\Delta m_{H^{\pm\pm}, H^\pm}|$: with increasing $m_{H^{\pm \pm}}$, the level of off-shellness remains the same while the phase space for $W^\pm W^\pm$ increases. 

It is, however, important to keep in mind that the dominance of $H^{\pm \pm} \rightarrow H^{\pm}W^{\pm *}$ in the lower part of the $H^\pm$ mass spectrum can potentially mitigate the exclusion limits on four-$W^\pm$ final states originating from doubly-charged Higgs pair decays studied in \cref{sec:HppHpp}. Indeed, in this case cascade decays of $H^{\pm}$ can occur dominantly, leading to  higher $W^\pm$, jet or lepton multiplicities. We discuss further this point in \cref{sec:further_discussions}.

%%%%%%%%%%%%%%%%%%%%%%%%%%%%%%%%%%%%%%%%%%%%%%%%
\subsubsection{The \texorpdfstring{\Hp}{} Higgs boson \label{sec:Hp}}

\begin{samepage}
\begin{figure}[!ht]
    \begin{center}
       {\includegraphics[width=0.49\textwidth]{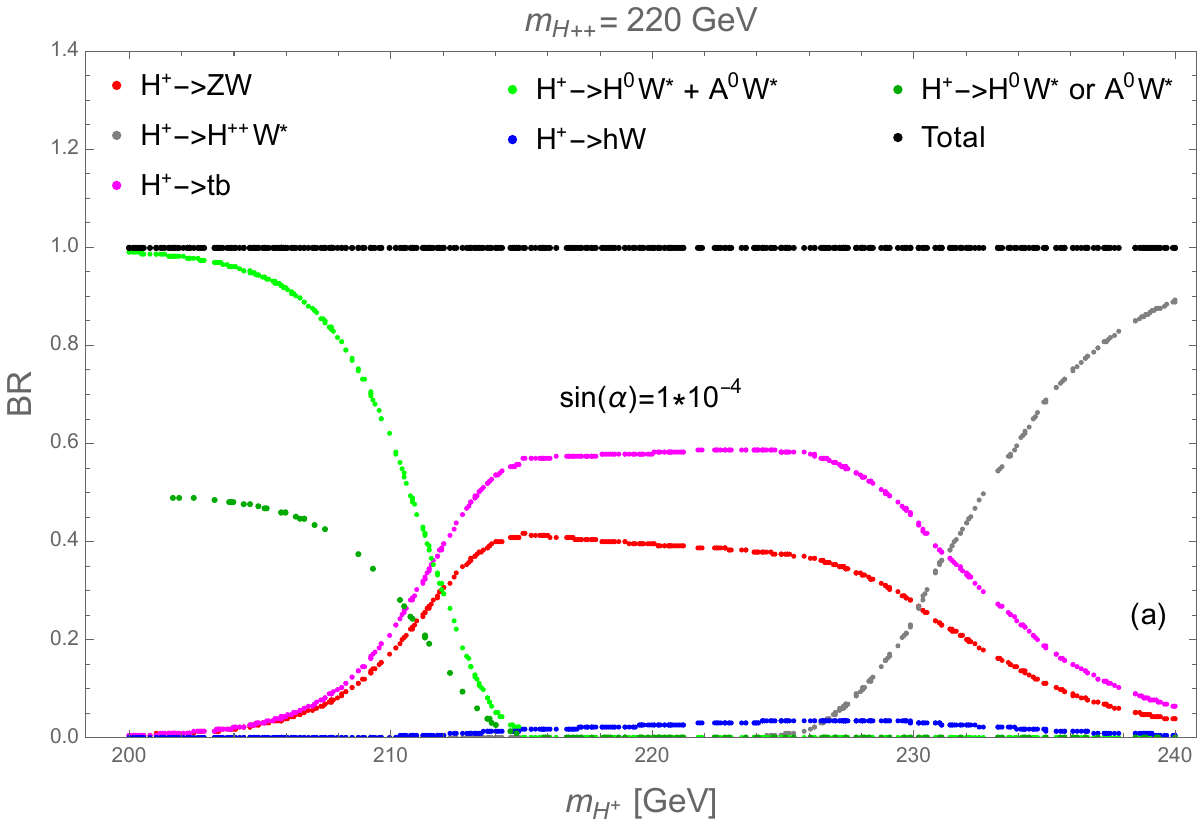}}
      {\includegraphics[width=0.49\textwidth]{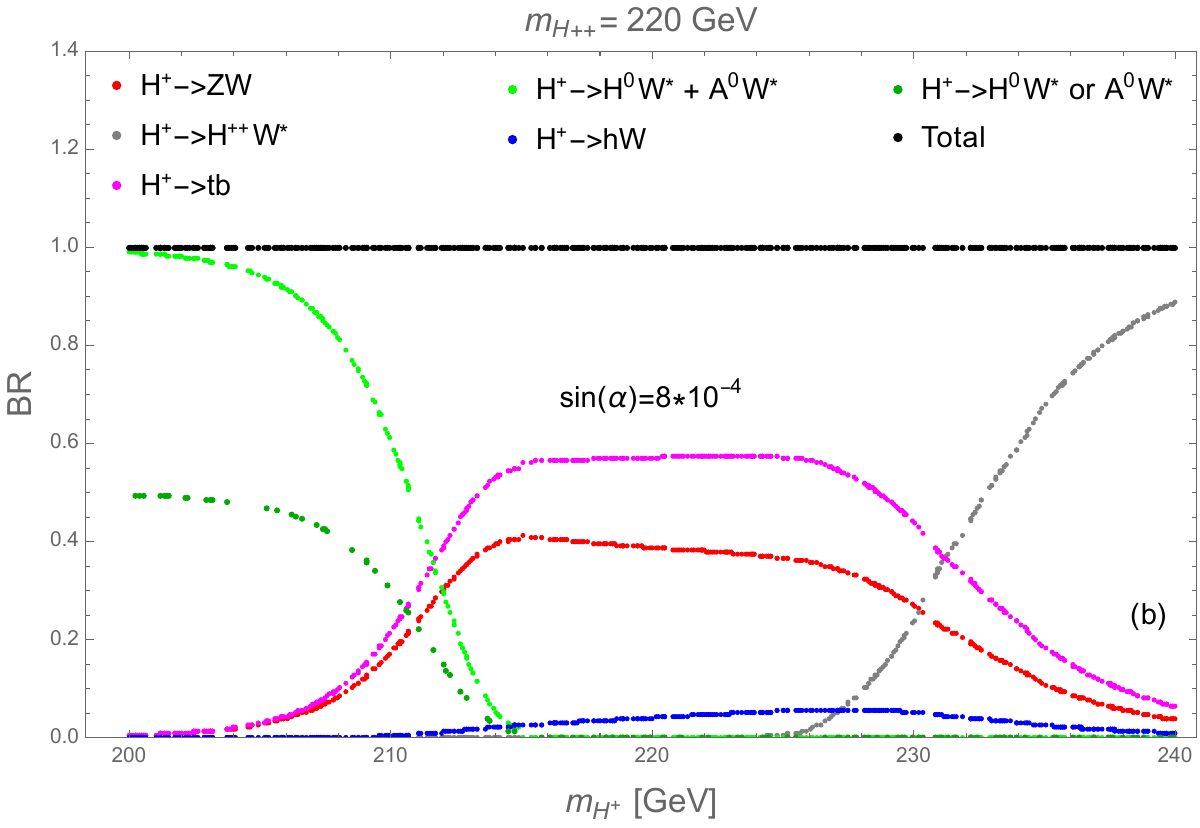}}   
      {\includegraphics[width=0.49\textwidth]{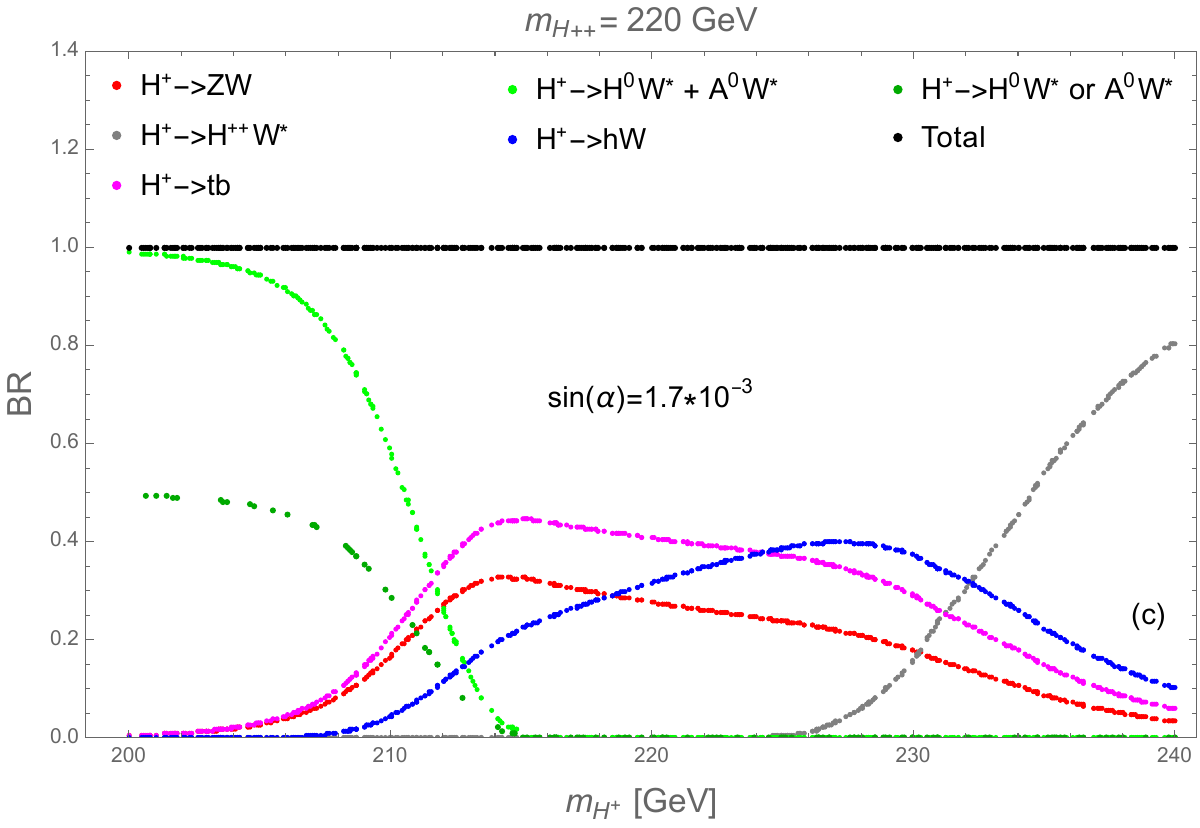}}
      \end{center}
\caption{\label{fig:BR_Hp1} Decay branching ratios of the singly-charged Higgs boson
as a function of its mass, 
for doubly-charged Higgs boson mass of 220~GeV, and $\sin\alpha\!\!=\!\!10^{-4},  8\times10^{-4}, 1.2\times10^{-3} $. }      
\end{figure}
\nopagebreak
\begin{figure}[!hb]
  \begin{center}
       {\includegraphics[width=0.49\textwidth]{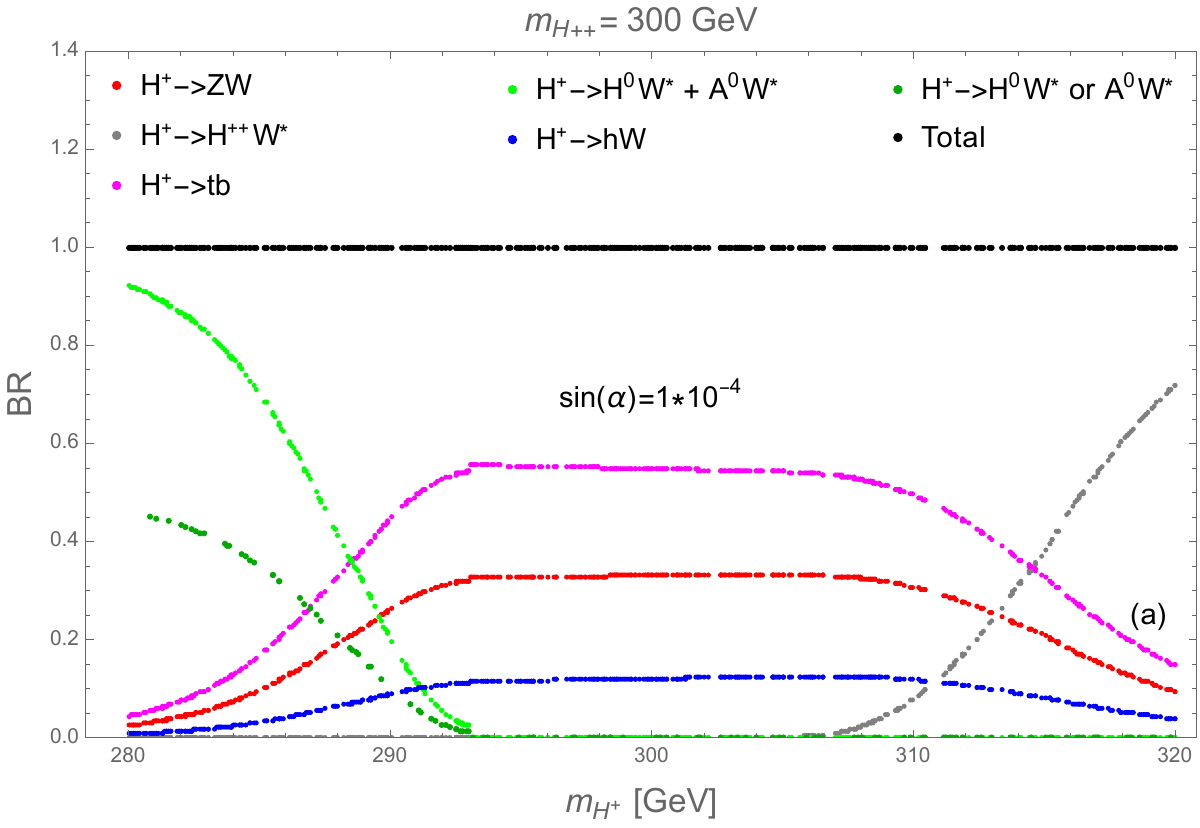}}
      {\includegraphics[width=0.49\textwidth]{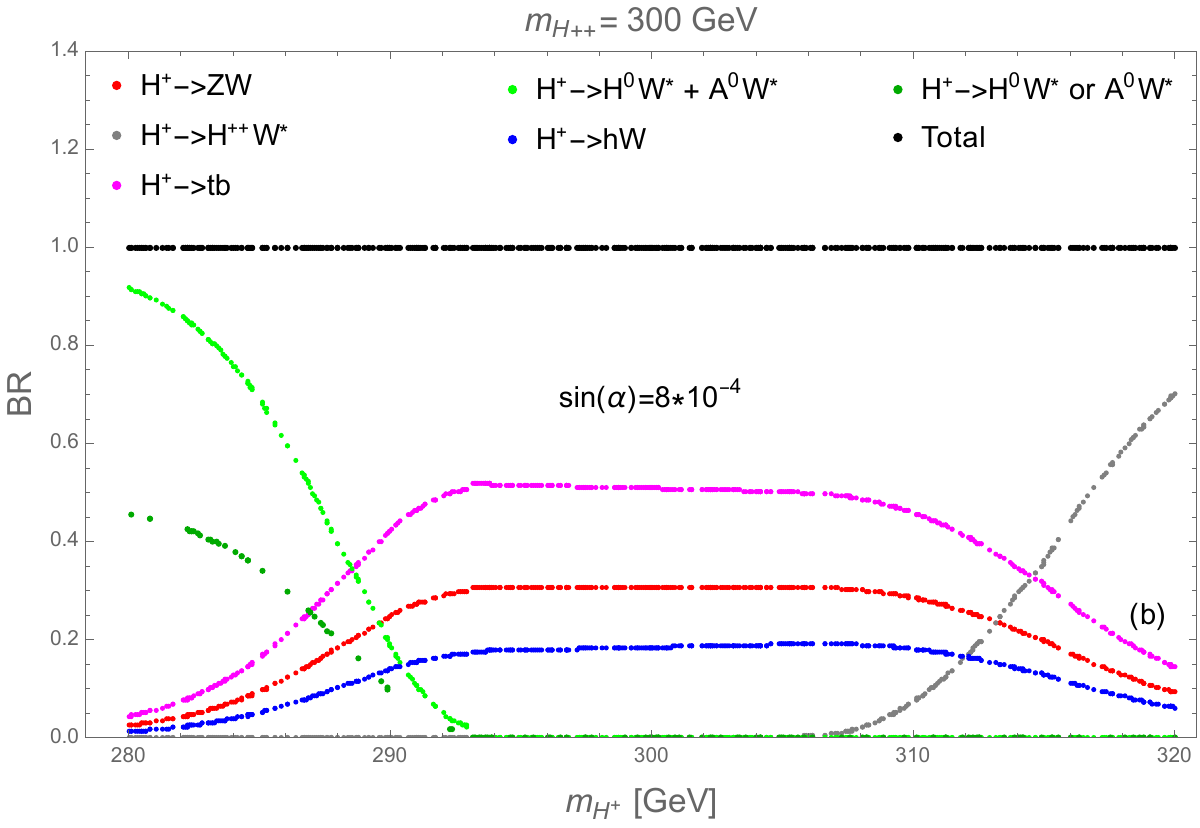}}   
      {\includegraphics[width=0.49\textwidth]{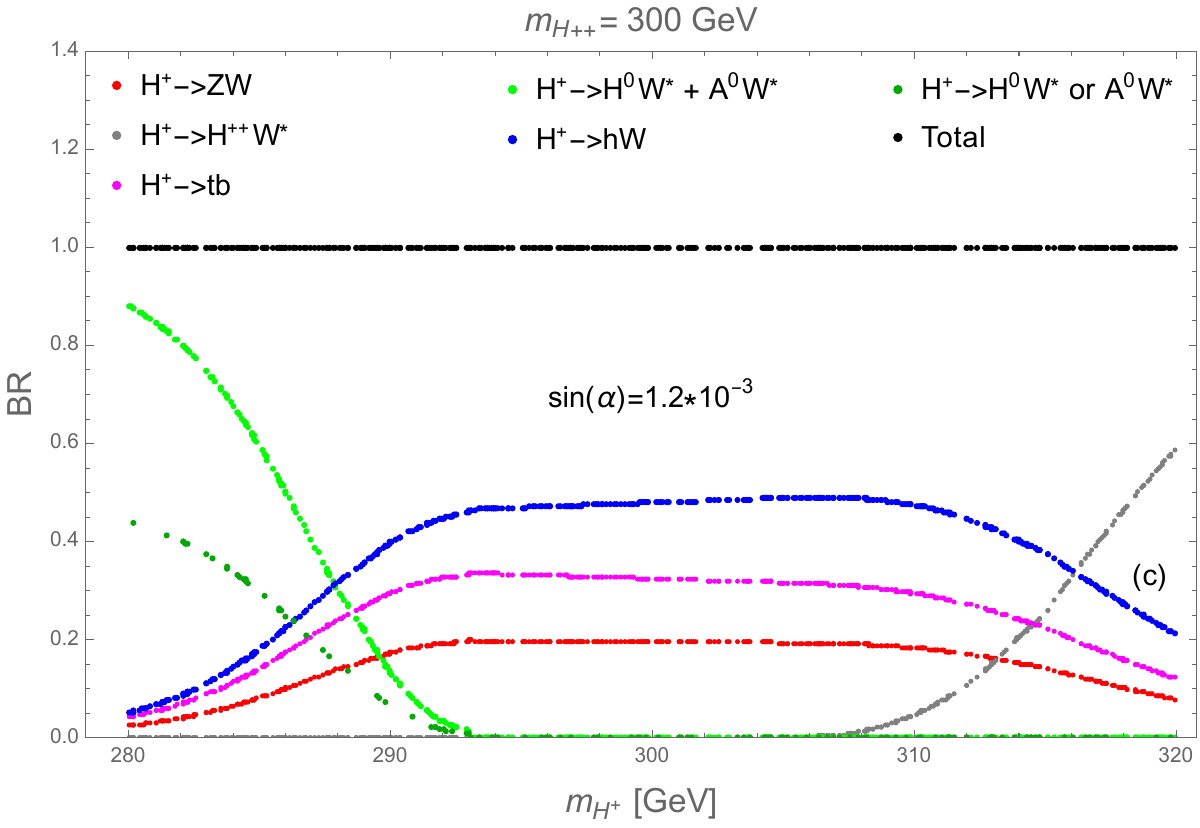}}
   \end{center}
\caption{\label{fig:BR_Hp2} Decay branching ratios of the singly-charged Higgs boson
as a function of its mass, 
for doubly-charged Higgs boson mass of 300~GeV, and $\sin\alpha\!\!=\!\!10^{-4},  8\times10^{-4}, 1.7\times10^{-3} $. }
\end{figure}
\end{samepage}
\begin{figure}[!ht]
\begin{center}
       {\includegraphics[width=0.49\textwidth]{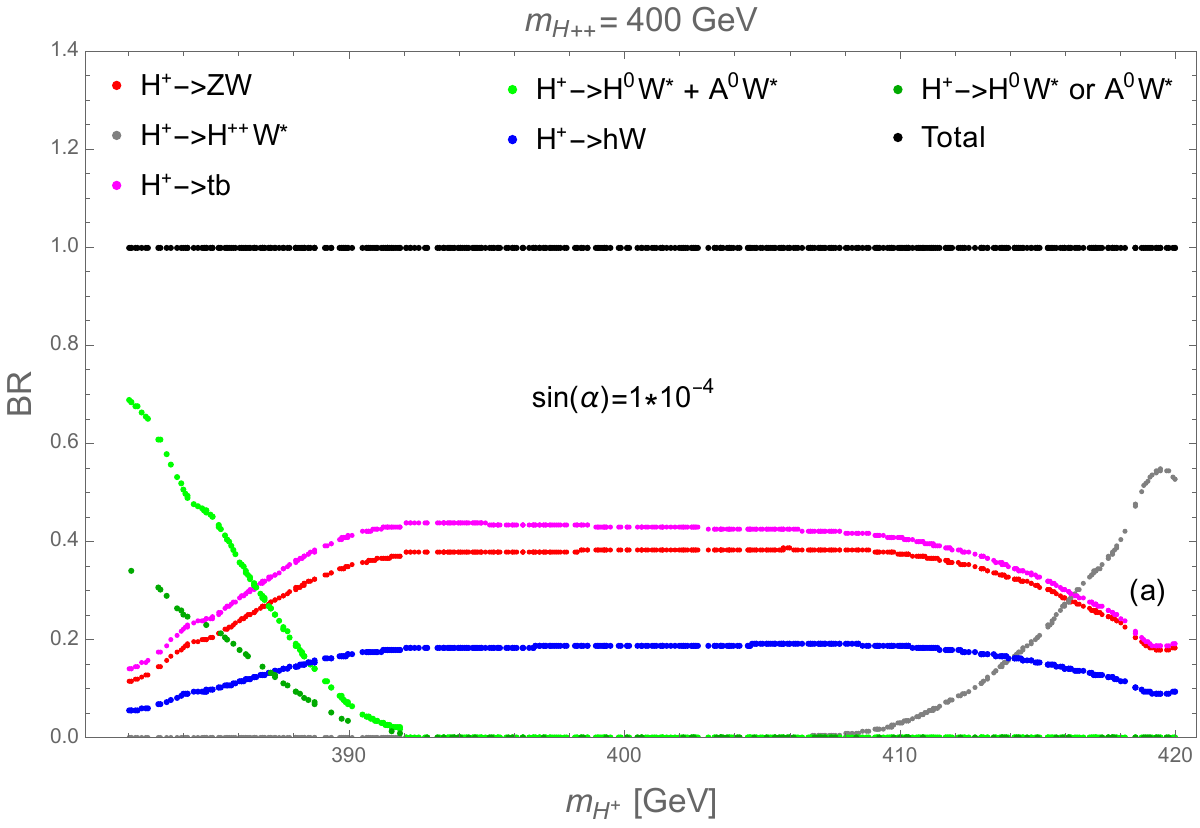}}
      {\includegraphics[width=0.49\textwidth]{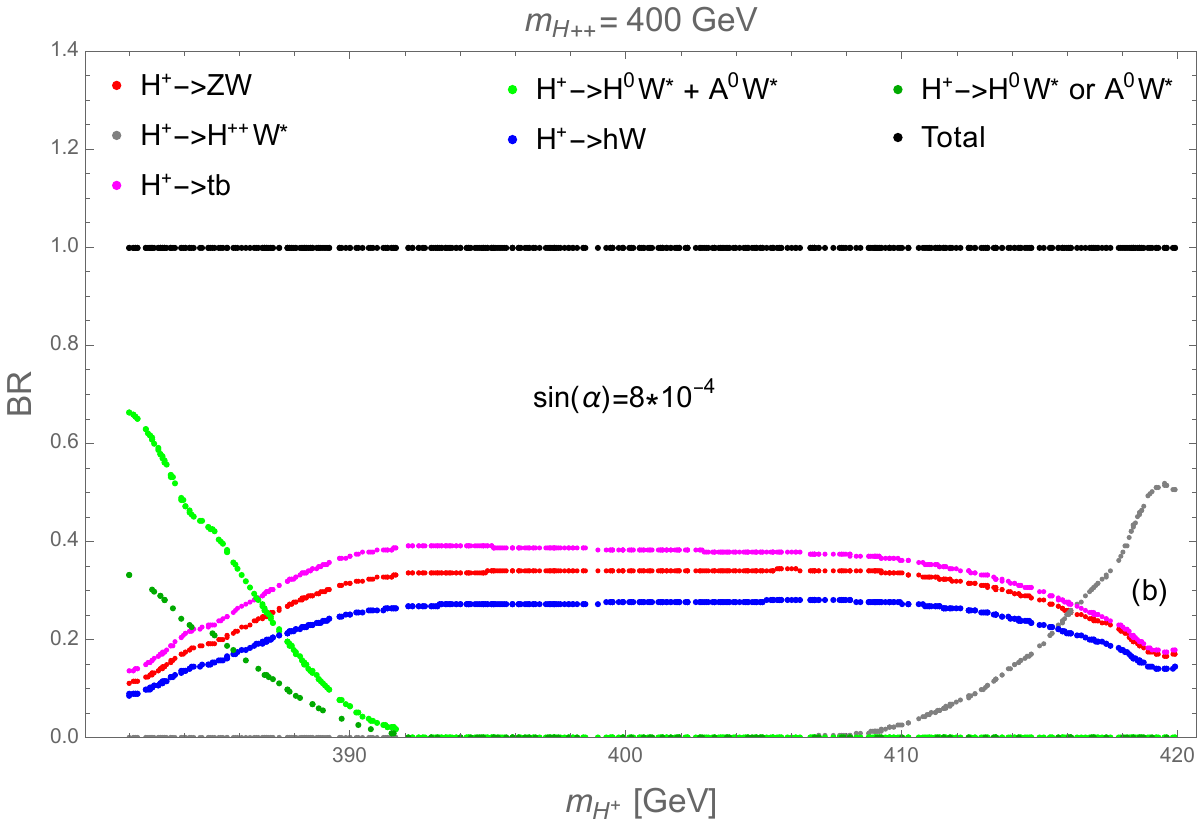}}   
      {\includegraphics[width=0.49\textwidth]{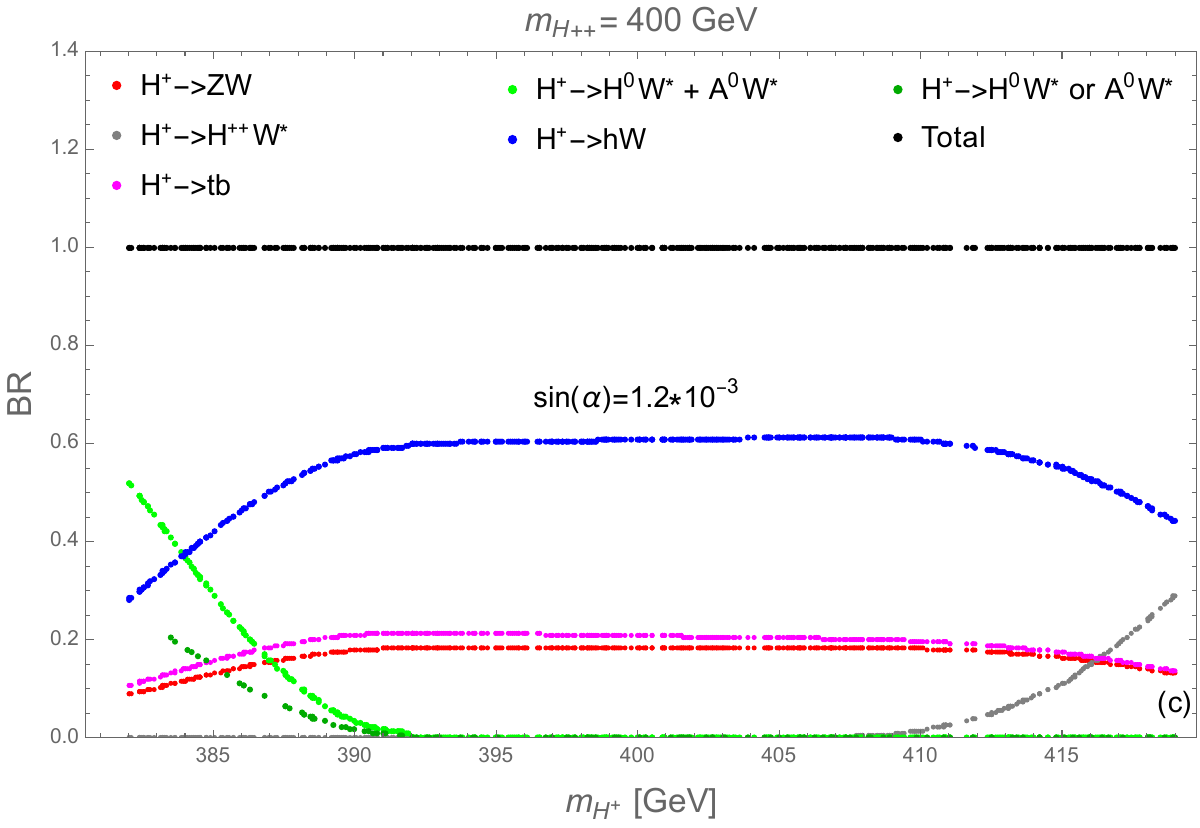}}
\end{center}
\caption{\label{fig:BR_Hp3} Decay branching ratios of the singly-charged Higgs boson
as a function of its mass, 
for doubly-charged Higgs boson mass of  400~GeV, and $\sin\alpha\!\!=\!\!10^{-4},  8\times10^{-4}, 1.2\times10^{-3} $. }
\end{figure}

The singly-charged Higgs boson shows more complex decay patterns. Note first that for the same reasons as in the previous section concerning the suppression of $H^{\pm\pm} \to \ell^\pm \ell^\pm$, here too the LNV decays $H^\pm \to \ell^\pm \nu_\ell$ are  highly suppressed for the relatively large chosen value $v_t=0.1$.

Close to the \Hp and \Hpp mass degeneracy, three channels compete whenever kinematically open: $W^\pm Z$, $tb$ and $h^0W^\pm$, as can be seen in the central regions of \cref{fig:BR_Hp1,fig:BR_Hp2,fig:BR_Hp3}. In addition, the decays  
$H^\pm \to H^0 W^{\pm *}$, $A^0 W^{\pm *}$ and $H^{\pm\pm} W^{\mp *}$, with one off-shell $W^\pm$, that would lead to further cascade decays, become dominant for lighter/heavier \Hp, but still within the allowed $\Delta m_{H^{\pm\pm}, H^\pm}$ range. However, for reasons similar to the ones noted in the previous subsection, the mass window where this occurs becomes smaller with increasing \mHpp. This variety of decay modes should be kept in mind when interpreting experimental exclusion limits based on one given decay mode.

Another important feature here is the high sensitivity to \sina, in particular for
the on-shell modes. For instance, a variation of \sina in a range of tiny values, such as $\left[10^{-4}, 10^{-3}\right]$ 
shown in the figures, can change significantly the relative contributions of the $W^\pm Z$ and $t b$ decay modes on one hand, as compared to the contribution of the $h^0 W^\pm$ decay mode on the other. This is illustrated in \cref{fig:BR_Hp1,fig:BR_Hp2,fig:BR_Hp3} for three values of \sina. 

Theoretically, the sensitivity of the branching fractions to \sina can be easily understood from \cref{eq:H+ZW+,eq:H+tb,eq:H+h0W+}. The $W^\pm Z$ and $t b$ channels have only
\mHp as a varying parameter, and they are both \sina independent and have the same suppression factor $v_t^2/v_d^2$ with respect to the $h^0 W^\pm$ channel. Thus their relative contributions  depend mainly on the available phase space for
$W^\pm Z$ and $tb$. Obviously the latter is always greater than the former and they tend to come close to each other for increasing \mHp,  irrespective of the chosen value of \sina. This is clearly seen from the trend on \cref{fig:BR_Hp2,fig:BR_Hp3}.
The $h^0W^\pm$ channel depends on \sina and \mHp. The squared coupling in \cref{eq:H+h0W+} vanishes for $\sin \! \alpha =\frac{v_t}{\sqrt{v_d^2 + v_t^2} }$ ($\simeq 4 \times 10^{-4}$ in our case), and is well approximated by $2 \sin^2 \! \alpha$ when   $\sin \! \alpha \gg \sin \! \beta'$ ($\simeq 5.7 \times 10^{-4}$ in our case). It follows that even for \sina as small as ${\cal O}(10^{-3})$
%$\sin \! \alpha \gtrsim 10^{-3}$ 
the squared coupling in $\Gamma_{H^+ \to h^0 W^+}$ starts overcoming the suppression factor $v_t^2/v_d^2$ present for the $W^\pm Z$ and $tb$ channels, leading to potential dominance of the $h^0 W^\pm$ channel. The effect should be further convoluted with the available phase space depending on \Hp as seen in the bottom plots of \cref{fig:BR_Hp1,fig:BR_Hp2,fig:BR_Hp3}.
For increasing \mHpp, one can also find \sina configurations where the branching ratios of the three channels evolve to a democratic balance 
of roughly $1/3$ each. \Cref{fig:BR_Hp3}(b) comes close to such a configuration which would occur for \sina slightly different from the one we considered, $\sin\alpha = 2 v_t/v_d \simeq 8.13 \times 10^{-4}$, as previously seen in~\cite{Perez:2008ha}, before a sudden drop of the $tb$ channel at high mass. 

In summary, \Hp features more decay channels than \Hpp. Its experimental search in the $W^\pm Z$ channel is prone to more uncertainties than is \Hpp in the $W^\pm W^\pm$ channel. Not only are there dominant off-shell decays both on the lighter and heavier part of the mass spectrum, but there is also
an important sensitivity to \sina in the intermediate mass spectrum. 
As previously stressed, this peculiar feature leads to an irreducible uncertainty as the sensitivity is in regions of \sina well below what can be triggered by the present ATLAS and CMS precision of the SM-like Higgs couplings.  More on this in  \cref{sec:further_discussions}

%%%%%%%%%%%%%%%%%%%%%%%%%%%%%%%%%%%%%%%%%%%%%%%%
\subsubsection{The CP-even Higgs boson \texorpdfstring{$H^{0}$}{}
\label{sec:H0}}

In the scenario under consideration,  the $H^0$ state is heavier than the SM-like Higgs boson. It can decay to pairs of gauge bosons, $h^0$'s, fermions, and possibly to a charged scalar in association with an off-shell $W^{\pm}$.  %leading to:
%\begin{equation}
%H^0 \to W^{\pm} W^{\mp}, Z Z, h^0 h^0, 
%H^{\pm} W^{\mp}, t \bar{t}
%\end{equation}
%and other fermion pairs. 
Note however that for reasons similar to the ones given in the previous sections,  the LNV  invisible decays $H^0 \to  \nu_\ell \nu_\ell$ are  highly suppressed for the relatively large benchmark value $v_t=0.1$ considered in the present work.

\begin{samepage}
\begin{figure}[!htb]
    \begin{center}
       {\includegraphics[width=0.49\textwidth]{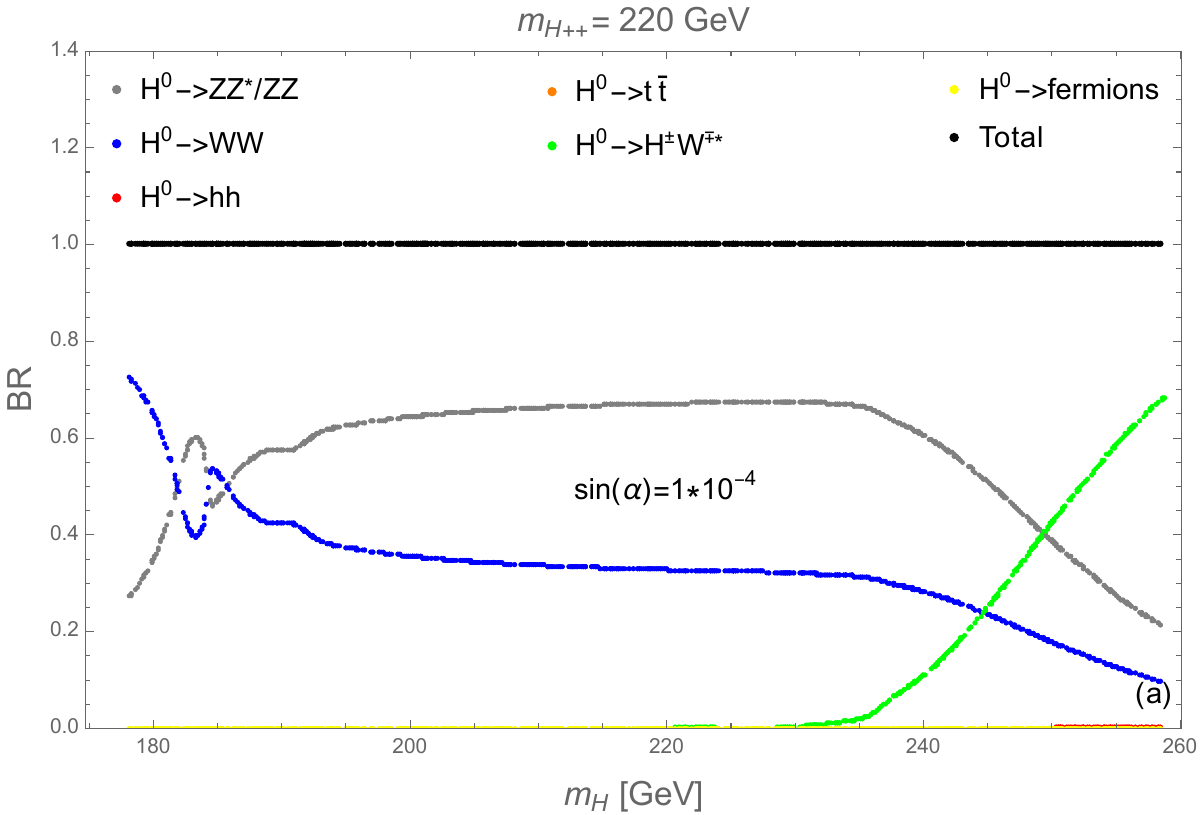}}
      {\includegraphics[width=0.49\textwidth]{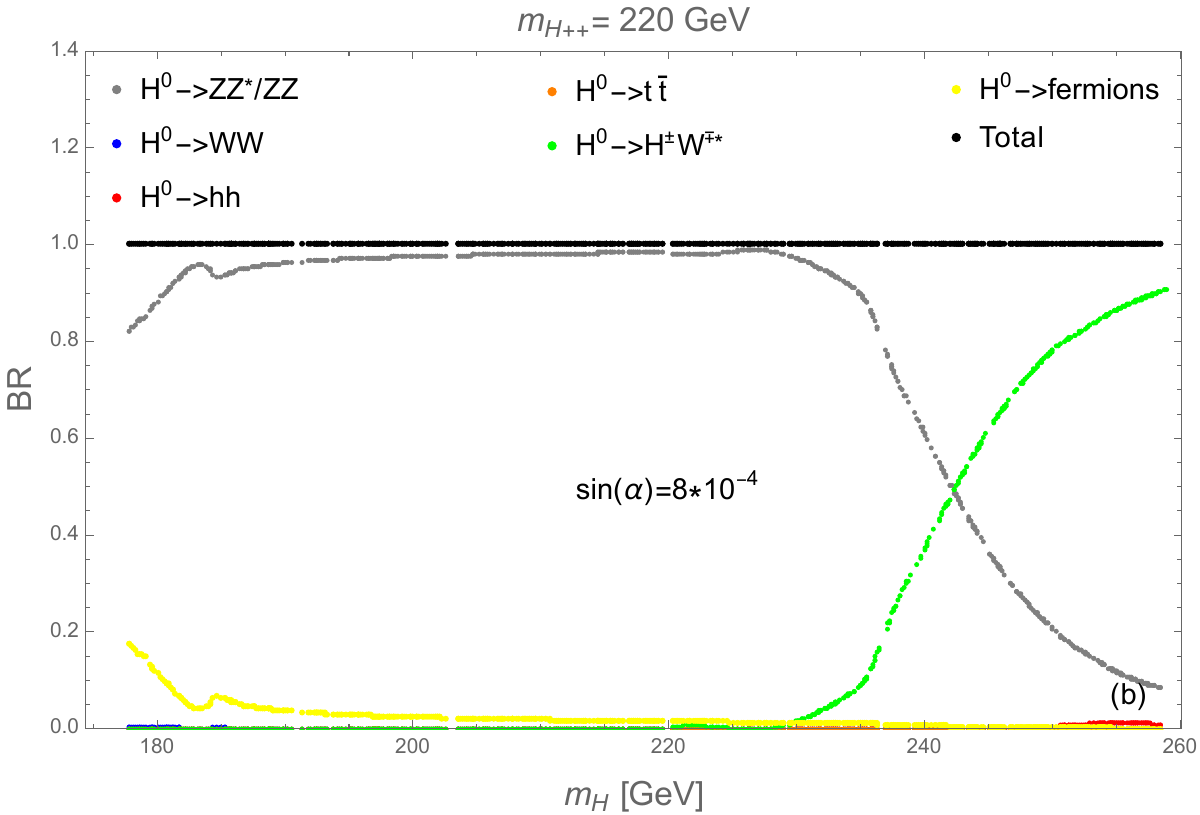}}   
      {\includegraphics[width=0.49\textwidth]{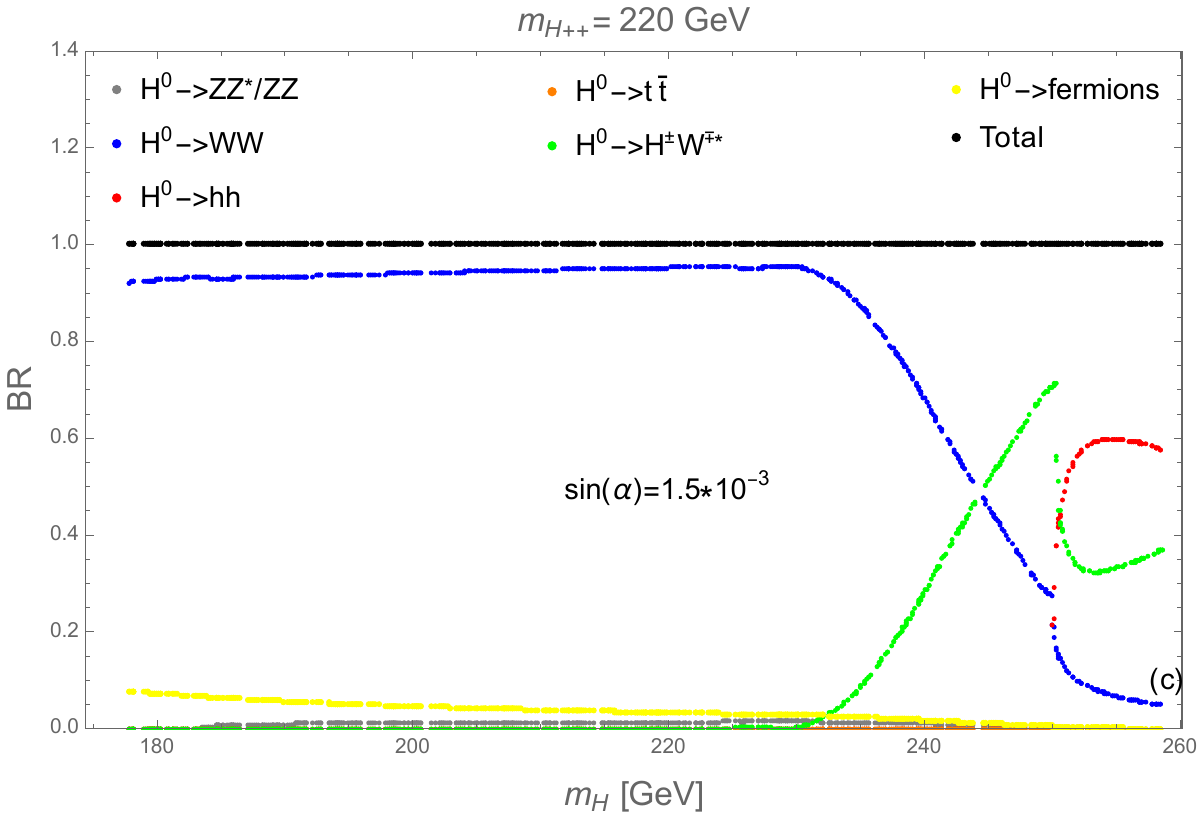}}
\end{center}
\caption{\label{fig:BR_H1}Decay branching ratios of the CP-even neutral Higgs boson
as a function of its mass, for doubly-charged Higgs boson mass  of 2 20~GeV and different $\sin\alpha$ values; `fermions' indicate the sum over all light fermions. }
\end{figure}
\nopagebreak
\begin{figure}[!htb]
\begin{center}
       {\includegraphics[width=0.49\textwidth]{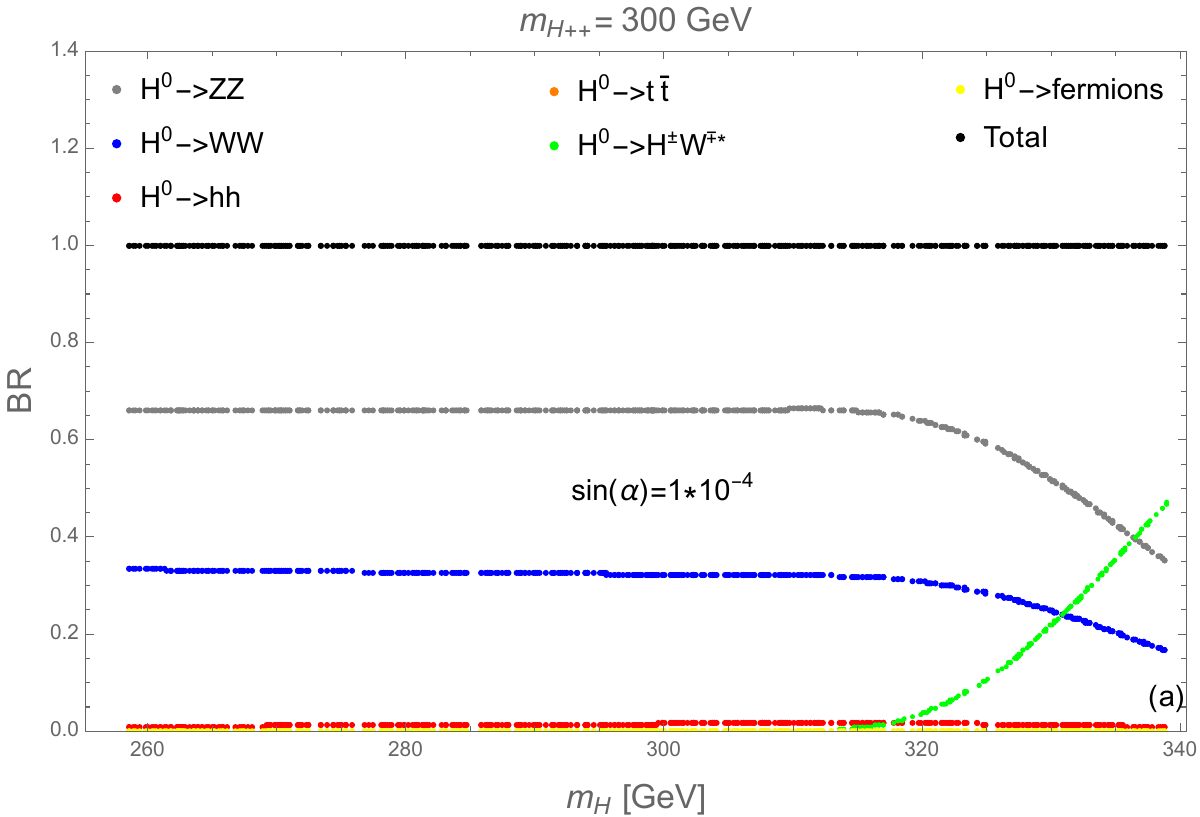}}
      {\includegraphics[width=0.49\textwidth]{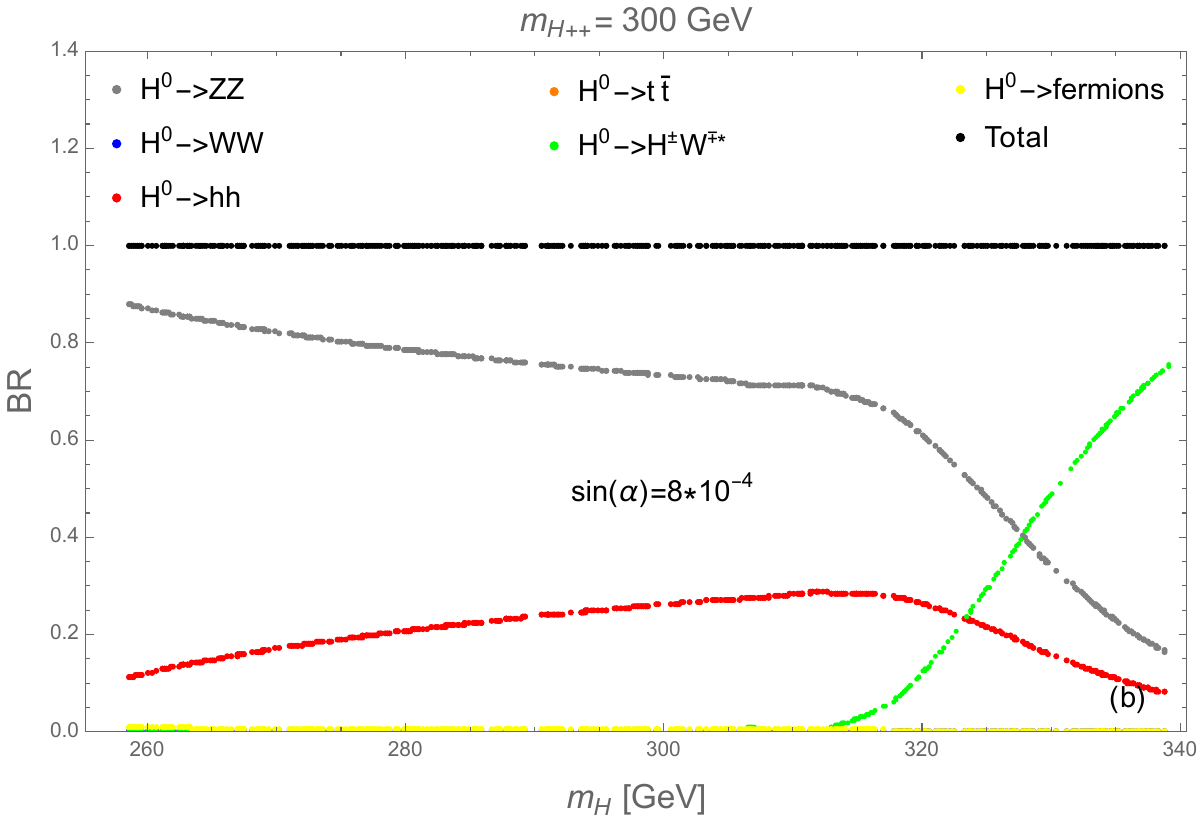}}   
      {\includegraphics[width=0.49\textwidth]{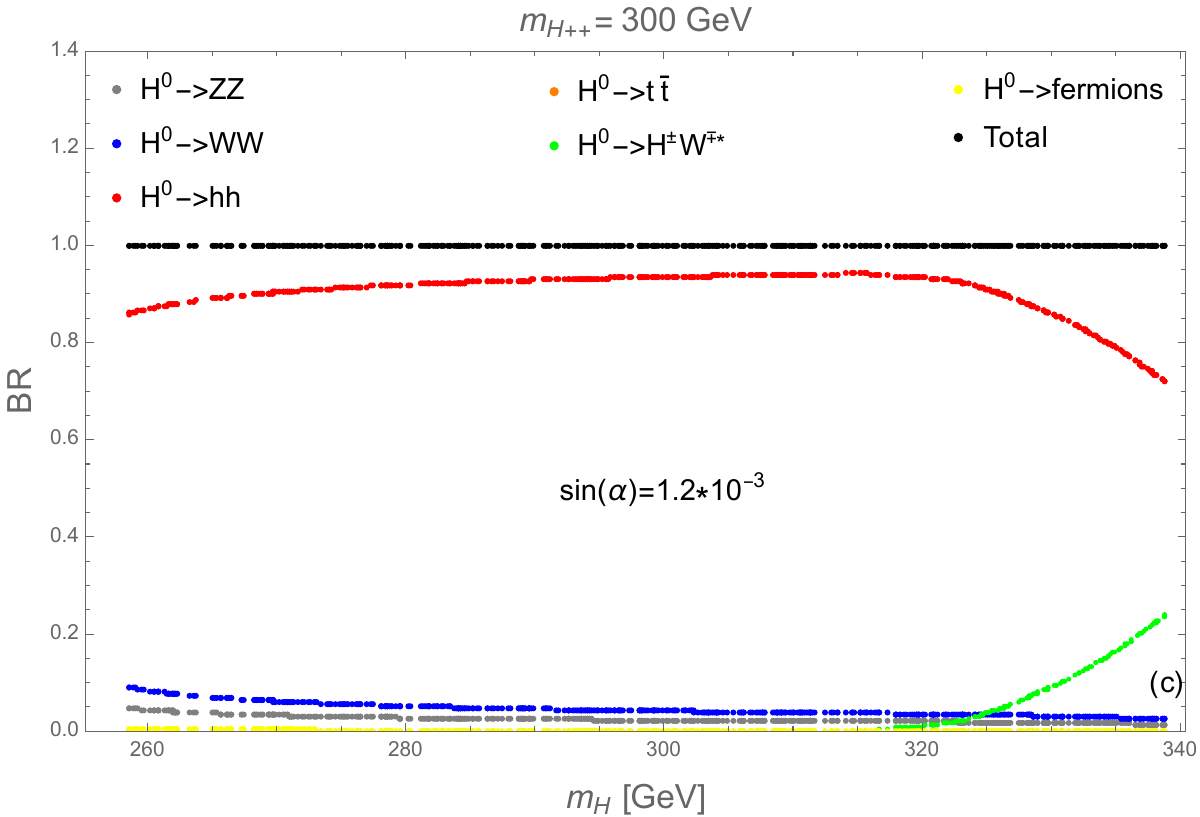}}
\end{center}
\caption{\label{fig:BR_H2}Decay branching ratios of the CP-even neutral Higgs boson
as a function of its mass, for doubly-charged Higgs boson mass  of 300~GeV and different \sina values; `fermions' indicate the sum over all light fermions.}
\end{figure}
\end{samepage}
\begin{figure}[!thb]
\begin{center}
      {\includegraphics[width=0.49\textwidth]{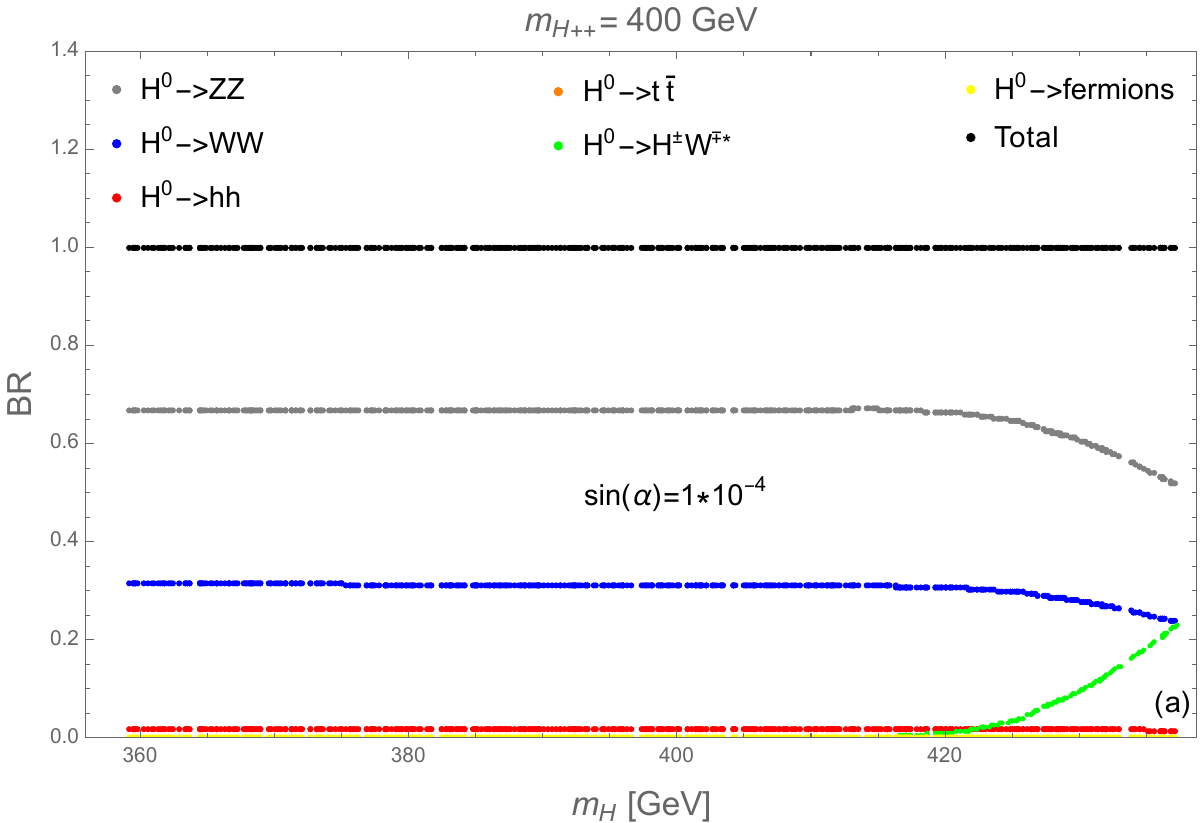}}
      {\includegraphics[width=0.49\textwidth]{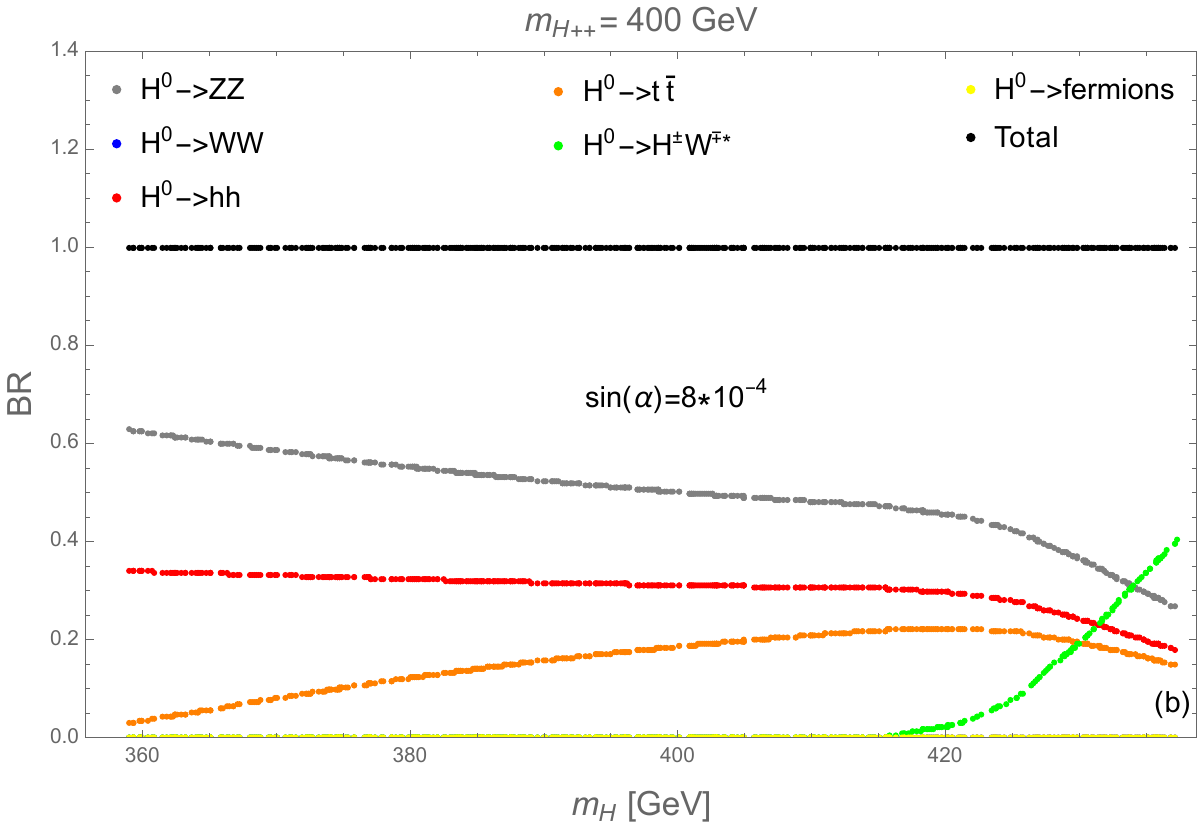}}   
      {\includegraphics[width=0.49\textwidth]{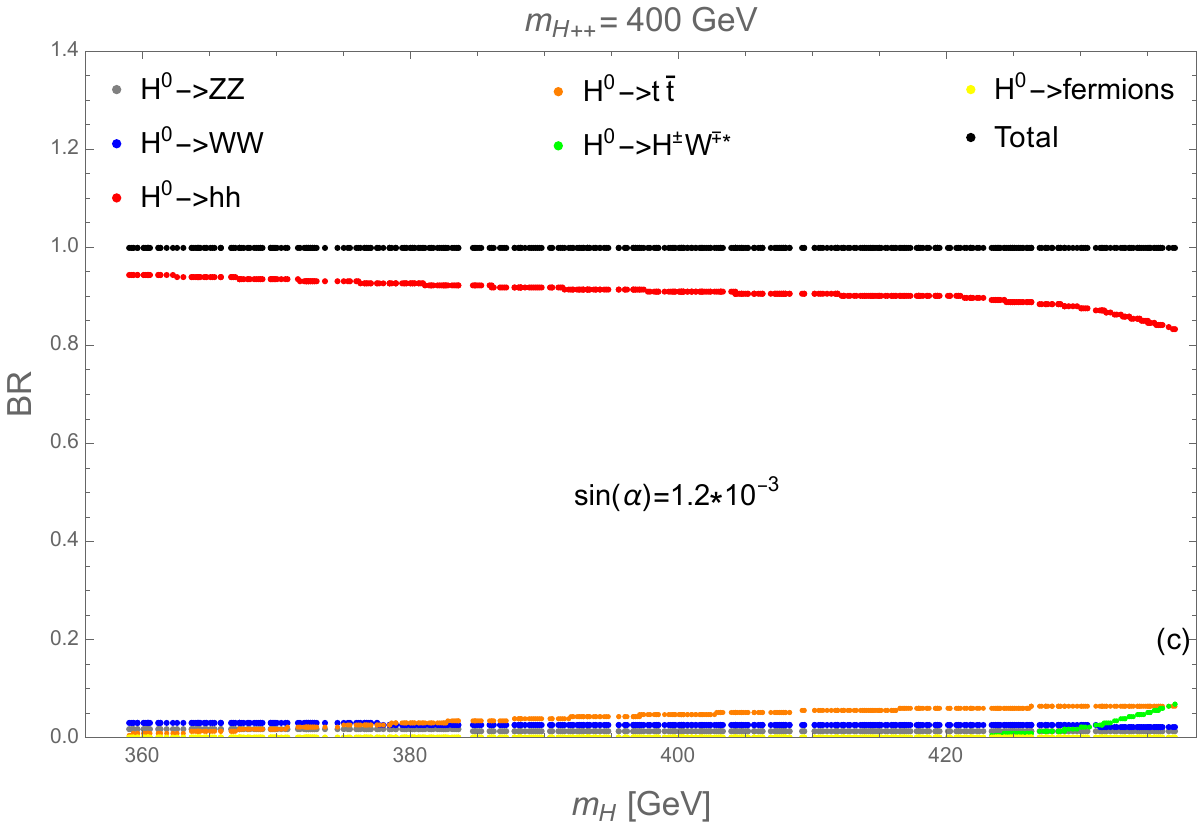}}
\end{center}
\caption{\label{fig:BR_H3}Decay branching ratios of the CP-even neutral Higgs boson
as a function its mass, for doubly-charged Higgs boson mass  of 400~GeV and different $\sin\alpha$ values; `fermions' indicate the sum over all light fermions.}
\end{figure}

The decay pattern of $H^0$, like that of \Hp, also varies with \sina 
(see \Cref{fig:BR_H1,fig:BR_H2,fig:BR_H3}). The dominant BRs are: decays to $ZZ$ for smaller \sina and for all \Hpp masses; decays to $W^\pm W^\mp$ for larger \sina at smaller \mHpp, and to $h^0 h^0$ at larger \mHpp  as soon as decays to on-shell SM Higgs pairs become available. The sensitivity  to \sina  of the relative contributions of the $ZZ$ versus $W^{\pm}W^{\pm}$ channels is easily understood from the structure of the coupling prefactors in \cref{eq:H0W+W-,eq:H0ZZ}. Sin$\,\alpha$ in the vicinity of $2\, v_t/v_d$ shuts off the $W^{\pm}W^{\pm}$ channel in favor of the $ZZ$ channel which then scales roughly as $4\, v_t^2$, while \sina in the vicinity of $4\, v_t/v_d$ would do exactly the opposite. This is clearly seen respectively in  \cref{fig:BR_H1}(b) and \cref{fig:BR_H1}(c), where we observe a lowering of the $H^0 \rightarrow W^\pm W^\mp$ as \sina increases, until a total cancellation happens at around \sina~=~$8 \times 10^{-4}$ (independent of $H^0$ mass, see also~\cref{fig:BR_H2,fig:BR_H3}). The cancellation of $H^0 \rightarrow Z Z $ occurs at  twice this value of \sina, as expected. For \sina$\ll 2\, v_t/v_d$ the width to $ZZ$ will always dominate over that to $WW$ by roughly a factor four. This tendency is shown in \cref{fig:BR_H1}(a) illustrated for a moderately small \sina. Note, however, the change when $2\, M_W < m_{H^0} < 2\, M_Z$ where at least one $Z$ is off-shell, leading to the suppression of the $ZZ^*$ channel.\footnote{\label{footnote:offshell}A good approximation is obtained when taking one $Z$ off-shell in this part of the parameter space. We evaluated this channel numerically using \texttt{MadGraph} for the three-body decay $H^0 \to f \bar{f} Z$, summing over all light fermions, rather than using the approximate analytical expressions for an off-shell $Z$ decay derived in \cite{PhysRevD.22.722,PhysRevD.30.248,Cahn:1990xc,Djouadi:1995gv,Aoki:2011pz}. Indeed the latter do not account for finite width effects that are important for a correct smooth transition when crossing the $ZZ$ threshold, as seen in \cref{fig:BR_H1}(a).} 

The relative contribution of $H^0 \to h^0 h^0$, when kinematically open, is much less straightforward to understand. Here the coupling depends on the scalar couplings of the potential \cref{eq:Vpot} which is  {\sl a priori} independent of the gauge coupling present in the two channels just discussed, \cref{eq:H0h0h0}. However, using the parameterization described in \cref{sec:parameters,app:paramstrategy}, one can re-express all the $\lambda$'s in terms of the scalar masses and \sina, leading to \cref{eq:H0h0h0prime} that allows direct comparisons. For instance, in the limit \mHpp$\simeq$ \mHp$\simeq m_{H^0} > m_{h^0}, M_Z,M_W$, and taking into account all the factors in \cref{eq:H0W+W-} and \cref{eq:H0h0h0prime}, one finds an enhancement of $H^0 \to h^0 h^0$ over $H^0 \to W^\pm W^\mp$ by a factor $9/2$ in the regime where the latter decay scales like $\sin^2\alpha$, i.e. for \sina $\gg 2\, v_t/v_d$. For somewhat lower \sina, partial cancellations in the couplings for the $WW$ and $ZZ$ channels take place as explained above, suppressing further their contributions with respect to the $h^0h^0$ channel. This is illustrated in \cref{fig:BR_H2} (c) and~\cref{fig:BR_H3} (c). 

For much lower \sina ($\ll 2\, v_t/v_d$), the $WW$ and $ZZ$ channels do not scale anymore as $\sin^2\alpha$, cf.~\cref{eq:H0W+W-,eq:H0ZZ},  while $h^0h^0$ still does, leading to a huge suppression of the latter as illustrated in \cref{fig:BR_H2}(a) and \cref{fig:BR_H3}(a). Now taking \sina in the vicinity of $2\, v_t/v_d$ where the $WW$ shuts off, induces a very large variation of the relative contributions of the $ZZ$ and $h^0h^0$ channels including a quick crossover, as can be extrapolated by comparing plots (b) and (c) of \cref{fig:BR_H2,fig:BR_H3}. This increased sensitivity to \sina is due to the accidental fact that at \sina$=2\, v_t/v_d$ the $ZZ$ channel scales like $\sin^2\alpha$ implying the onset of the $h^0h^0$ dominance. Then, increasing further \sina, the $ZZ$ channel shuts off around $4\, v_t/vd$ where now $WW$ scales as $\sin^2\alpha$ and thus remains overwhelmed by $h^0h^0$.   

 There is also the $ H^0 \rightarrow t \bar t$ channel, which scales as $\sin^2\alpha$ and opens up for $H^{\pm \pm}$  mass values around 350-400~GeV, but reaches a branching fraction of at most 0.2 for some values of $\sin\alpha$. The subdominance of this channel with respect to the $h^0h^0$, $WW$ and $ZZ$ channels, even in the regime where the last two scale also like $\sin^2\alpha$, is due on the one hand to the smaller phase space available for a pair of top quarks, and on the other hand to an extra suppression factor $m_{H^0}^2 - 4 m_t^2$ induced by the Dirac nature of the top quark. Moreover, even though these effects tend to be washed out for increasingly heavy $H^0$, the conclusion remains the same due to mass enhancement in the widths; comparing \cref{eq:H0h0h0prime,eq:H0qq},  again in the limit \mHpp$\simeq$ \mHp$\simeq m_{H^0} > m_{h^0}, m_t, M_Z,M_W$, one finds the $h^0h^0$ channel to be enhanced by a factor $(3/4)\times (m_{H^0}/m_t)^2$ with respect to the $t\bar t$ channel.  This is more so regarding decays to lighter fermions which remain negligible across all the mass range.

Apart from the $ZZ^*$ channel mentioned previously, the main $H^0$ decay with one off-shell particle is found to be $H^0 \to H^\pm W^{\mp *}$. This occurs when $m_{H^\pm} <m_{H^0}$, knowing that  $\left|m_{H^0} - m_{H^\pm} \right| < M_W $ is always satisfied as a consequence of \cref{eq:Deltam_20} and the generic mass hierarchy shown in \cref{fig:spectrum-evol}. Note that in this case one also has $m_{H^{\pm\pm}} < m_{H^\pm} $,  but due to the upper bounds on the mass splitting decays of $H^0$ to \Hp\!\!\Hm or \Hpp\!\!\Hmm  will always have at least one off-shell particle. Moreover, for relatively light $H^0$ there would also be off-shell decays to $h^0h^{0*}$ or $tt^*$. For all these cases, the off-shell particles are heavier than the $W^\pm$, and the decay widths have either $\sin^2\alpha$ suppression factors~---~in the $h^0h^0$ and $tt$ cases~---~or 
$(v_t/m_{H^0})^2$ and $(v_t/v_d) \times \sin\alpha$ suppression factors~---~in the \Hp\!\!\Hm and \Hpp\!\!\Hmm cases\footnote{This is obtained by re-expressing the $\lambda_i$'s in terms of the input parameters defined in \cref{sec:parameters} as done previously for $H^0 \to h^0h^0$. }~---~while $H^0 \to H^\pm W^{\mp *}$ has none of these suppressions, cf.~\cref{eq:H0H+W-}. They remain thus highly suppressed with respect to $H^0 \to H^\pm W^{\mp *}$ until the opening of the on-shell $h^0h^0$ and $t\bar t$ channels where the off-shell $W^\pm$- and $\sin^2\!\alpha$-suppression effects, and the enhancement effects due to phase space broadening for increasing masses, start competing. Moreover, 
 since the allowed mass splitting between $H^0$ and \Hp is bounded, the level of $W^\pm$ off-shellness saturates while the phase space for $h^0h^0$ or $t \bar t$ increases.
 These features are well illustrated on \cref{fig:BR_H1} where all off-shell contributions discussed above remain invisible, as well as the on-shell $h^0h^0$ for smaller \sina, the latter becoming the leading channel for higher \sina. Increasing the mass scale, as in \cref{fig:BR_H2,fig:BR_H3} where $h^0h^0$ then $t\bar t$ are on-shell, one clearly sees the interplay among the \sina, $W^\pm$ off-shellness and phase-space effects.
Thus, when considering the $H^0 \rightarrow H^\pm W^{\mp *}$ decay mode and when combining different production possibilities, as in associated neutral production ($A^0 H^0$), we have to take into account also the decay branching fractions of \Hp (see \cref{fig:BR_Hp1,fig:BR_Hp2,fig:BR_Hp3}).
Finally, let us note that the loop-induced decay widths of $H^0$ into $\gamma\gamma$, $Z\gamma$ or $gg$, not considered here, are all suppressed by either $\sin^2\alpha$ or $(v_t/v_d)^2$ due to the essentially triplet content of $H^0$. They thus remain {\sl rare decays} as compared to the other channels discussed in this section.

%%%%%%%%%%%%%%%%%%%%%%%%%%%%%%%%%%%%%%%%%%%%%%%%
\subsubsection{The CP-odd Higgs boson \texorpdfstring{$A^{0}$}{}
\label{sec:A0}}

The decay pattern of $A^0$ is quite different from that of $H^0$ even though having essentially the same mass. In contrast to $H^0$, the CP-odd state has no tree-level decays to $W^{\pm}W^{\mp}$  or $ZZ$ but can decay to $h^0Z$.\footnote{For some points $A^{0}$ can become slightly heavier than $H^{0}$, in which case the channel $A^0 \to H^0 Z^{*} \to H^0 \nu \bar \nu$ becomes kinematically open. The corresponding width remains however extremely suppressed. The same argument holds for $H^0$ decays, when slightly heavier than $A^0$.}  The main decay channels for smaller scalar mass scales are thus to pairs of light fermions, principally to $b \bar b$, 
then to $h^0 Z^{(*)}$ or $H^\pm W^{\mp *}$ when kinematically and/or \sina favored~---~note that similarly to the case of the CP-even state,  the LNV invisible  decays $A^0 \to \nu_\ell \nu_\ell$ are totally suppressed
for the $v_t$ value under consideration. Larger scalar mass scales favor $t\bar t$ over $b \bar b$ once the former is kinematically open,  still sharing with the $h^0Z$ and $H^\pm W^{\mp *}$ channels for generic \sina. The dependence on \sina is present only in the partial width of $A^0 \to h^0 Z$. As evident from \cref{eq:A0hZ}, this channel shuts off at the critical value of \sina\!$\simeq v_t/v_d$. It scales with $\sin^2\!\alpha$ for much higher \sina than this value, and scales with $(v_t/v_d)^2$ for much lower values of \sina. The remaining partial widths of $A^0$ are always \sina-independent, but either suppressed by a $(v_t/v_d)^2$ factor, \cref{eq:A0qqbar}, or with no coupling suppression, \cref{eq:A0H+W}. Decays into singly-charged Higgs boson with an off-shell $W^{\mp *}$  or into $h^0Z$  are prevalent whenever
 $m_{A^0} > m_{H^\pm}$ or whenever \sina is far from the critical value, respectively. The relative contribution of $A^0 \rightarrow H^{\pm} W^{\mp *}$ tends to decrease for increasing $m_{A^0}$. This results from the fact that the $W^\pm$ off-shellness  saturates due to the bound on $\Delta m_{H^{\pm\pm}, H^\pm}$, while the phase space for the $A^0 \to h^0 Z$ channel increases progressively with $m_{A^0}$. For $m_{h^0} <  m_{A^0} < m_{h^0} + M_Z$ the $A^0 \to h^0 Z^*$ channel competes with $A^0 \to b \bar b$, the relative contributions being balanced by the effects of \sina and $Z$ off-shellness (see~\cref{fig:BR_A1}).

\begin{samepage}
\begin{figure}[!ht]
    \begin{center}
       {\includegraphics[width=0.49\textwidth]{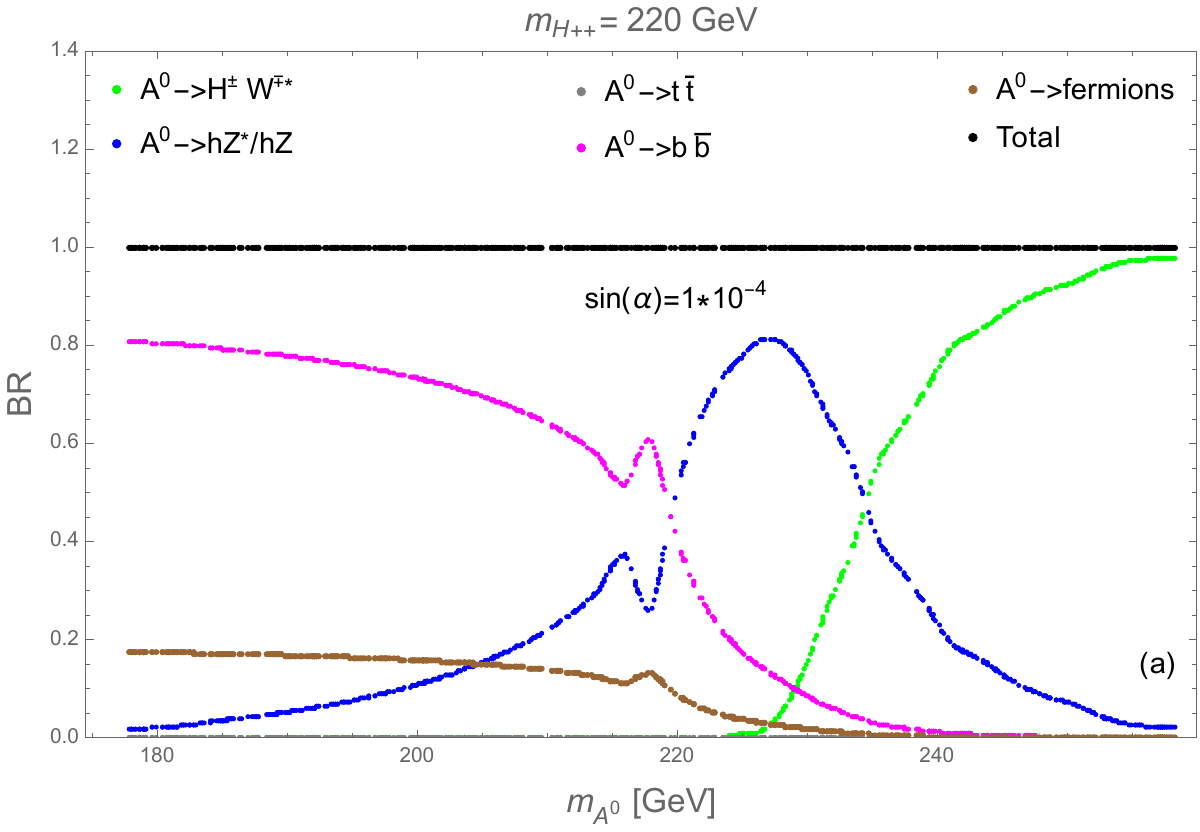}}
       {\includegraphics[width=0.49\textwidth]{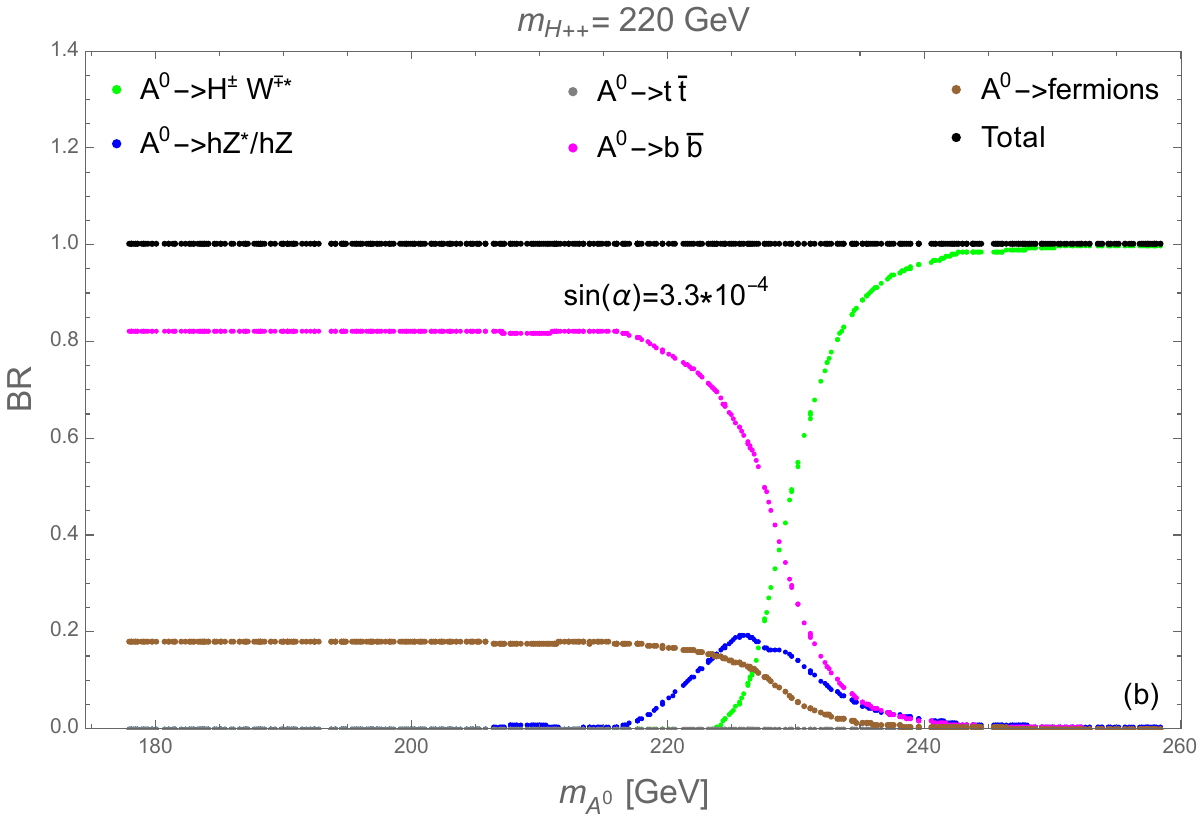}}
      {\includegraphics[width=0.49\textwidth]{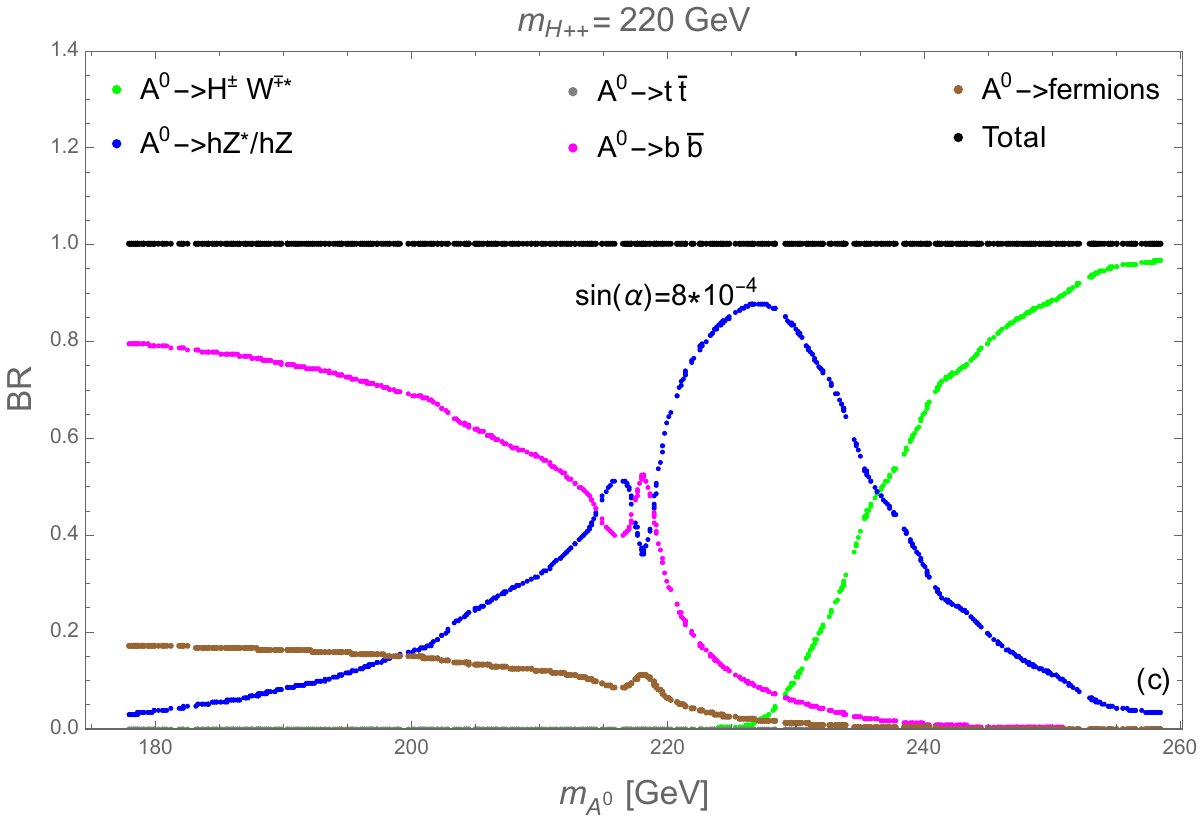}}
      {\includegraphics[width=0.49\textwidth]{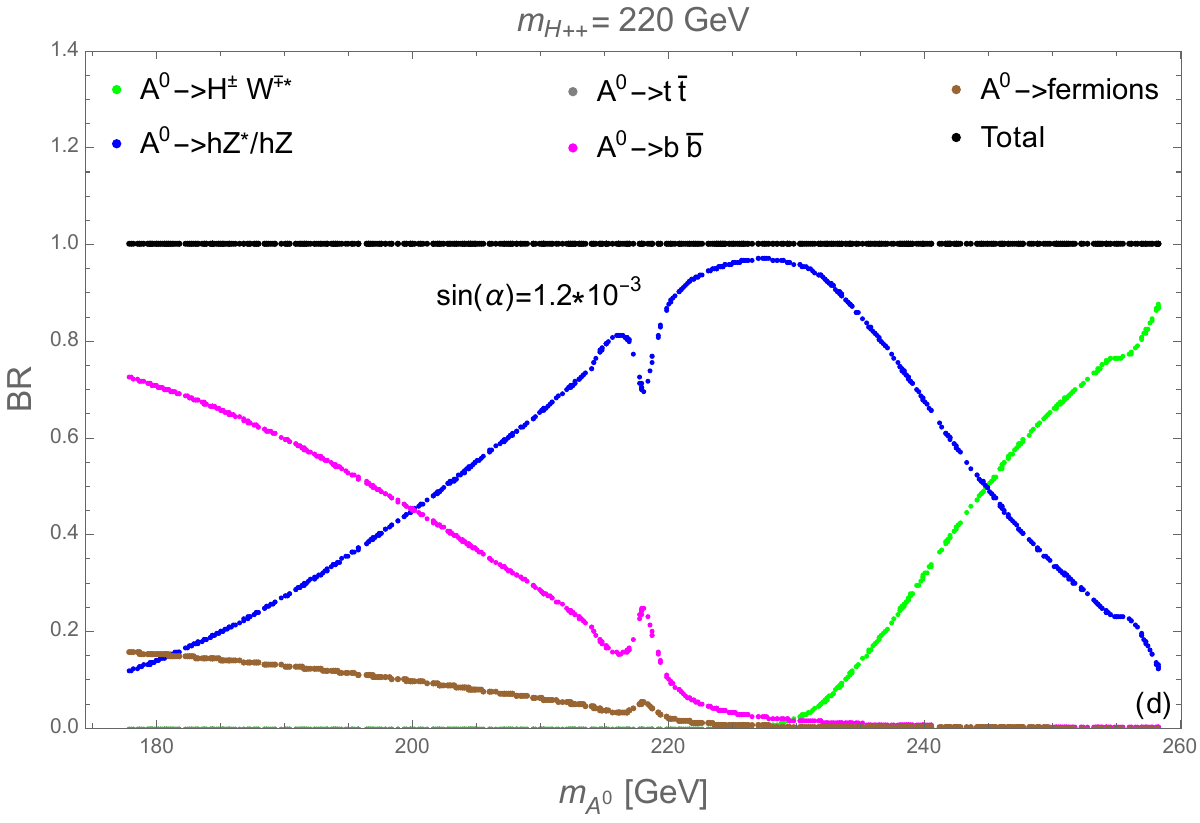}}
\end{center}
\caption{\label{fig:BR_A1}Decay branching ratios of the CP-odd neutral Higgs boson
as a function of its mass, for doubly-charged Higgs boson mass  of 220~GeV and different $\sin\alpha$ values; `fermions' indicate the sum over all light fermions other than the b-quark. }
\end{figure}
\nopagebreak
\begin{figure}[!th]
\begin{center}
        {\includegraphics[width=0.49\textwidth]{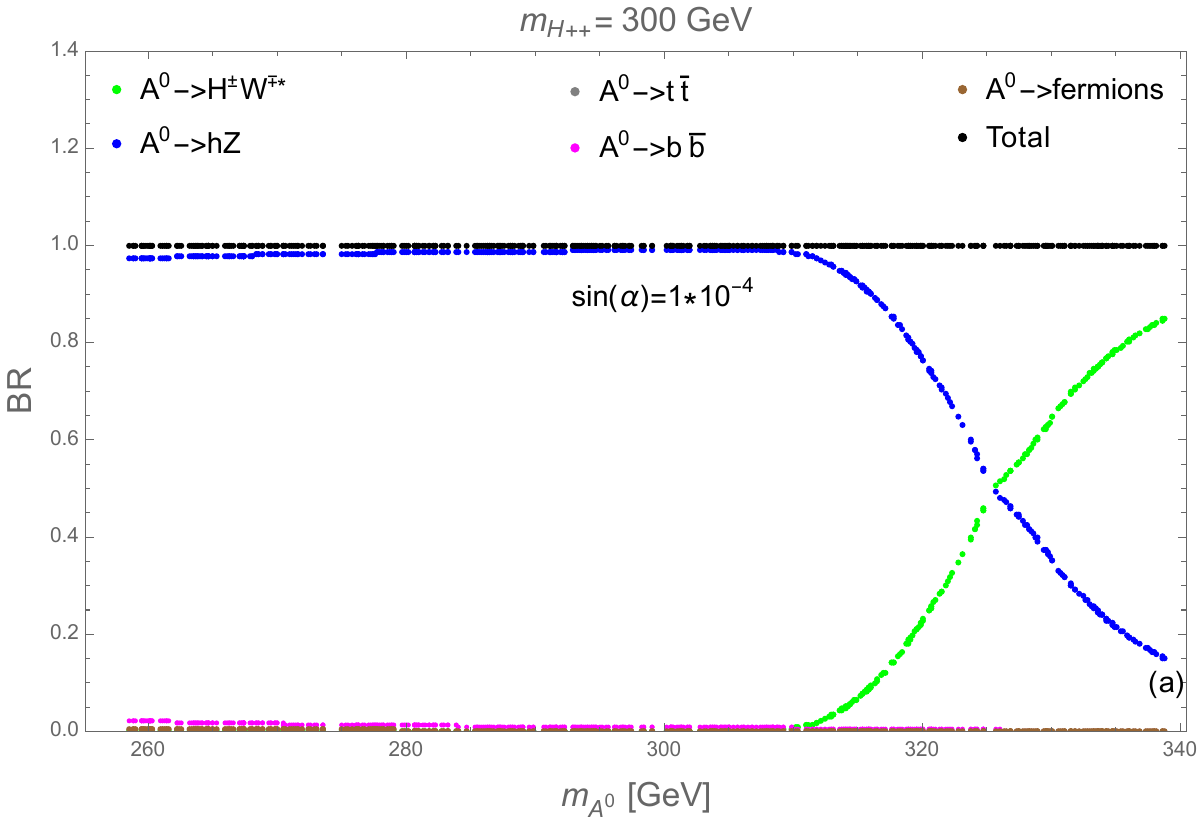}}
      {\includegraphics[width=0.49\textwidth]{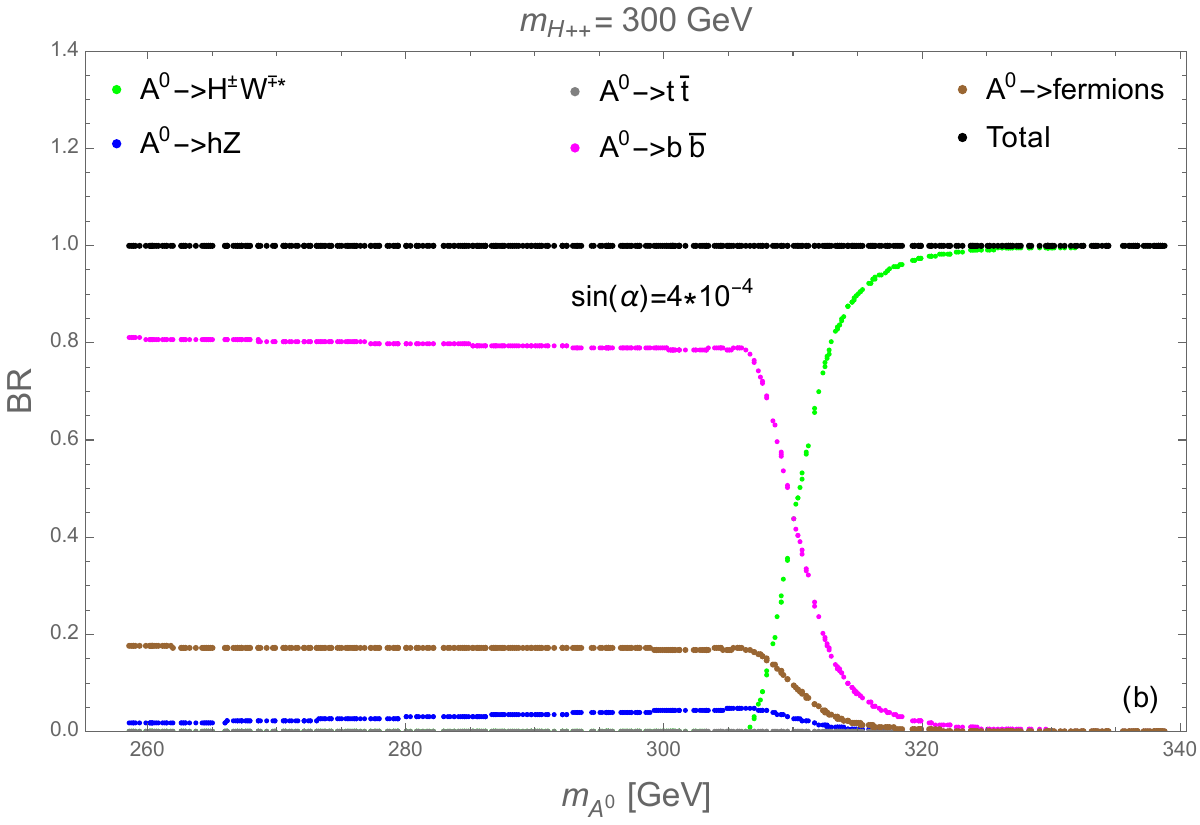}}

      {\includegraphics[width=0.49\textwidth]{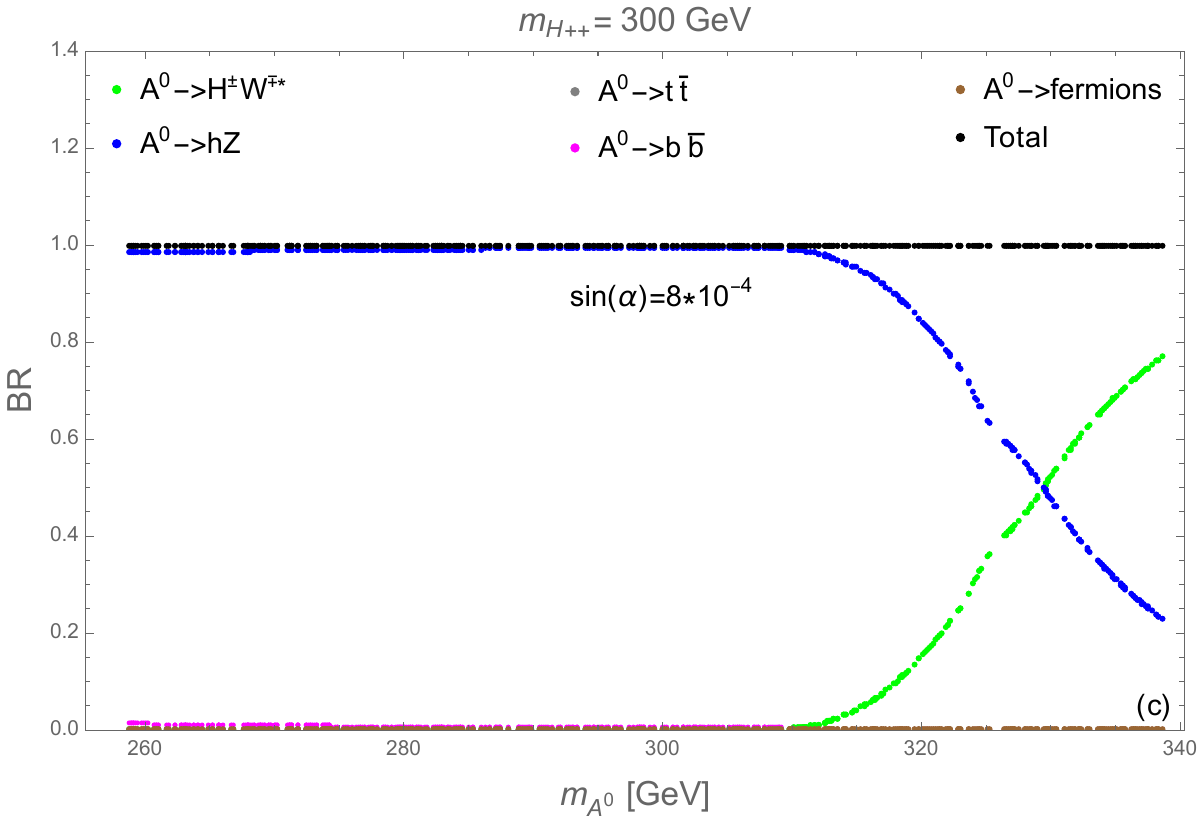}}
      {\includegraphics[width=0.49\textwidth]{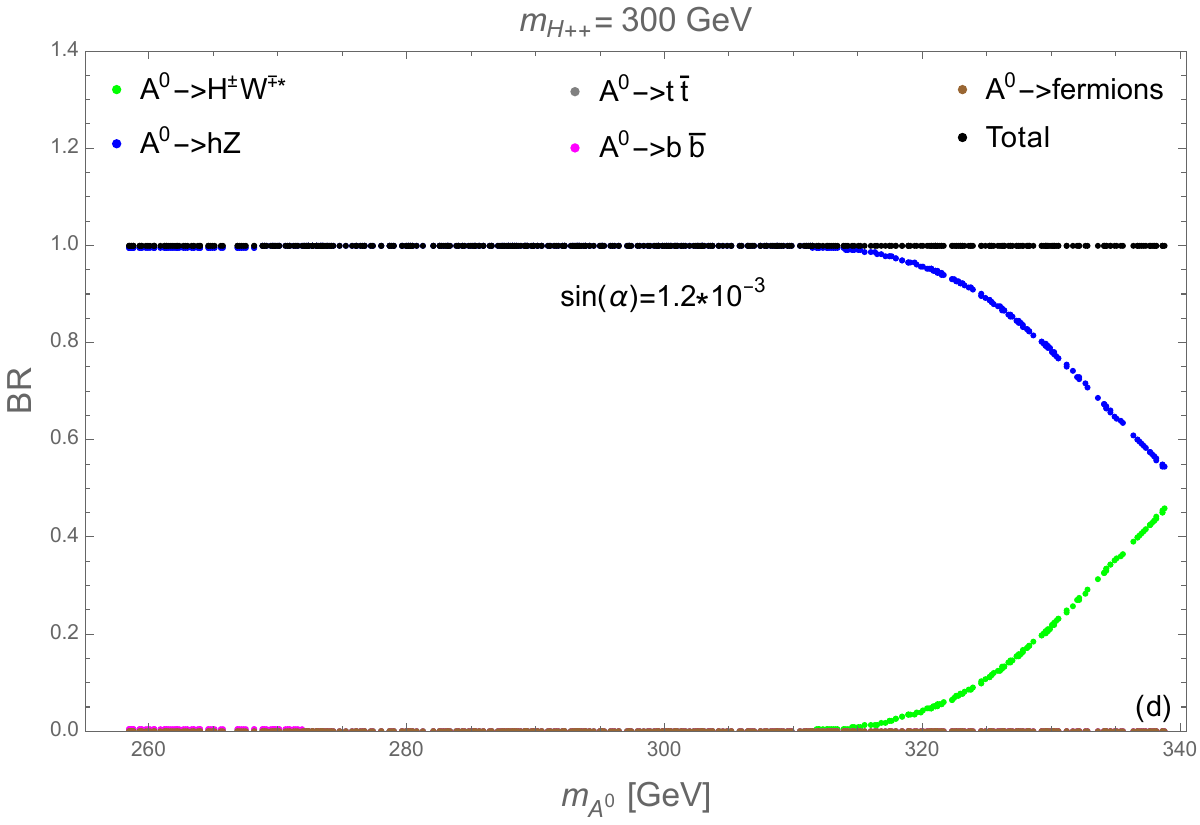}}
\end{center}
\caption{\label{fig:BR_A2}Decay branching ratios of the CP-odd neutral Higgs boson
as a function of its mass, for doubly-charged Higgs boson mass  of 300~GeV and different $\sin\alpha$ values; `fermions' indicate the sum over all light fermions other than the b-quark.}
\end{figure}
\end{samepage}
\begin{figure}[!ht]
\begin{center}
      {\includegraphics[width=0.49\textwidth]{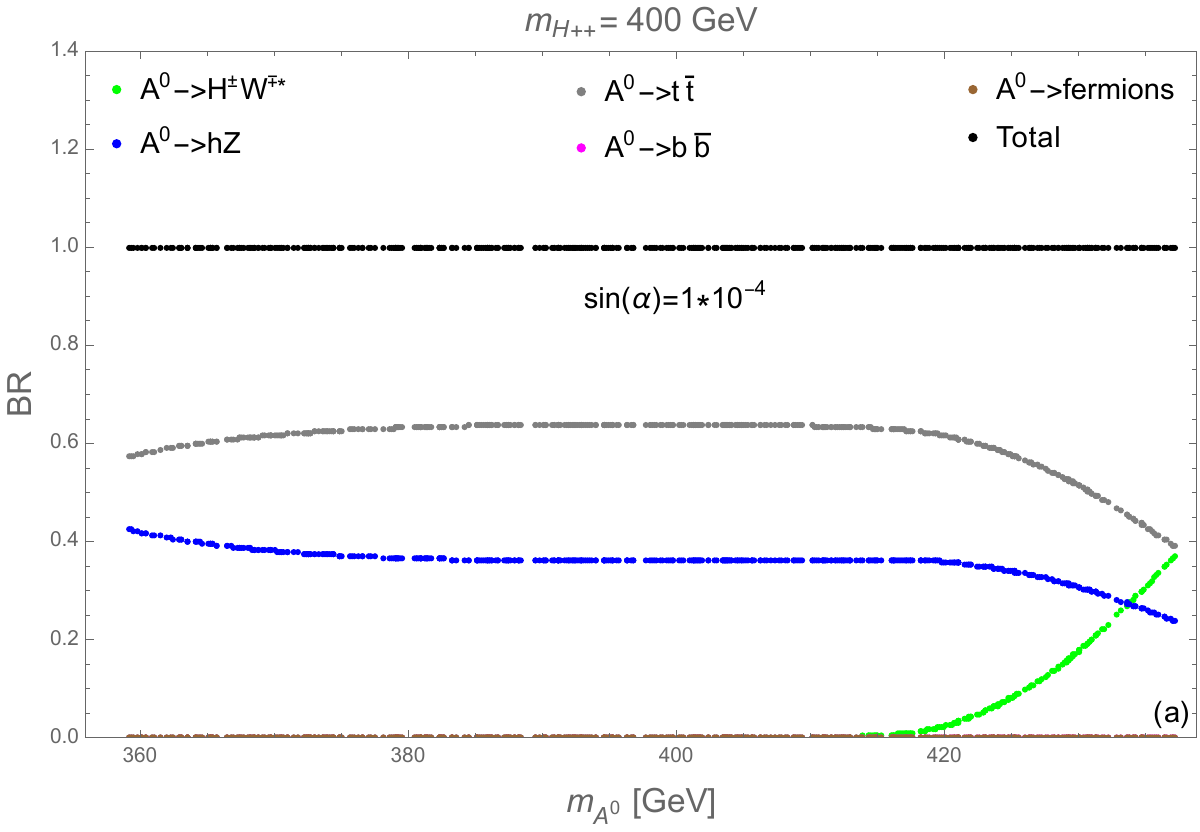}}
      {\includegraphics[width=0.49\textwidth]{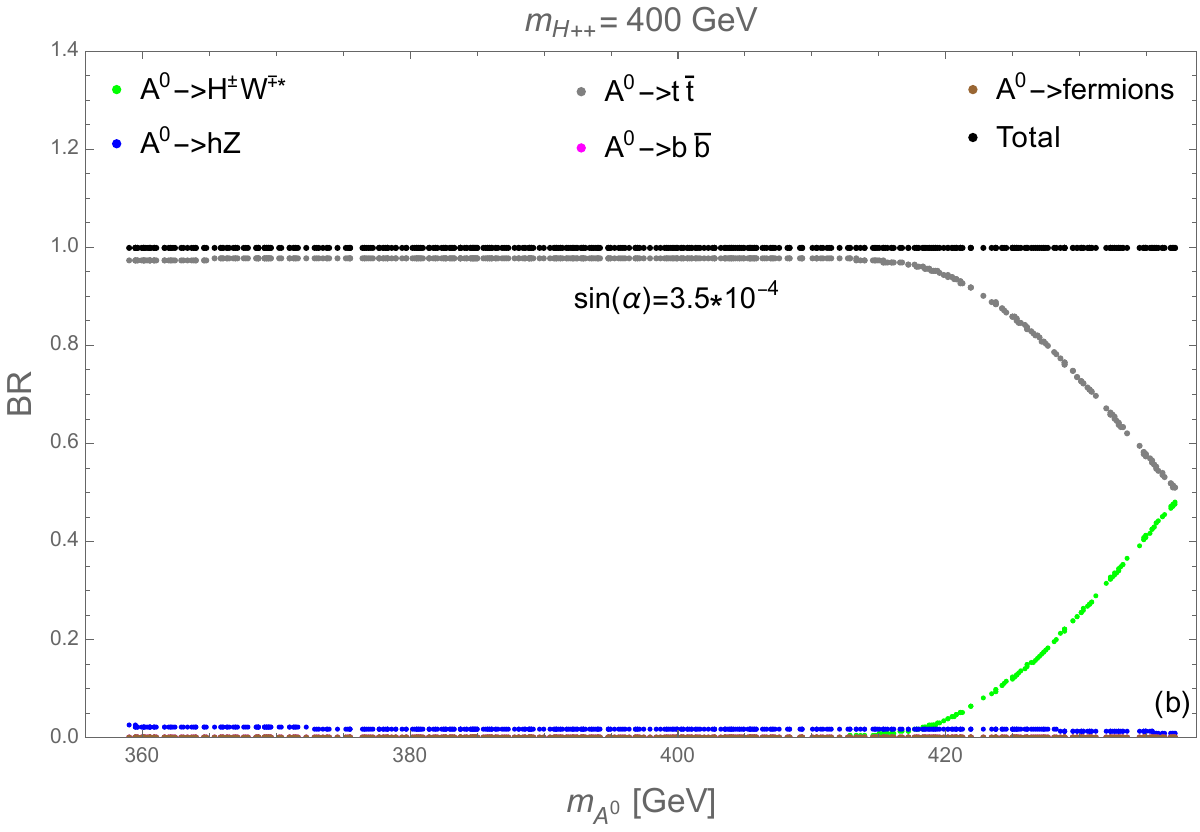}}
      {\includegraphics[width=0.49\textwidth]{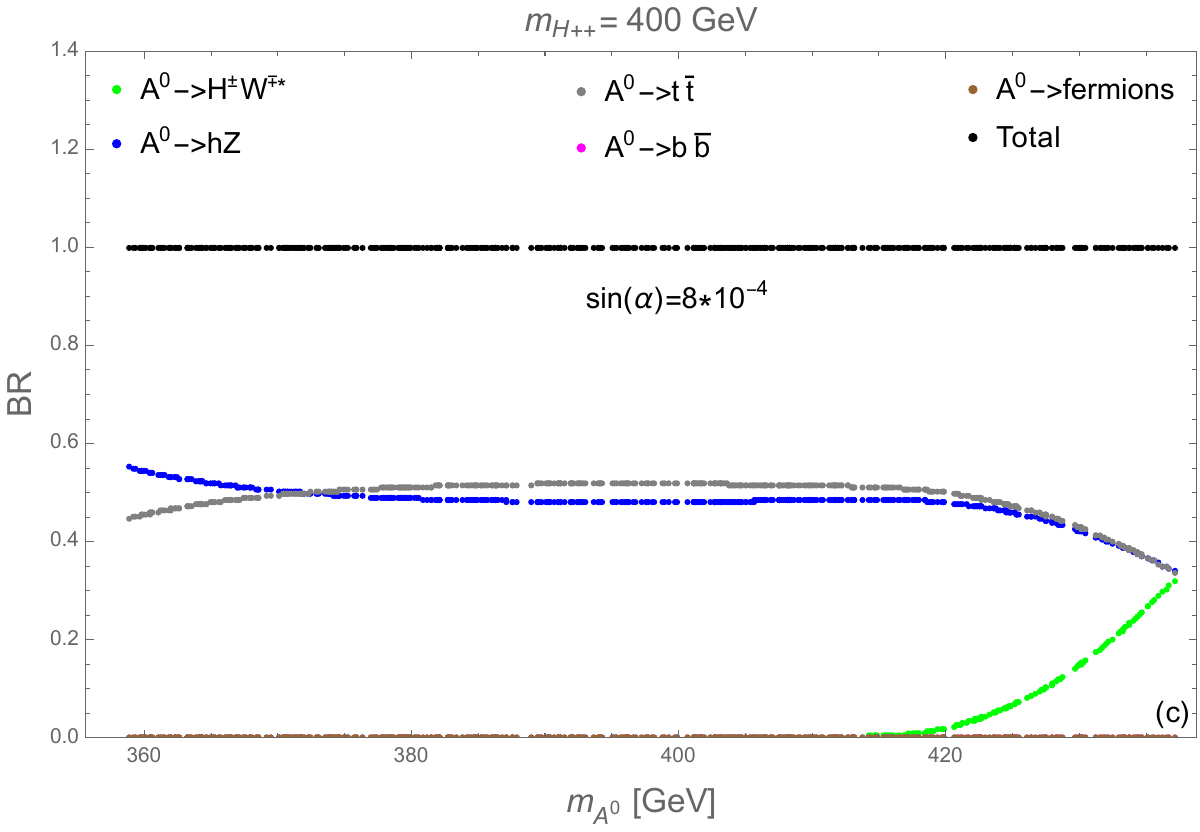}}
      {\includegraphics[width=0.49\textwidth]{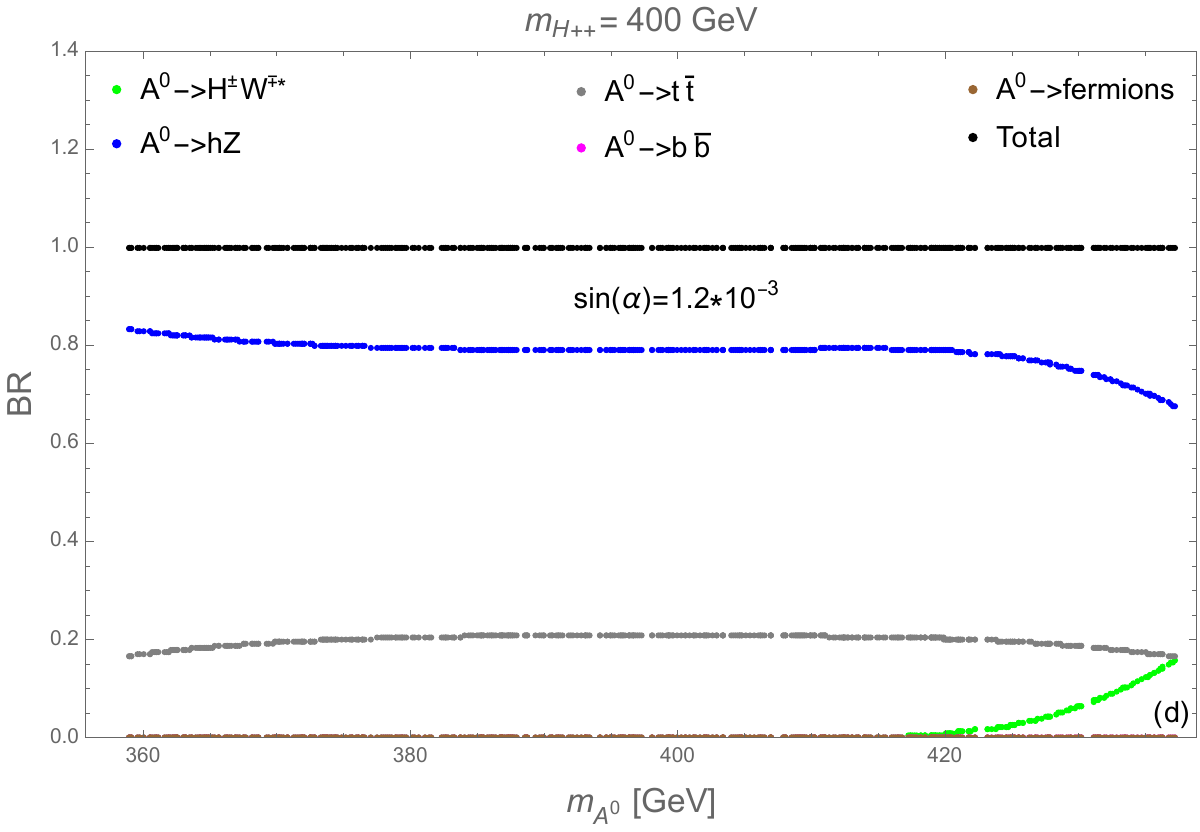}}
\end{center}
\caption{\label{fig:BR_A3}Decay branching ratios of the CP-odd neutral Higgs boson
as a function of its mass, for doubly-charged Higgs boson mass  of 400~GeV and different $\sin\alpha$ values; `fermions' indicate the sum over all light fermions other than the b-quark.}
\end{figure}

\Cref{fig:BR_A1,fig:BR_A2,fig:BR_A3} illustrate quantitatively the above generic features.
%$A^{0}$ having the same mass as $H^{0}$, no cascades from $H^0$ can happen since there's no available phase space, so the $A^0 \to H^0 Z^{*} $ channel is completely suppressed.
 The $A^0 \to h^0 Z^{(*)}$ width,  being the only one dependent on \sina, dictates the trend of the \sina sensitivities of the various branching ratios. 
This  channel is dominant for many of the cases presented here, but becomes vanishingly small for all benchmark \Hpp masses, when \sina takes the critical value of $\simeq 4 \times 10^{-4}$, as illustrated in plots (b) of \cref{fig:BR_A1,fig:BR_A2,fig:BR_A3}. When approaching this value from below or from above, $A^0 \to h^0 Z^*$ gives way to
 $A^0 \to b \bar b $ for lower \mHpp and $m_{A^0}$, as seen in  \cref{fig:BR_A1}(a) and (c). When moving to \sina above $1\times 10^{-3}$, $A^0 \to h^0 Z^*$ dominates in a significant fraction of the $m_{A^0}$ range. The same behavior is seen when the $t \bar{t}$ channel opens up, as illustrated for \mHpp$=400$~GeV in \cref{fig:BR_A3}: $A^0 \to t \bar{t}$ is leading for \sina values around the point where $A^0 \to h^0 Z$ is totally suppressed, but then, as \sina goes far away from it, $A^0 \to h^0 Z$  becomes again dominant. As far as the expected general behavior of $A^0 \rightarrow H^{\pm} W^{\mp *}$ is concerned, its branching ratio indeed shrinks with increasing \mHpp and \sina, being almost halved for higher \sina values and \mHpp$=300$~GeV, and 5 times smaller for \mHpp$=400$~GeV.

As a last comment, the total contribution of fermion pairs other than $b\bar b$ and $t\bar t$, is an almost factor 4 smaller than $A^0 \to b \bar b$, due to their small masses (see \cref{eq:A0qqbar}).

%%%%%%%%%%%%%%%%%%%%%%%%%%%%%%%%%%%%%%%%%%%%%%%%
\subsubsection{Closing remarks on the heavy scalars decay channels}
%\label{sec:A0}
While all the channels discussed above lead most of  the time to prompt decays, the decay widths of $H^\pm \to h^0 W^{\pm}$, $H^0\to W^+W^-$, $H^0\to ZZ$ and $A^0 \to h^0 Z$ can become very small if \sina lies in the close vicinity of some multiples of $v_t/v_d$, as noted in the three last subsections. The ensuing suppression of their branching ratios could be compensated by looking for displaced vertices, as already suggested for \Hpp  \cite{BhupalDev:2018tox,Antusch:2018svb}, which was however relevant in a different part of the parameter space than the one considered here. Note that we did not illustrate cases where $v_t$ and \sina would have opposite signs. These can however be easily inferred from our general discussion.

We end this section by stressing that having fixed $v_t$ to $0.1$~GeV comes without loss of generality as far as the high sensitivity to \sina is concerned. The latter is quite complementary to studies where $v_t$ is varied and \sina correlated to it in a fixed way, see e.g. Refs.~\cite{Melfo:2011nx,Primulando:2019evb}, which tends to conceal the possible coexistence 
of different decay channels even for given masses and $v_t$, as we repeatedly illustrated.

%%%%%%%%%%%%%%%%%%%%%%%%%%%%%%%%%%%%%%%%%%%%%%%%
%%%%%%%%%%%%%%%%%%%%%%%%%%%%%%%%%%%%%%%%%%%%%%%%
%%%%%%%%%%%%%%%%%%%%%%%%%%%%%%%%%%%%%%%%%%%%%%%%
\section{The intermediate states } 
\label{sec:allFinalStates}

The aim of this section is to provide a compendium of all possible final states that can result from the leading DY two-scalar production modes discussed in \cref{sec:prodModes}. Rather than presenting the exhaustive list, tedious and not very practical, we will give general guides that allow to probe the existence of a given final state.  This is facilitated by noting three main features discussed at length in \cref{sec:mass_spectrum,sec:decayModes}, that we summarize as follows:
\begin{enumerate}[(a)]
    \item \label{itm:a}The trend of the mass splitting among the scalar states is the same {\sl independently} of the actual overall mass scale. Moreover, it always satisfies one of the two mass hierarchies, see also \cite{Melfo:2011nx},
    \begin{align}
      & m_{H^{\pm\pm}} \geq m_{H^{\pm}} \geq m_{H^0} = m_{A^0}, \label{eq:h1} \tag{h1} \\
      & m_{H^{\pm\pm}} \leq m_{H^{\pm}} \leq m_{H^0} = m_{A^0}, \label{eq:h2} \tag{h2}
    \end{align}
    thus only two possible configurations of decay chains of one state to another.
    Moreover, the degeneracy of \Hpp with any other state implies the degeneracy of all; it follows that a smaller  $\Delta m_{H^{\pm\pm}, H^\pm}$ disfavors all long cascades as compared to direct decays to SM states.
    \item \label{itm:c}For given $v_t$, an absolute mass scale, a mass hierarchy and mass splittings of the scalars, a high sensitivity to \sina can occur. 
    \item \label{itm:b}
    %\otiliacom{ Gilbert, please update the text here, accounting for the feedback sent by Adam :D } 
    %\gilbert{Adam was suggesting "Smaller mass scales of the scalars favor direct decays to SM"; but this is not what I wanted to say. It is not between SM on one side and non-SM on the other. It is within the SM contributions themselves irrespective of whether the mass scale is small or large; e.g. compare the upper-left plot of \Cref{fig:BR_A1,fig:BR_A2} or the upper-right plot of \Cref{fig:BR_H1,fig:BR_H2}.
    %I have swapped items (b) and (c) and modified slightly below. I hope it is clearer (?)}
    For a given \sina, the absolute mass scale of the scalars influences the relative contributions of the various direct decays to SM states (except for the LNV final states that are always highly suppressed in our case). Note also that here $h^0$ is counted among the SM states since it has SM-like Higgs properties. 
    \end{enumerate}

From the above points one is lead to consider three decay categories: long-chain decays (LC) where three scalar states are involved in the cascade, intermediate-chain decays (IC) where only two are involved, and finally  direct-chain decays (DC) where the scalar state decays directly to SM states, $W^\pm$'s, $Z$'s, tops, bottoms and $h^0$'s. We dub the latter `intermediate states'  as they would decay further, leading ultimately to final states containing leptons, light jets, photons and missing transverse energy. It follows from (\ref{eq:h1}) and (\ref{eq:h2}) that \Hpp, $H^0$ and $A^0$ can decay within all three categories, while \Hp can have only IC and DC decays. We summarize in \cref{tab:general_decay_pattern} all possible patterns of decay chains. Combined with the DC decays listed in \cref{tab:decays_to_SM}, one obtains explicitly all possible SM content of the intermediate state decays for each of the scalar states of the model. The total content of decay products for a given DY two-scalar production mode, \Hpp~\!\!\Hmm, \Hpp~\!\!\Hm, \Hp$\!\!H^0$, \Hp$\!\!A^0$, $H^0A^0$, is then trivially obtained by combining the ones for each of the two scalars, with the same mass hierarchy. Recall that the DY \Hp\!\!\Hm pair production is always suppressed due to $\gamma^*/Z^*$ destructive interference.

\FloatBarrier
%%%%%
%%%%%
%%%%%
\begin{table}[htb!]
\centering
\def\arraystretch{1.2}
 \caption{ 
    General decay patterns of the scalars;  \Hpp, $H^0$ and $A^0$ can have LC, IC and DC decays. They are grouped on the upper line of the table; \Hp can have IC and DC decays listed on the left column. The blocks indicate the SM-particles content of intermediate states corresponding to a given decay chain. The relevant mass hierarchy is also indicated. The decaying mother scalar depends on the context and is uniquely identified either on the upper line or on the left column.
    The actual content of the decays to SM particles, $[\cdots \to \rm SM]$, is given in \cref{tab:decays_to_SM}.
    \xmark-marks indicate kinematically forbidden decay chains of scalars, and dashes the absence of such chains.
   }
\label{tab:general_decay_pattern}
\scalebox{0.72}{
\begin{tabular}{ l |c c c c c c }
\hline \hline
& $\scriptstyle H^{\pm\pm} \to \rm W^{\pm} W^{\pm}$ & $\scriptstyle H^{\pm\pm} \to H^\pm W^{\pm *}$ & $\scriptstyle H^0 \to \rm SM$ &  $\scriptstyle H^0 \to H^\pm W^{\mp *}$ & $\scriptstyle A^0 \to \rm SM$ & $\scriptstyle A^0 \to H^\pm W^{\mp *}$   \\
\hline 
$\scriptstyle H^\pm \to \rm SM$  & -- & $\scriptstyle (\ref{eq:h1}):~1W^{\pm} + [H^\pm \to \rm SM]$ & -- & $\scriptstyle (\ref{eq:h2}):~1W^{\pm} + [H^\pm \to \rm SM]$ & -- & $\scriptstyle (\ref{eq:h2}):~1W^{\pm} + [H^\pm \to \rm SM]$\\
$\scriptstyle H^\pm \to H^ {\pm\pm}W^{\mp *}$ & $\scriptstyle (\ref{eq:h2}):~3W$ & \xmark & -- & $\scriptstyle (\ref{eq:h2}):~4W$  &  -- &  $\scriptstyle (\ref{eq:h2}):~4W$ \\
$\scriptstyle H^\pm \to H^0 W^{\pm *}$ & -- & $\scriptstyle (\ref{eq:h1}):~2W + [H^0 \to \rm SM]$& $\scriptstyle (\ref{eq:h1}):~1W + [H^0 \to \rm SM]$& \xmark & -- & \xmark  \\
$\scriptstyle H^\pm \to A^0 W^{\pm *}$ & -- & $\scriptstyle (\ref{eq:h1}):~2W + [A^0 \to \rm SM]$ & -- & \xmark &$\scriptstyle (\ref{eq:h1}):~1W + [A^0 \to \rm SM]$ & \xmark  \\
 \hline \hline

\end{tabular}
}
\end{table}

%%%%%
%%%%%
%%%%%
\begin{table}[ht!]
\centering
\def\arraystretch{1.2}
 \caption{ 
   Two-body decays to SM particles.
}
\label{tab:decays_to_SM}
\scalebox{0.8}{
\begin{tabular}{ l |l l l l l }
\hline \hline
&     &  &  & \\
\hline 
$[\scriptstyle H^{\pm\pm} \to \rm SM]$ & $\scriptstyle H^{\pm\pm} \to W^\pm W^\pm$ & & &\\
$[\scriptstyle H^\pm \to \rm SM]$  & $\scriptstyle H^\pm \to Z W^\pm$    & $\scriptstyle H^\pm \to t b$ & $\scriptstyle H^\pm \to h^0 W^\pm$  &  \\
$[\scriptstyle H^0 \to \rm SM]$ &  $\scriptstyle H^0 \to ZZ$   & $\scriptstyle H^0 \to W^{\pm}W^{\mp}$ & $\scriptstyle H^0 \to b \bar b, t \bar t $ & $\scriptstyle H^0 \to h^0h^0$  \\ 
$[\scriptstyle A^0 \to \rm SM]$ &  $\scriptstyle A^0 \to h^0 Z$   & $\scriptstyle A^0 \to b \bar b$  & $\scriptstyle A^0 \to t \bar t$  & \\
\hline
\multicolumn{6}{c}{ $h^0$ counted as SM Higgs; all LNV decays suppressed in our case }\\
 \hline \hline

\end{tabular}
}
\end{table}

Let us focus first on the consequences of \ref{itm:a} in the case where at least one of the two produced scalars has an LC or IC decay. \Cref{tab:general_decay_pattern} shows that in this case $W^\pm$-bosons are always present in the intermediate state decay products, a feature common to all channels.
Although some of these $W^{\pm}$'s are off-shell, it is useful to organize the discussion in terms of their multiplicity.
At the level of intermediate states the $W^{\pm}$  multiplicity  ranges from at least $1W^{\pm}$ in the case of \Hpp~\!\!\Hm production with (\ref{eq:h1})-hierarchy or \Hp$\!\!H^0$ and \Hp$\!\!A^0$ productions with both hierarchies (\ref{eq:h1}) and (\ref{eq:h2}), up to a maximum of $8W^{\pm}$'s in the case of \Hpp~\!\!\Hmm production with (\ref{eq:h1})-hierarchy or $H^0A^0$ production with (\ref{eq:h2})-hierarchy. All intermediate $nW^{\pm}$ multiplicities with $1\leq n \leq 8$ can occur. Moreover, a close inspection of \cref{tab:general_decay_pattern,tab:decays_to_SM} allows to pinpoint several selection rules of which we list a few below, that could help define experimental search strategies:
\begin{enumerate}
    \item   Requiring $ n=7$ or $8$ excludes SM  particles other than $W^{\pm}$'s from the intermediate state. It then follows from  charge conservation that:
    \begin{itemize}
        \item $n=8$ selects the production mode \Hpp~\!\!\Hmm with  (\ref{eq:h1})-hierarchy, or the production mode $H^0A^0$ with (\ref{eq:h2})-hierarchy,
        \item $n=7$ selects the production mode  \Hpp~\!\!\Hm with (\ref{eq:h1})-hierarchy, or the production modes \Hp$\!\!H^0$ and \Hp$\!\!A^0$, with (\ref{eq:h2})-hierarchy. 
    \end{itemize}
    \item For $n=5$ or $6$, configurations with or without SM particles  other than $W^{\pm}$'s in the intermediate states can occur. \vspace{0.3cm} \\ 
     If only $W^{\pm}$'s are present:
     \begin{itemize}
     \item $n=5$ corresponds either to \Hpp~\!\!\Hm production mode with (\ref{eq:h2})-hierarchy or to \Hp$\!\!H^0$ production mode with (\ref{eq:h1})-hierarchy. 
     \item $n=6$ corresponds only to the production mode \Hpp~\!\!\Hmm with  (\ref{eq:h1})-hierarchy. This is similar to the $n=8$ case but involves shorter decay chains. Note that the production mode $H^0A^0$ is not selected in this case. 
     \end{itemize}
     If particles other than $W^\pm$'s are also present:
     \begin{itemize}
     \item $n=5$ corresponds  either to \Hpp~\!\!\Hm production mode with (\ref{eq:h1})-hierarchy and the presence of one pair of $Z$-bosons or $t$'s or $b$'s or $h^0$'s, or one $Z$ and one $h^0$, or to 
    $H^\pm H^0$ and  $H^\pm A^0$ production modes with (\ref{eq:h2})-hierarchy and the presence of a single $Z$ boson or $h^0$. 
    %The multiplicities of $t$'s and $b$'s thus allow to distinguish between the \Hpp~\!\!\Hm and $H^0A^0$. Note also that the \Hpp~\!\!\Hmm production mode does not occur at all in this case.
     \item $n=6$ corresponds  either to \Hpp~\!\!\Hmm production mode with (\ref{eq:h1})-hierarchy and the presence of one pair of $Z$-bosons or $t$'s or $b$'s or $h^0$'s, or one $Z$ and one $h^0$, or to 
    $H^0 A^0$ production mode with (\ref{eq:h2})-hierarchy and the presence of a single $Z$ boson or $h^0$. 
     \end{itemize}
\item For $1\leq n \leq 4$, 
%and except for $4W$'s from DC decays of the \Hpp~\!\!\Hmm production mode,  
there will always be SM particles other than and on top of the $W^{\pm}$'s in the intermediate states, with increasing multiplicity for decreasing $n$.
\end{enumerate}

The above features do not apply if the two produced scalars have both DC decays. In this case the discussion is straightforward; one only needs to refer to \cref{tab:decays_to_SM}. It should be kept in mind, though, that the description based so far on the two mass hierarchies should be further convoluted with  \cref{itm:b,itm:c} in order to reach comparative and quantitative assessments of the different production and decay patterns. For this, specific mass scales and \sina values should be considered, as exemplified in \cref{fig:BR_A1,fig:BR_A2,fig:BR_A3,fig:BR_H1,fig:BR_H2,fig:BR_H3,fig:BR_Hp1,fig:BR_Hp2,fig:BR_Hp3,fig:BR_Hpp}.

Finally, depending on the applied selection criteria and SM particle taggers, a given true final state composed of leptons, light jets, photons and missing transverse energy can originate from more than one intermediate state configuration.

Of all the possible BRs combinations presented in 
\cref{fig:BR_Hpp,fig:BR_Hp1,fig:BR_Hp2,fig:BR_Hp3,fig:BR_H1,fig:BR_H2,fig:BR_H3,fig:BR_A1,fig:BR_A2,fig:BR_A3}, of all achievable decay modes discussed in \cref{tab:general_decay_pattern,tab:decays_to_SM}, and of all viable values of the \sina parameter,  only those that ensure maximum values for the production cross-section in a given mass interval for the charged and neutral associated production modes are selected. Thus, after decaying the Higgs bosons according to their most promising BRs in a specific mass range, one ends up with the most promising final states, some of which will further be analysed in the subsequent sections.

To illustrate the above, let us take as an example the $H^{\pm \pm} H^{\mp}$ production mode and the \mHpp = 220~GeV benchmark mass point. In the $m_{H^\pm} < 212$~GeV  mass range, $H^\pm \rightarrow H^0 W^{\pm *}$ or $H^\pm \rightarrow A^0 
W^{\pm *}$  decays happen with equal branching ratios since $H^0$ and $A^0$ are degenerate and the couplings are the 
essentially the same for $v_t \ll v_d$, (see \cref{fig:BR_Hp1,fig:BR_Hp2,fig:BR_Hp3}, and 
\cref{eq:H+H0W+,eq:H+A0W+}). If \Hp goes for the $H^0$ decay,  looking for $m_H$ values lower than 212~GeV (see 
\cref{fig:BR_H1,fig:BR_H2,fig:BR_H3}), the preferred decay modes become evident: $H^0 \rightarrow Z Z^*$ for mostly 
low $\sin\alpha$ values, and $H^0 \rightarrow W^{\pm } W^{\mp *}$ for high $\sin\alpha$ values. 
Considering that 
$H^{\pm 
\pm}$ goes to $W^{\pm}W^{\pm}$ (see \cref{fig:BR_Hpp}), the two 
intermediate states are $3 W^\pm$  + $2 Z$  and $5 W^{\pm}$ bosons, respectively. If, on the other hand, the selected
decay mode is $A^0$, then this pseudo-scalar Higgs boson can go to $h^0Z^*$ or 
$b \bar b$, hence, the intermediate states are $3W^\pm$+$1h^0$+$1Z$ or $3W^\pm$+$2b$. 
From all these intermediate states, one could consider as experimental final states those with one or several leptons, jets, photons, and missing transverse energy, as such signatures are highly targeted at the LHC.

%%%%%%%%%%%%%%%%%%%%%%%%%%%%%%%%%%%%%%%%%%%%%%%%
%%%%%%%%%%%%%%%%%%%%%%%%%%%%%%%%%%%%%%%%%%%%%%%%
%%%%%%%%%%%%%%%%%%%%%%%%%%%%%%%%%%%%%%%%%%%%%%%%
\FloatBarrier
\section{ Illustrative benchmark points for a  potential experimental search at the LHC}
\label{sec:benchmark-exp}

In this section, the experimental search potential at the LHC is assessed by examining three production modes: the pair production of doubly-charged Higgs bosons, the associated production of both doubly- and singly-charged Higgs bosons, and the associated production of the neutral bosons $A^0$ and $H^0$. Among all the production modes discussed in \cref{sec:prodModes}, these were selected primarily due to their higher production cross-sections spanning a wide range of masses. 
These processes are collectively referred to as signals.
From all the final states outlined in \cref{sec:allFinalStates}, a subset has been chosen for detailed cutflow analysis. The focus is on multi-lepton experimental signatures, as the multi-jet or single-lepton ones are subject to a high level of background from SM processes such as QCD or $W^\pm/Z$+jets production.
Thus, three distinct final states, or channels, are being studied in more detail: two leptons of the same electric charge (\llSC), three leptons (\lll), or four leptons (\llll).
%https://www.researchgate.net/figure/ATLAS-results-of-the-SM-cross-section-measurements_fig9_330009152
Although the branching ratios are relatively small, searches in final states with \llSC, \lll{} and \llll{} remain highly interesting as the characteristics of these signatures can be exploited to achieve sufficient background suppression.

The signal samples are generated with \texttt{MadGraph}~\cite{Alwall:2014hca} version \verb|MG5_aMC_v3.5.3|, interfaced with Pythia8~\cite{Bierlich:2022pfr} for showering and hadronization. The \texttt{MadSpin} module~\cite{Artoisenet:2012st} was also included, to ensure that both the off-shell and spin correlation effects are retained in the signal generation. These proton-proton collision event samples have 50$\times 10^3$ events generated, and are processed through \texttt{DELPHES}~\cite{deFavereau:2013fsa} framework to simulate a fast and realistic detector response.
The \texttt{DELPHES} parameter card designed for the ATLAS~\cite{PERF-2007-01} detector is utilized with modifications to align with the event selection and object definitions from Ref.~\cite{ATLAS:2021pairbosons}. 
Notably, for jets the anti-$k_\mathrm{T}$ algorithm~\cite{Cacciari:2008gp} with a radius parameter $R = 0.4$ is employed, 
the electron and muon selection efficiencies are updated according to Refs.~\cite{ATLAS:2023dxj,ATLAS:2020auj}, 
the isolation identification working points are revised as in Ref.~\cite{ATLAS:2021pairbosons}, and the $b$-tagged jets efficiency is adjusted to 70\%.
Only leptons that satisfy the \pt$>10$~GeV and $|\eta|<2.47$ (for electrons) and $|\eta|<2.5$ (for muons) requirements are kept.
Unless otherwise specified, the energy in the center of mass $\sqrt{s}$ is 13~TeV. 

Finally, the signal samples are analysed using the \texttt{SimpleAnalysis}~\cite{ATLAS:2022yru} framework. 
In this framework, the analysis outlined in Ref.~\cite{ATLAS:2021pairbosons} is implemented to obtain yields at various selection stages, the region acceptance $A$\footnote{The acceptance $A$ is the ratio between the number of events passing the region definition and the total number of events.}, and to assess the statistical significance of the signal within the defined signal regions. 
Each event is taken with a weight that accounts for the \texttt{MadGraph} generator weight, production cross-section, decays BRs, and the ATLAS total integrated luminosity (140~fb$^{-1}$, 300~fb$^{-1}$ and 3000~fb$^{-1}$). The production cross-section times BR is given by \texttt{MadGraph}, and a k-factor of 1.25 is considered 
\cite{Muhlleitner:2003me,Fuks:2019clu} to account for the NLO effects.
The signal significance $Z$ is computed with the following formula~\cite{Cowan:2010js}:
\begin{equation}
    Z = \pm\sqrt{2} \times \sqrt{ n \mathrm{ln} \frac{n(b+\sigma^2)}{b^2+n\sigma^2} - \frac{b^2}{\sigma^2}\mathrm{ln}\frac{b^2+n\sigma^2}{b(b+\sigma^2)}},
    \label{eqn:Zn_function}
\end{equation}
where $n$ is the total number of events, and $b$ represents the number of SM and detector background events\footnote{The detector background collectively denotes the electron charge flip and fake/non-prompt lepton background sources~\cite{ATLAS:2022swp}.}; $\sigma$ denotes the uncertainty on the background, and a 30\% flat systematic uncertainty is considered in the computation. The combined signal significance of various orthogonal signal regions is simply their quadratic sum.

As discussed in Ref.~\cite{Cowan:2010js}, in particle physics the signal significance $Z$ is often used to quantify the rejection of a background hypothesis. 
A $Z$ value of 1.64 is equivalent to a $p$-value of 0.05 at 95\% confidence level, thus roughly enough to exclude a signal hypothesis.\footnote{The $p$-value is used to quantify the level of disagreement between data and a tested hypothesis.}
A $Z$ value of 5 corresponds to a $p$-value of $2.87 \times 10^{-7}$, an appropriate level to constitute a discovery. 
The prospects studies presented in this paper will consider these values for any quantification of the exclusion or discovery potential at LHC, or HL-LHC.

%CMS~\cite{CMS-TDR-08-001} .  
% $H^{\pm\pm}$

%%%%%%%%%%%%%%%%%%%%%%%%%%%%%%%%%%%%%%%%%%%%%%%%
\FloatBarrier
\subsection{ Search for charged Higgs bosons } 
\label{sec:charged_Sector_Prospects}
% https://gitlab.cern.ch/atlas-physics-office/HDBS/ANA-HDBS-2019-06/ANA-HDBS-2019-06-PAPER

Guided by the search presented in Ref.~\cite{ATLAS:2021pairbosons}, this section presents prospect studies for the \Hpp pair production sector, 
via $pp\rightarrow \gamma^{*}/Z^{*} \rightarrow H^{\pm\pm}H^{\mp\mp}$,
and for the \Hpp\Hm associated production sector, via $pp\rightarrow W^{\pm *}\rightarrow H^{\pm\pm}H^{\mp}$. The considered decay modes are $H^{\pm \pm} \rightarrow W^\pm W^\pm$ (100\% BR), and $H^{\mp} \rightarrow W^\mp Z$ (with a maximum BR varying between 30\%~-~40\%, depending on the boson mass) or $H^{\mp} \rightarrow tb$ (with the same maximum BR variation for the selected \sina value). 
Three benchmark \Hpp hypothetical mass points (\mHpp) are studied: 220~GeV, 300~GeV and 400~GeV.
\cref{tab:charged_ap_pp_nominals} in \cref{app:NominalP_Details} provides additional details on the parameters used in the event sample generation.
For the three signal mass points, the production cross-section ($\sqrt s = 13$~TeV) times the BRs times the k-factor values are:
\begin{itemize}
    \item \Hpp pair production: 62.1~fb, 18.1~fb  and 5.2~fb, respectively.
    \item \Hpp\Hm associated production: 
    \begin{itemize}
     \item When $H^{\mp} \rightarrow W^\mp Z$: 49.2~fb, 10.7~fb and 3.3~fb.
     \item When $H^{\mp} \rightarrow t b$: 59.5~fb, 15.6~fb and 3.3~fb.
    \end{itemize}
\end{itemize}

\begin{table}[t!]
\begin{center}
	\def\arraystretch{1.5}
	\caption{
	    Table showing the event pre-selection criteria for the \llSC, \lll, \llll{} channels from Ref.~\cite{ATLAS:2021pairbosons}. In the \llSC{} and \llll{} channels the leptons are ordered by decreasing \pt{}, while in \lll{} channel by increasing \pt{}, respectively. 
        Here, $\ell_{0}$ stands for the lepton that has an opposite charge with respect to the total lepton charge ($\sum Q_\ell$), and $\ell_{1}$ and $\ell_{2}$ represent the two same-charge leptons.
        SFOC is the same-flavour opposite-charge lepton pairs. $N_\ell$ denotes the number of leptons and  $N_{(b\textrm{-tagged) jets}}$ the number of ($b$-tagged) jets. 
        The type T, L$^*$ and L leptons stands for different lepton collections, and are discussed in Ref.~\cite{ATLAS:2021pairbosons}.
        The symbol ``~--~" means no requirement is applied.
        The bold and slashed criteria show the differences with respect to Ref.~\cite{ATLAS:2021pairbosons}. 
    }
    \label{Tab:preselection}
    \scriptsize
	\resizebox{0.88\textwidth}{!}
	{
	    \begin{tabular}{l||c|c|c}
		\hline\hline		
			Selection criteria & \llSC & \lll & \llll \\
			\hline\hline
			\multicolumn{4}{c}{At least one lepton with \pTlep{} $ >30$~GeV} \\
			\hline
		    $N_\ell$ (type T)   & =2 &  \textbf{$\boldsymbol{\geq}$3} &  \textbf{$\boldsymbol{\geq}$4} \\ 
			$N_\ell$ (type L$^*$) & -- & -- &  \textbf{$\boldsymbol{\geq}$4}\\ 
			$N_\ell$ (type L)   & =2 & =3 &  \textbf{$\boldsymbol{\geq}$4}\\ 
			$|\sum Q_\ell|$ & =2 & =1& $\neq$4 \\
		    Lepton \pTlep{} & \pTleplep{} $>30, 20$~GeV  & 
		    \pTlepleplep{}$>10, 20, 20$~GeV  & 
		    \pTleplepleplep{}$>10$~GeV \\  
			\hline
 
			\hline
			$\met{}$ & $ >70$~GeV & $>30$~GeV &  $>30$~GeV  \\
			\hline
			$N_{\textrm{jets}}$  &  $\ge 3$ & $\ge 2$ &  -- \\
			\hline
			$N_{b\textrm{-tagged jets}}$  & \multicolumn{3}{c}{ =0 } \\
			\hline
			Low SFOC $m_{\ell\ell}$ veto & --  & \multicolumn{2}{c}{$m_{\ell\ell}^\text{oc}>15$~GeV}  \\
			\hline
			$Z$ boson decay veto &  $|m_{ee}^\text{sc}- m_{Z}| > 10$~GeV &
			{$|m_{\ell\ell}^\text{oc}- m_{Z}| > 10 $~GeV} & 	{\cancel{\textbf{$\boldsymbol{|m_{\ell\ell}^\text{oc}- m_{Z}| > 10} $~GeV}}}\\
		\hline\hline
		\end{tabular}
	}
\end{center}
\end{table}

The event pre-selection criteria from Ref.~\cite{ATLAS:2021pairbosons} are shown in \cref{Tab:preselection}.~(\cref{app:Rel21Paper_AdditionalDetails} presents the various variables used in the analysis.)
Motivated by the lack of sensitivity in the \llll{} channel in Ref.~\cite{ATLAS:2021pairbosons}, the selection was adjusted to be more inclusive. 
This adjustment is expected to increase the signal yields with only a negligible increase in the number of SM backgrounds.
Naturally, this should be verified with a more realistic approach. However, for the purposes of this paper, this assumption is good enough.
The three channels, \llSC, \lll{} and \llll{}, are defined to be mutually exclusive, with exactly two, exactly three, and at least four leptons that satisfy the looser (type L) lepton selection criteria, respectively.
To decrease the amount of detector backgrounds, requirements are also placed on the number of leptons defined with tighter definitions (type L$^*$ and T), that have a better fake/non-prompt lepton background rejection~\cite{ATLAS:2021pairbosons,ATLAS:2023dxj,ATLAS:2020auj}. 
Additionally, to further mitigate sources of SM and detector backgrounds, criteria based on the leptons' \pt{}, $\met$, the number of ($b$-tagged) jets and $m_{\ell\ell}^{oc}$ invariant mass are applied, as detailed in \cref{Tab:preselection}.
For this signal model experimental signature, the requirement of zero $b$-tagged jets in the event is removing most of the $t\bar t$ background.

Finally, following the event pre-selection criteria, the requirements for the signal regions (SRs) are applied. 
The SRs are discussed in detail in Ref.~\cite{ATLAS:2021pairbosons}, and presented in \cref{Tab:SRselection} from \cref{app:Rel21Paper_AdditionalDetails}. 
They emerge from an extensive optimization done to maximize the sensitivity to \Hpp\Hmm pair production processes with $H^{\pm \pm} \rightarrow W^\pm W^\pm$ decays. 
Thus, to further increase the sensitivity to \Hpp\Hm{} associated production signals, certain SR selections were omitted in this paper, as highlighted in the aforementioned table.
According to Ref.~\cite{ATLAS:2021pairbosons}, main background sources in the SRs are $W^\pm Z$ SM processes, with lower contributions from detector background~\cite{ATLAS:2022swp} and rarer SM processes like $\ttbar Z$ and $\ttbar W^\pm$.

%%%%%%%%%%%%%%%%%
%%%%%%%%%%%%%%%%%
\FloatBarrier
\subsubsection{ \texorpdfstring{$H^{\pm \pm} H^{\mp \mp}$}{HppHmm} pair production sector \label{sec:HppHpp}}

\begin{table}[t!]
\centering
    \def\arraystretch{1.5}
    \caption{
        The signal yields, and the acceptance, obtained for the \Hpp\Hmm pair production processes with $H^{\pm \pm} \rightarrow W^\pm W^\pm$ decays. 
        Three steps are considered: no-selection, lepton selection and event pre-selection.
        Only the MC statistical uncertainty is shown, and the considered integrated luminosity is 140~fb$^{-1}$, $\sqrt s = 13$~TeV. 
    }
    \label{tab:Cutlfow_PreSel_HdbHdb_WWWW}
	\setlength{\tabcolsep}{0.0pc}
	\resizebox{0.8\textwidth}{!}{
		\begin{tabular*}{\textwidth}{@{\extracolsep{\fill}}llcccc}
			\noalign{\smallskip}\hline\noalign{\smallskip}
			&  & & \Hpp \Hmm 220~GeV & \Hpp \Hmm 300~GeV  & \Hpp \Hmm 400~GeV \\
			& Selection & & N events ($A$) & N events ($A$) & N events ($A$) \\
			\noalign{\smallskip}\hline\noalign{\smallskip}
			%%%%%
		 & All &  &              8693.93 $\pm$ 38.90 (100.00 \%)         & 2532.67 $\pm$ 11.34  & 729.83 $\pm$ 3.27 \\
			\noalign{\smallskip}\hline\hline\noalign{\smallskip}
			%%%%%
			\multirow[c]{6}{*}[0in]{\rotatebox{90}{$\ell$ selection}}
		 & $0\ell$ & &                   4606.20 $\pm$ 28.32 (52.98 \%)  & 1284.34 $\pm$ 8.08 (50.71 \%)  & 353.04 $\pm$ 2.27 (48.37 \%)  \\
         & $1\ell$ & &                   2634.45 $\pm$ 21.41 (30.30 \%)  & 827.48 $\pm$ 6.48 (32.67 \%)   & 252.88 $\pm$ 1.93 (34.65 \%)  \\
         & $2\ell$ &  &                  720.64 $\pm$ 11.20 (8.29 \%)    & 236.57 $\pm$ 3.46 (9.34 \%)    & 75.47 $\pm$ 1.05 (10.34 \%)   \\
         & $2\ell^{\mathrm{SC}}$ &  &    242.36 $\pm$ 6.49 (2.79 \%)     & 77.71 $\pm$ 1.99 (3.07 \%)     & 25.35 $\pm$ 0.61 (3.47 \%)    \\
         & $3\ell$ & &                   90.12 $\pm$ 3.96 (1.04 \%)      & 27.44 $\pm$ 1.18 (1.08 \%)     & 10.19 $\pm$ 0.39 (1.40 \%)    \\
         & $4\ell$ &  &                  6.44 $\pm$ 1.06 (0.07 \%)       & 1.83 $\pm$ 0.30 (0.07 \%)      & 0.64 $\pm$ 0.10 (0.09 \%)     \\			
			\noalign{\smallskip}\hline\hline\noalign{\smallskip}
			%%%%%	
			\multirow[c]{3}{*}[0in]{\rotatebox{90}{pre-selection}}                           
         & $2\ell^{\mathrm{SC}}$ &  &    104.04 $\pm$ 4.25 (1.20 \%)     & 42.05 $\pm$ 1.46 (1.66 \%) & 15.01 $\pm$ 0.47 (2.06 \%) \\
         & $3\ell$ &  &                  29.58 $\pm$ 2.27 (0.34 \%)      & 9.23 $\pm$ 0.68 (0.36 \%)  &  3.63 $\pm$ 0.23 (0.50 \%) \\
         & $4\ell$ &  &                  5.92 $\pm$ 1.01 (0.07 \%)       & 1.52 $\pm$ 0.28 (0.06 \%)  &  0.54 $\pm$ 0.09 (0.07 \%) \\
			\noalign{\smallskip}\hline\hline\noalign{\smallskip}
			%%%%%
    	\end{tabular*}}
\end{table}	

\cref{tab:Cutlfow_PreSel_HdbHdb_WWWW} shows signal yields at different selection stages, for the three considered signal benchmark points, for the \Hpp\Hmm pair production with $H^{\pm \pm} \rightarrow W^\pm W^\pm$ decays. 
The first set is the total number of weighted events in the signal sample.
The second set, the lepton selection, shows the number of signal events after $=0\ell$, $=1\ell$ or $=2\ell$ requirements~\footnote{For the $=0\ell$, $=1\ell$ and $=2\ell$ selections, the leptons are required to satisfy \pt$>10$~GeV, and $|\eta|<2.47$ (for electrons) or $|\eta|<2.5$ (for muons).}, 
and after the \llSC, \lll{} and~\llll channel selections. 
For the latter, the requirements on the type T and L$^*$ lepton counting from \cref{Tab:preselection} are applied. The type L lepton criteria are not imposed at this step, and the leptons must satisfy the \pt{} thresholds used at the pre-selection level. 
One can see the drastic decrease in statistics once moving to a multi-lepton final state, as well as the typical selection acceptance for the considered signal mass points. 
The last set of results, the event pre-selection, shows the signal yields after all the pre-selection requirements are applied, as well as the associated selection acceptance.
As expected, the signal yields decrease with the \Hpp{} boson mass, mainly because of the decrease in the \Hpp\Hmm pair production cross-section. 
Generally, at this stage the acceptance is below 2.1\%. An increase in acceptance with the signal mass point is expected, as the events tend to be more energetic and contain fewer soft objects, thus are more likely to pass the lepton or jet selection criteria.

\begin{samepage}
\begin{table}[t!]
\centering
    \def\arraystretch{1.5}
    \caption{
         The signal (sig) yields and the acceptance obtained for the \Hpp\Hmm pair production processes with $H^{\pm \pm} \rightarrow W^\pm W^\pm$ decays, in the defined SRs.
         The background (bkg) yields, and the associated uncertainties, are taken from Ref.~\cite{ATLAS:2021pairbosons}.
         For signal, only the MC statistical uncertainty is shown. The considered integrated luminosity is 140~fb$^{-1}$, $\sqrt s = 13$~TeV.
    }
    \label{tab:Cutlfow_SRs_HdbHdb_WWWW}
	\setlength{\tabcolsep}{0.0pc}
	\resizebox{0.75\textwidth}{!}{
		\begin{tabular*}{\textwidth}{@{\extracolsep{\fill}}llcccc}
			\noalign{\smallskip}\hline\hline\noalign{\smallskip}			
			& & \Hpp \Hmm 220~GeV & \Hpp \Hmm 300~GeV  & \Hpp \Hmm 400~GeV \\
			\noalign{\smallskip}\hline\noalign{\smallskip}
			& & &{$\llSC$ SRs}\\
		 	\noalign{\smallskip}\hline\noalign{\smallskip}
		 	%%%%%
			%\multirow[c]{3}{*}[0in]{\rotatebox{90}{$\llSC$ SRs}}
			& N sig (A) &    23.14 $\pm$ 2.01 (0.27 \%)     &   6.85 $\pm$ 0.59 (0.27 \%)       & 2.06 $\pm$ 0.17 (0.28 \%)      & \\
			& N bkg  &  14.40 $\pm$ 1.9  &  9.89 $\pm$ 1.32     & 5.46 $\pm$ 0.77 & \\
		 %	& 400 GeV  &  15.93 &  --      & -- & 0.07 $\pm$ 0.05  \\
			\noalign{\smallskip}\hline\hline\noalign{\smallskip}
			%%%%%
		    & & &{$3\ell$  SRs}\\
		    \noalign{\smallskip}\hline\noalign{\smallskip}
			& N sig (A) &  14.61 $\pm$ 1.59 (0.17 \%)      &   3.55 $\pm$ 0.42 (0.14 \%)     &    1.27 $\pm$ 0.14 (0.17 \%)   & \\
			& N bkg  &  17.97 $\pm$ 1.88 &  20.08  $\pm$ 1.89    & 15.93$\pm$ 1.42  & \\
			\noalign{\smallskip}\hline\hline\noalign{\smallskip}
			%%%%%
			& & &{$4\ell$  SRs}\\
		    \noalign{\smallskip}\hline\noalign{\smallskip}
			& N sig (A) &     3.13 $\pm$ 0.74 (0.04 \%)    &       1.12 $\pm$ 0.24 (0.04 \%)        &  0.35 $\pm$ 0.007 (0.05 \%)  & \\
			& N bkg  & 0.51  $\pm$ 0.17 &  2.10  $\pm$ 0.30   & 2.90  $\pm$ 0.40 & \\
			\noalign{\smallskip}\hline\hline\noalign{\smallskip}
			%%%%%
			%%%%%
    	\end{tabular*}}
\end{table}
\nopagebreak
\begin{table}[t!]
\centering
    \def\arraystretch{1.5}
    \caption{
        Combined signal significance $Z$ obtained for the \Hpp\Hmm pair production processes with $H^{\pm \pm} \rightarrow W^\pm W^\pm$ decays. The values are from combining the signal significance computed in the \llSC{} SR, \lll{} SR and \llll{} SR.
     }
    \label{tab:Signif_HdbHdb_WWWW}
   \resizebox{0.95\textwidth}{!}{
        \begin{tabular}{ l l c c c c c c c c c}
        \hline \hline
        	&         & \multicolumn{3}{c}{$H^{\pm\pm} H^\mp$ 220 GeV} & \multicolumn{3}{c}{$H^{\pm\pm} H^\mp$ 300 GeV}  & \multicolumn{3}{c}{$H^{\pm\pm} H^\mp$ 400 GeV} \\
        	& & \multicolumn{3}{c}{ signal significance $Z$} & \multicolumn{3}{c}{ signal significance $Z$} & \multicolumn{3}{c}{ signal significance $Z$} \\ 
        	\hline 
        	& \multirow[c]{2}{*}[0in]{Luminosity} & \multicolumn{3}{c}{ $\sqrt s$ } & \multicolumn{3}{c}{ $\sqrt s$ } & \multicolumn{3}{c}{ $\sqrt s$ } \\ 
        	 &  &  13 TeV &  13.6 TeV & 14 TeV &   13 TeV  & 13.6 TeV & 14 TeV &   13 TeV & 13.6 TeV & 14 TeV \\
        	\hline \hline
        	
	        & 140~fb$^{-1}$  &  4.43 & 3.74 & 3.79 & 1.59 & 1.66 & 1.67  & 0.72 & 0.72 & 0.66 \\
        	& 300~fb$^{-1}$  &   6.55 & 5.54 & 5.63 & 2.35 & 2.46 & 2.48   &  1.07 & 1.07 & 0.98 \\
        	& 3000~fb$^{-1}$ & 16.69 & 13.34 & 13.45 & 5.76 & 5.88 & 5.86  &  2.73 & 2.71 & 2.47   \\
        	\hline \hline
        \end{tabular}}
\end{table}

\end{samepage}

The signal yields for \Hpp\Hmm pair production with $H^{\pm \pm} \rightarrow W^\pm W^\pm$ decays, in the \llSC, \lll{} and \llll signal regions are shown in \cref{tab:Cutlfow_SRs_HdbHdb_WWWW}. 
The number of background events taken from Ref.~\cite{ATLAS:2021pairbosons} is also shown for completeness. 
The signal yields are generally lower than in Ref.~\cite{ATLAS:2021pairbosons}, as here the mass difference between the \Hpp{} and \Hm{} bosons is much smaller, thus the signal events kinematics differ. Moreover, other model parameters are also changed, as discussed earlier in this paper.

Finally, the combined signal significance $Z$ computed using the results in the \llSC, \lll{} and \llll signal regions is shown in \cref{tab:Signif_HdbHdb_WWWW}.
The results are shown for an energy in center of mass $\sqrt{s}$ of 13~TeV (LHC Run-2), 13.6~TeV (LHC Run-3) or 14~TeV (predicted for HL-LHC). The considered integrated luminosity is 140~fb$^{-1}$ (achieved at end of LHC Run-2), 300~fb$^{-1}$ (probably at the end of the LHC Run-3) or 3000~fb$^{-1}$ (possibly at HL-LHC).
To compute $Z$, the uncertainty on the background is lowered from 30\% to 20\% for the 300~fb$^{-1}$ case, and to 10\% for the 3000~fb$^{-1}$ case, respectively.
A decrease in the uncertainty is expected, as the large increase in luminosity will ensure enough statistics for a more precise background estimation.
For the backgrounds, factors of 1.1 and 1.2 are used to account for the increase in the production cross-section from 13~TeV to 13.6~TeV and 14~TeV, respectively. 
If we consider a signal mass point excluded if the signal significance $Z$ is 1.64, then the 200~GeV \Hpp mass point can be excluded and the 300~GeV mass point cannot, when using a dataset corresponding to an integrated luminosity of 140~fb$^{-1}$.
Compared to Ref~\cite{ATLAS:2021pairbosons}, the exclusion mass limits could be weaker, very likely because of the different model parametrization and constraints applied for the signal production. The sensitivity could be regained with dedicated signal regions optimization studies. For a luminosity of 3000~fb$^{-1}$, all three mass points could be excluded.

\cref{tab:Signif_HdbHdb_WWWW} shows also the signal significance $Z$ computed for $\sqrt s$ of 13.6~TeV or 14~TeV. At 13.6~TeV or 14~TeV, for a luminosity of 140~fb$^{-1}$ the 300~GeV \Hpp mass point could be excluded with no additional changes in the analysis from Ref~\cite{ATLAS:2021pairbosons}. 
With more data, for example with 3000~fb$^{-1}$, all considered benchmark signal mass points could be excluded. However, if such a model exists and the predicted \Hpp particle is real, and with a mass below or around 300~GeV, then it could be discovered.
In general, with a more realistic analysis, with a dedicated signal regions optimization and with improvements in the background estimation methodology, the discovery potential or the exclusion power can be significantly improved.

%%%%%%%%%%%%%%%%%
%%%%%%%%%%%%%%%%%
\FloatBarrier
\subsubsection{ \texorpdfstring{\Hpp\!\!\Hm}{HppHm} associated production sector 
\label{sec:HppHp}}

\begin{table}[t!]
\centering
    \def\arraystretch{1.5}
    \caption{
        The signal yields, and the acceptance, obtained for the \Hpp\Hm associated production processes with $H^{\pm \pm} \rightarrow W^\pm W^\pm$ and $H^{\mp} \rightarrow W^\mp Z$ decays. 
        Three steps are considered: no-selection, lepton selection and event pre-selection.
        Only the MC statistical uncertainty is shown, and the considered integrated luminosity is 140~fb$^{-1}$, $\sqrt s = 13$~TeV. 
    }
    \label{tab:Cutlfow_PreSel_HdbHsg_WWWZ}
	\setlength{\tabcolsep}{0.0pc}
	\resizebox{0.8\textwidth}{!}{
		\begin{tabular*}{\textwidth}{@{\extracolsep{\fill}}llcccc}
			\noalign{\smallskip}\hline\noalign{\smallskip}
			&  & & $H^{\pm\pm} H^\mp$ 220~GeV & $H^{\pm\pm} H^\mp$ 300~GeV  & $H^{\pm\pm} H^\mp$ 400~GeV \\
			& Selection & & N events ($A$) & N events ($A$) & N events ($A$) \\
			\noalign{\smallskip}\hline\noalign{\smallskip}
			%%%%%
    		& All &  &              6891.88 $\pm$ 30.86 (100.00 \%)         & 1500.06 $\pm$ 6.72  & 464.03 $\pm$ 2.08 \\
			\noalign{\smallskip}\hline\hline\noalign{\smallskip}
			%%%%%
			\multirow[c]{6}{*}[0in]{\rotatebox{90}{$\ell$ selection}}
			 & $0\ell$ & &                   3970.63 $\pm$ 23.42 (57.61 \%)          & 828.72 $\pm$ 5.00 (55.25 \%) & 248.20 $\pm$ 1.52 (53.49 \%) \\
             & $1\ell$ & &                   1835.86 $\pm$ 15.93 (26.64 \%)          & 434.75 $\pm$ 3.62 (28.98 \%) & 141.53 $\pm$ 1.15 (30.50 \%) \\
             & $2\ell$ &  &                  486.19 $\pm$ 8.20 (7.05 \%)             & 118.32 $\pm$ 1.89 (7.89 \%)  & 40.83 $\pm$ 0.62 (8.80 \%)   \\
             & $2\ell^{\mathrm{SC}}$ &  &    136.62 $\pm$ 4.34 (1.98 \%)     & 34.31 $\pm$ 1.02 (2.29 \%) & 11.39 $\pm$ 0.33 (2.46 \%) \\
             & $3\ell$ & &                   82.25 $\pm$ 3.37 (1.19 \%)      & 20.75 $\pm$ 0.79 (1.38 \%) & 7.60 $\pm$ 0.27 (1.64 \%)  \\
             & $4\ell$ &  &                  12.56 $\pm$ 1.32 (0.18 \%)      & 2.89 $\pm$ 0.30 (0.19 \%)  & 1.32 $\pm$ 0.11 (0.28 \%)  \\			
			\noalign{\smallskip}\hline\hline\noalign{\smallskip}
			%%%%%	
			\multirow[c]{3}{*}[0in]{\rotatebox{90}{pre-selection}}                           
             & $2\ell^{\mathrm{SC}}$ &  &    50.92 $\pm$ 2.65 (0.74 \%)      & 16.60 $\pm$ 0.71 (1.11 \%) & 6.07 $\pm$ 0.24 (1.31 \%)  \\
             & $3\ell$ &  &                  12.28 $\pm$ 1.30 (0.18 \%)      & 3.28 $\pm$ 0.32 (0.22 \%)  & 1.23 $\pm$ 0.11 (0.26 \%)  \\
             & $4\ell$ &  &                  11.59 $\pm$ 1.26 (0.17 \%)      & 2.59 $\pm$ 0.29 (0.17 \%)  & 1.18 $\pm$ 0.10 (0.25 \%)  \\
			\noalign{\smallskip}\hline\hline\noalign{\smallskip}
			%%%%%
    	\end{tabular*}}
\end{table}	

Results for the \Hpp\Hm associated production sector with $H^{\pm \pm} \rightarrow W^\pm W^\pm$ and $H^{\mp} \rightarrow W^\mp Z$ decays are discussed in the following. The signal yields at different selection stages, for the three signal mass points, is presented in \cref{tab:Cutlfow_PreSel_HdbHsg_WWWZ}. The pre-selection region acceptance is generally smaller than for the \Hpp\Hmm pair production case, less than 1.4\%, given the stringent criteria applied to reject the various sources of background. In addition, the pre-selection requirements in Ref.~\cite{ATLAS:2021pairbosons} were optimized considering only \Hpp\Hmm signals. 
Compared to \Hpp\Hmm signal, the \llll{} channel has a significant increase in statistics thanks to presence of the $Z$ boson in the decay chain.

\begin{samepage}
\begin{table}[t!]
\centering
    \def\arraystretch{1.5}
    \caption{
         The signal (sig) yields and the acceptance obtained for the \Hpp\Hm associated production processes with $H^{\pm \pm} \rightarrow W^\pm W^\pm$ and $H^{\mp} \rightarrow W^\mp Z$ decays, in the defined SRs.
         The background (bkg) yields, and the associated uncertainties, are taken from Ref.~\cite{ATLAS:2021pairbosons}.
         For signal, only the MC statistical uncertainty is shown. The considered integrated luminosity is 140~fb$^{-1}$, $\sqrt s = 13$~TeV.
    }
    \label{tab:Cutlfow_SRs_HdbHsg_WWWZ}
	\setlength{\tabcolsep}{0.0pc}
	\resizebox{0.75\textwidth}{!}{
		\begin{tabular*}{\textwidth}{@{\extracolsep{\fill}}llcccc}
			\noalign{\smallskip}\hline\hline\noalign{\smallskip}			
			& & $H^{\pm\pm} H^\mp$ 220~GeV & $H^{\pm\pm} H^\mp$ 300~GeV  & $H^{\pm\pm} H^\mp$ 400~GeV \\
			\noalign{\smallskip}\hline\noalign{\smallskip}
			& & &{$\llSC$ SRs}\\
		 	\noalign{\smallskip}\hline\noalign{\smallskip}
		 	%%%%%
			%\multirow[c]{3}{*}[0in]{\rotatebox{90}{$\llSC$ SRs}}
			& N sig (A) &   8.97$\pm$ 1.11 (0.13 \%)   &   1.98$\pm$ 0.24 (0.13 \%)       & 0.61 $\pm$ 0.08(0.13 \%) & \\
				& N bkg  &  14.40 $\pm$ 1.9  &  9.89 $\pm$ 1.32     & 5.46 $\pm$ 0.77 & \\
		 %	& 400 GeV  &  15.93 &  --      & -- & 0.07 $\pm$ 0.05  \\
			\noalign{\smallskip}\hline\hline\noalign{\smallskip}
			%%%%%
		    & & &{$3\ell$  SRs}\\
		    \noalign{\smallskip}\hline\noalign{\smallskip}
			& N sig (A) &   4.14$\pm$ 0.76 (0.06 \%)   &   0.81$\pm$ 0.16 (0.05 \%)       & 0.34 $\pm$ 0.06(0.07 \%) & \\
			& N bkg  &  17.97 $\pm$ 1.88 &  20.08  $\pm$ 1.89    & 15.93$\pm$ 1.42  & \\
			\noalign{\smallskip}\hline\hline\noalign{\smallskip}
			%%%%%
			& & &{$4\ell$  SRs}\\
		    \noalign{\smallskip}\hline\noalign{\smallskip}
			& N sig (A) &   4.00$\pm$ 0.74 (0.06 \%)   &   1.53$\pm$ 0.22 (0.10 \%)       & 0.68 $\pm$ 0.08(0.15 \%) & \\
				& N bkg  & 0.51  $\pm$ 0.17 &  2.10  $\pm$ 0.30   & 2.90  $\pm$ 0.40 & \\
			\noalign{\smallskip}\hline\hline\noalign{\smallskip}
			%%%%%
			%%%%%
    	\end{tabular*}}
\end{table}
\nopagebreak
\begin{table}[t!]
\centering
    \def\arraystretch{1.5}
    \caption{
        Combined signal significance $Z$ obtained for the \Hpp\Hm associated production processes with $H^{\pm \pm} \rightarrow W^\pm W^\pm$ and $H^{\mp} \rightarrow W^\mp Z$ decays. The values are from combining the signal significance computed in the \llSC{} SR, \lll{} SR and \llll{} SR.
    }
    \label{tab:Signif_HdbHsg_WWWZ}
    \resizebox{0.95\textwidth}{!}{
        \begin{tabular}{l l c c c c c c c c c}
        \hline \hline
        	&         & \multicolumn{3}{c}{$H^{\pm\pm} H^\mp$ 220 GeV} & \multicolumn{3}{c}{$H^{\pm\pm} H^\mp$ 300 GeV}  & \multicolumn{3}{c}{$H^{\pm\pm} H^\mp$ 400 GeV} \\
        	& & \multicolumn{3}{c}{ signal significance $Z$} & \multicolumn{3}{c}{ signal significance $Z$} & \multicolumn{3}{c}{ signal significance $Z$} \\ 
        	\hline 
        	& 	\multirow[c]{2}{*}[0in]{Luminosity} & \multicolumn{3}{c}{ $\sqrt s$ } & \multicolumn{3}{c}{ $\sqrt s$ } & \multicolumn{3}{c}{ $\sqrt s$ } \\ 
        	& &  13 TeV &  13.6 TeV & 14 TeV &   13 TeV  & 13.6 TeV & 14 TeV &   13 TeV & 13.6 TeV & 14 TeV \\
        	\hline \hline
	        & 140~fb$^{-1}$  &  3.55 & 3.85 & 3.61 & 0.97 & 1.11 & 1.10  & 0.40 & 0.45 & 0.57 \\
        	& 300~fb$^{-1}$  &   5.22 & 5.66 & 5.32 & 1.43 & 1.63 & 1.63   &  0.59 & 0.67 & 0.84   \\
        	& 3000~fb$^{-1}$ & 14.89 & 16.03 & 14.90 & 3.91 & 4.41 & 4.36  &  1.61 & 1.76 & 2.23   \\
        	\hline \hline
        \end{tabular}}
\end{table}

%\begin{table}[t!]
%\centering
%    \def\arraystretch{1.5}
%    \caption{
%        Combined signal significance $Z$ obtained for the \Hpp\Hm %associated production processes with $H^{\pm \pm} \rightarrow W^\pm %W^\pm$ and $H^{\mp} \rightarrow W^\mp Z$ decays, in the defined SRs. 
%        \otilia{Table caption 50\% DONE},
%        \textcolor{red}{For Alexandra: TO UPDATE the info in this %table; cross-check the N bkg events and N sig events used in the Z %computation. I'd also improve its style, as we want to show numbers %also for many $\sqrt{s}$ values}
%     }
%    \label{tab:Signif_HdbHsg_WWWZ}
%    \setlength{\tabcolsep}{0.0pc}
%    \resizebox{0.99\textwidth}{!}{
%        \begin{tabular*}{\textwidth}{@{\extracolsep{\fill}}l l l l c %c c}
%	       \noalign{\smallskip}\hline\hline\noalign{\smallskip}
%        	& & &         & $H^{\pm\pm} H^\mp$ 220 GeV & $H^{\pm\pm} %H^\mp$ 300 GeV  & $H^{\pm\pm} H^\mp$ 400 GeV \\
%        	& Selection &  & Luminosity &  $Z$ 13 / 13.6 / 14 TeV &  % $Z$ 13 / 13.6 / 14 TeV &   $Z$ 13 / 13.6 / 14 TeV \\
%	        \noalign{\smallskip}\hline\hline\noalign{\smallskip}
%        	\multirow[c]{3}{*}[0in]{\rotatebox{90}{All SRs combined}} %
%	        & & & 140~fb$^{-1}$  &  3.55 / 4.41 / 4.16 & 0.97 / 1.33 %/ 1.1  & 0.41 / 0.43 / 0.45 \\
%        	& & & 300~fb$^{-1}$  &   5.22 / 6.48 / 6.12 & 1.43 /1.96 %/ 1.63   &  0.61 / 0.63 /0.66   \\
%        	& & & 3000~fb$^{-1}$ & 14.89 / 18.47 / 17.36 & 3.91 / %5.31 /4.36  &  1.64 / 1.69 / 1.74   \\
%        
%        	\noalign{\smallskip}\hline\hline\noalign{\smallskip}
%        \end{tabular*}}
%\end{table}
\end{samepage}

Results in the signal regions are shown in \cref{tab:Cutlfow_SRs_HdbHsg_WWWZ} and \cref{tab:Signif_HdbHsg_WWWZ}. As expected, also the \llll{} SR acceptance is higher than for the \Hpp\Hmm signal. According to the obtained signal significance $Z$, the 200~GeV \Hpp{} signal mass point could be excluded at $\sqrt s = 13$~TeV, as in Ref.~\cite{ATLAS:2021pairbosons}. 
For $\sqrt s = 14$~TeV and for a luminosity of 3000~fb$^{-1}$, mass points up to 400~GeV could be excluded, and mass points up to 220~GeV could be discovered, with no changes in the analysis. However, by conducting a more pragmatic analysis, optimizing dedicated signal regions, and enhancing the methodology for background estimation, one can substantially increase the likelihood of discovery or the capability to exclude these signals.

%\FloatBarrier
%%%%%%%%%%%%%%%%%%
In the following discussion we focus on signals with $H^{\mp} \rightarrow t b$ decays, remaining within the context of the \Hpp\Hm associated production sector.
To obtain the results, the lepton selection requirements remain the same as for the studies performed for the $H^{\mp} \rightarrow W^\mp Z$ decay mode.
However, the pre-selection from \cref{Tab:preselection} was altered to include events with $b$-tagged jets ($N_{b\textrm{-tagged jets}} \geq 0$), as these events now dominate the experimental final states. In a more realistic analysis, one should replace this inclusive selection with a requirement of $\geq 1$ or $\geq 2$ $b$-tagged jets, to remove a higher fraction of background events. 
This is not done here, as the excellent $b$-tagging performance~\cite{ATLAS:2022qxm} from ATLAS experiment is not well represented when using the \texttt{DELPHES} framework. 
To increase the signal statistics in the \llSC{} channel, the $\met$ cut was relaxed to 50~GeV, down from 70~GeV.
In the \lll{} channel, the leading (sub-leading) lepton \pt{} is now $>50$~GeV ($>20$~GeV) instead of $>30$~GeV ($>10$~GeV).
Tightening these criteria will help to reduce the detector backgrounds without loosing too much signal.
The $Z$ boson decay veto was also removed, as a requirement on the number of $b$-tagged (or not) jets will remove most of the $Z \to \ell \ell$ plus one fake/non-prompt lepton background processes.
Some potential signal regions are proposed in \cref{Tab:MySRselection_HdbHsg_HsgTOtb}, starting from Ref~\cite{ATLAS:2021pairbosons} signal regions.
The pre-selection and signal region requirements are optimized by examining only the signal distributions, starting from the selections referenced in Ref~\cite{ATLAS:2021pairbosons}. Naturally, this is a simplistic approach, and a more realistic optimization should include an examination of the background as well. Nonetheless, this provides an estimate of the signal yields expected using a very straightforward approach, without significant alterations to the analysis strategy outlined in Ref~\cite{ATLAS:2021pairbosons}.

\begin{table}[!ht]
\centering
\def\arraystretch{1.5}
    \caption{
        The signal yields, and the acceptance, obtained for the \Hpp\Hm associated production processes with $H^{\pm \pm} \rightarrow W^\pm W^\pm$ and $H^{\mp} \rightarrow t b$ decays. 
        For these results, the cut on the number of $b$-tagged jets is completely removed. 
        Three steps are considered: no-selection, lepton selection and event pre-selection.
        Only the MC statistical uncertainty is shown, and the considered integrated luminosity is 140~fb$^{-1}$, $\sqrt s = 13$~TeV.
    }
    \label{tab:Cutlfow_PreSel_HdbHsg_WWtb}
    \setlength{\tabcolsep}{0.0pc}
    \resizebox{0.8\textwidth}{!}{
    \begin{tabular*}{\textwidth}{@{\extracolsep{\fill}}llccc}
    \noalign{\smallskip}\hline\noalign{\smallskip}
    &  & $H^{\pm\pm} H^{\mp}$ 220 GeV & $H^{\pm\pm} H^{\mp}$ 300 GeV  & $H^{\pm\pm} H^{\mp}$ 400 GeV \\
    & Selection & N events ($A$) & N events ($A$) & N events ($A$) \\
    \noalign{\smallskip}\hline\noalign{\smallskip}
    %%%%%
          & All &                 8331.88 $\pm$ 37.31 (100.00 \%)         & 2186.12 $\pm$ 9.80  & 464.70 $\pm$ 2.08 \\
        \noalign{\smallskip}\hline\hline\noalign{\smallskip}
        %%%%%
        \multirow[c]{6}{*}[0in]{\rotatebox{90}{$\ell$ selection}}
          & $0\ell$ &             4882.84 $\pm$ 28.56 (58.60 \%)  & 1238.07 $\pm$ 7.38 (56.63 \%)  & 257.43 $\pm$ 1.55 (55.40 \%) \\
         & $1\ell$ &             2144.03 $\pm$ 18.92 (25.73 \%)  & 599.88 $\pm$ 5.13 (27.44 \%)   & 136.08 $\pm$ 1.13 (29.28 \%) \\
         & $2\ell$ &             433.61 $\pm$ 8.51 (5.20 \%)     & 133.31 $\pm$ 2.42 (6.10 \%)    & 30.32 $\pm$ 0.53 (6.53 \%)   \\
         & $2\ell^{\mathrm{SC}}$&158.16 $\pm$ 5.14 (1.90 \%)     & 49.83 $\pm$ 1.48 (2.28 \%)     & 11.91 $\pm$ 0.33 (2.56 \%)   \\
         & $3\ell$ &             32.87 $\pm$ 2.34 (0.39 \%)      & 11.70 $\pm$ 0.72 (0.54 \%)     & 2.77 $\pm$ 0.16 (0.60 \%)    \\
         & $4\ell$ &             3.34 $\pm$ 0.75 (0.04 \%)       & 0.57 $\pm$ 0.16 (0.03 \%)      & 0.21 $\pm$ 0.04 (0.04 \%)    \\
        \noalign{\smallskip}\hline\hline\noalign{\smallskip}       
        %%%%%                                                      
        \multirow[c]{3}{*}[0in]{\rotatebox{90}{pre-selection}}     
         & $2\ell^{\mathrm{SC}}$&80.08 $\pm$ 3.66 (0.96 \%)      & 31.25 $\pm$ 1.17 (1.43 \%) & 7.95 $\pm$ 0.27 (1.71 \%) \\
         & $3\ell$ &             13.85 $\pm$ 1.52 (0.17 \%)      & 5.83 $\pm$ 0.51 (0.27 \%)  & 1.47 $\pm$ 0.12 (0.32 \%) \\
         & $4\ell$ &             0.00 $\pm$ 0.00 (0.00 \%)       & 0.00 $\pm$ 0.00 (0.00 \%)  & 0.00 $\pm$ 0.00 (0.00 \%) \\
        \noalign{\smallskip}\hline\hline\noalign{\smallskip}
        %%%%%
         & $\llSC$ SR  &  5.34 $\pm$ 0.94 (0.06 \%)      & 2.19 $\pm$ 0.31 (0.10 \%)  & 0.61 $\pm$ 0.08 (0.13 \%) \\
        \noalign{\smallskip}\hline\hline\noalign{\smallskip}
        %%%%%
        & $3\ell$ SR  & 7.01 $\pm$ 1.08 (0.08 \%)  & 3.29 $\pm$ 0.38 (0.15 \%) & 0.94 $\pm$ 0.09 (0.20 \%) \\
        \noalign{\smallskip}\hline\hline\noalign{\smallskip}
        %%%%%
    \end{tabular*}}
\end{table}

The signal yields obtained are presented in \cref{tab:Cutlfow_PreSel_HdbHsg_WWtb}.
The final state is expected to consist of three leptons and two $b$-tagged jets. 
However, due to inefficiencies in lepton selection and $b$-jet tagging, fewer leptons or less $b$-tagged jets  may be observed experimentally. Additionally, more than three leptons can be observed due to photon conversions.
After the pre-selection, the \llSC{} channel exhibits the highest acceptance for all the signal mass points (from 1.0\% to 1.7\%).
However, this channel is expected to be heavily dominated by detector background, primarily from $\ttbar$ processes, making the \lll{} channel remain  competitive. 
The \llll{} channel is not considered worthwhile for study.
The results in the signal regions, particularly when compared with the ones presented in  \cref{tab:Cutlfow_SRs_HdbHsg_WWWZ},
indicate that it would be beneficial to investigate this signal production and decay mode with 140~fb$^{-1}$ of $\sqrt s = 13$~TeV data, 
and certainly with the larger dataset anticipated for LHC Run-3.

%%%%%%%%%%%%%%%%%%%%%%%%%%%%%%%%%%%%%%%%%%%%%%%%
\FloatBarrier
\subsection{ Search for neutral Higgs bosons \label{sec:H0A0}}

The neutral sector of the Type-II Seesaw Model has never been investigated in the light of LHC phenomenology despite giving sizeable production cross-sections. As seen in \cref{fig:prod_xs3}, pair produced neutral scalars do not yield enough to be seen experimentally, the only exception being the $A^{0}H^{0}$ process. On the contrary, in association with the charged sector, some processes like $H^{\pm}A^{0}$ or $H^{\pm}H^{0}$ remain dominant. However, it is important to note that considering these processes can further reduce the cross-section times branching ratios, unlike the case with the doubly-charged Higgs boson. Irrespective of this, $H^{\pm}H^{0}$ production mode can give very complex final states through the SM Higgs decay of $H^{0}$, as discussed in \cref{sec:allFinalStates}.
Interesting experimental final states are also observed for the $A^0H^0$ production mode.

In this paper, we scrutinize only the $ p p \rightarrow \gamma^{*}/ Z^* \rightarrow A^{0}H^{0}$ associated production mode in the context of an LHC search.  
%\footnote{Because of time constraints, the production of a neutral boson in association with a charged boson will be covered in a future paper.}. 
Several decay modes are possible for both neutral bosons, as illustrated and discussed in \cref{sec:H0,sec:A0} and \cref{sec:allFinalStates}. 
Among these, only a subset is selected and considered for the generation of the signal samples---see \cref{sec:further_discussions} for a qualitative discussion of the other decay modes.
The $H^0$ boson is assumed to decay into a pair of $Z$ bosons with a maximum BR ranging from 0.6 to almost 1 for the selected \sina value, depending on the chosen values of \mHpp (see \cref{fig:BR_H1,fig:BR_H2,fig:BR_H3}), while for the $A^0$ boson, two decay modes are under consideration (see \cref{fig:BR_A1,fig:BR_A2,fig:BR_A3}):
\begin{itemize}
    \item $A^0 \to H^\pm W^{\mp *}$, with a BR decreasing progressively from almost 1 to 0.4 as \mHpp increases, and $H^\pm \to W^{\pm }Z$ decays. For the latter decay mode the maximum BR varies between 0.3 and 0.4, depending on \mHpp, for the selected \sina value (see \cref{fig:BR_Hp1,fig:BR_Hp2,fig:BR_Hp3}).  
    \item $A^0 \to H^\pm W^{\mp * }$ (BR as above), with $H^\pm \to H^{\pm\pm}W^{\mp *}$  and $H^{\pm\pm} \to W^{\pm }W^{\pm }$ decays. For the latter decay, a BR of 100\% is assumed, while for the $H^\pm \to H^{\pm\pm}W^{\mp *}$ decay mode, the maximum BR is reached around 0.5 to 0.9 values respectively, constant with \sina variations, but changing with \mHpp (see \cref{fig:BR_Hp1,fig:BR_Hp2,fig:BR_Hp3}). 
\end{itemize}
Three benchmark $H^0$ hypothetical mass points are selected: 243~GeV, 325~GeV and 437~GeV. 
The $A^0$ boson mass is set to be equal to the $H^0$ boson mass. Additional details on the parameters used for generating the signal event samples are provided in \cref{tab:neutral_ap_nominals}, in \cref{app:NominalP_Details}.
The cross-section~$\times$~BRs times the k-factor are 6.14~fb, 0.66~fb and 0.086~fb when the $H^{\pm\pm}$ boson is absent from the decay, and 10.86~fb, 0.51~fb and 0.17~fb when $H^{\pm\pm}$ is present, respectively.
This increase in the cross-section is attributed to the presence of the $H^{\pm\pm}$ boson and the fact that BR($H^{\pm} \rightarrow H^{\pm \pm} W^{\mp *}$) is higher or significantly higher than BR(\hpwz) in the first and third case respectively, but a little bit smaller in the second (see \cref{fig:BR_Hp1,fig:BR_Hp2,fig:BR_Hp3}). 

To conduct the studies, the lepton selection criteria from the prospect studies discussed in \cref{sec:charged_Sector_Prospects} are applied.
At the pre-selection level, the requirements detailed in \cref{Tab:preselection} are adjusted to align with the signal characteristics representative for the $A^0H^0$ production mode. 
Specifically, the missing transverse energy threshold is relaxed to 10~GeV, reflecting the typically low $\met$ due to the presence of only two neutrinos in the case when the $H^{\pm\pm}$ boson is not in the decay chain.
For the \lll{} channel, the condition to exclude events beneath the $Z$ mass peak is dropped, as is the constraint on the absolute sum of the lepton electric charges, $|\sum Q_\ell|$; this latter criterion is also removed in the \llll{} channel.
This approach is motivated by the abundance of $W^\pm$ and $Z$ bosons present in the decay process.
Furthermore, a minimum of three jets is now required for each event.
Some potential signal regions for the two decay modes of the $A^0$ boson are presented in Tables~\ref{Tab:MySRselection_A0H0_withoutHppInTheDecayCh} and~\ref{Tab:MySRselection_A0H0_withHppInTheDecayCh}. 
The SRs selections applied are quite similar, and independent of the boson mass.

The optimization of pre-selection criteria and signal regions is based on the work outlined in Ref.~\cite{ATLAS:2021pairbosons}, and currently relies solely on signal samples, excluding the consideration of the various background processes. 
This approach, while straightforward, is overly simplistic.
A more exhaustive optimization is essential, and it should include at least the main background sources.
Leveraging the distinctive shape differences between the multiple sources of background and the signal can significantly enhance the signal-to-background ratio. 
Nevertheless, the results obtained from applying the aforementioned selection criteria provide an estimate of the potential signal event yield in typical signal regions defined in an actual analysis.
This estimate could be achieved with minimal adjustments to the analysis presented in Ref.~\cite{ATLAS:2021pairbosons}.

%%%%%
%%%%%
%%%%%
\begin{table}[!ht]
\centering
    \def\arraystretch{1.5}
    \caption{
        The signal yields, and the acceptance, obtained for the $A^{0}H^{0}$ associated production processes with $H^0 \to ZZ$, $A^0 \to H^\pm W^{\mp *}$ and $H^\pm \to W^{\pm }Z$ decays.  
        Three steps are considered: no-selection, lepton selection and event pre-selection.
        Only the MC statistical uncertainty is shown, and the considered integrated luminosity is 140~fb$^{-1}$, $\sqrt s = 13$~TeV.
    }
\label{tab:Cutlfow_PreSel_H0A0_ZZ_WHptoWZ}
\setlength{\tabcolsep}{0.0pc}
\resizebox{0.8\textwidth}{!}{
\begin{tabular*}{\textwidth}{@{\extracolsep{\fill}}llccc}
                \noalign{\smallskip}\hline\noalign{\smallskip}
                &  & $H^0 / A^0$ 243~GeV & $H^0 / A^0$ 325~GeV & $H^0 / A^0$ 437~GeV \\
                & Selection & N events ($A$) & N events ($A$) & N events ($A$) \\
                \noalign{\smallskip}\hline\noalign{\smallskip}
%%%%%
  & All &                 860.91 $\pm$ 3.86 (100.00 \%)   & 93.01 $\pm$ 0.42 & 13.80 $\pm$ 0.06  \\
\noalign{\smallskip}\hline\hline\noalign{\smallskip}
%%%%%
\multirow[c]{6}{*}[0in]{\rotatebox{90}{$\ell$ selection}}
 & $0\ell$ &             560.72 $\pm$ 3.11 (65.13 \%)    & 58.93 $\pm$ 0.33 (63.36 \%) & 8.46 $\pm$ 0.05 (61.31 \%)  \\
 & $1\ell$ &             147.02 $\pm$ 1.59 (17.08 \%)    & 16.79 $\pm$ 0.18 (18.05 \%) & 2.59 $\pm$ 0.03 (18.76 \%)  \\
 & $2\ell$ &             64.56 $\pm$ 1.06 (7.50 \%)      & 7.88 $\pm$ 0.12 (8.47 \%)   & 1.28 $\pm$ 0.02 (9.31 \%)   \\
 & $2\ell^{\mathrm{SC}}$&7.31 $\pm$ 0.36 (0.85 \%)       & 0.90 $\pm$ 0.04 (0.97 \%)   & 0.14 $\pm$ 0.01 (1.02 \%)   \\
 & $3\ell$ &             13.73 $\pm$ 0.49 (1.59 \%)      & 1.83 $\pm$ 0.06 (1.97 \%)   & 0.30 $\pm$ 0.01 (2.20 \%)   \\
 & $4\ell$ &             4.10 $\pm$ 0.27 (0.48 \%)       & 0.51 $\pm$ 0.03 (0.55 \%)   & 0.10 $\pm$ 0.01 (0.69 \%)   \\
\noalign{\smallskip}\hline\hline\noalign{\smallskip}
%%%%%
\multirow[c]{3}{*}[0in]{\rotatebox{90}{pre-selection}}
 & $2\ell^{\mathrm{SC}}$&3.10 $\pm$ 0.23 (0.36 \%)       & 0.40 $\pm$ 0.03 (0.43 \%) & 0.08 $\pm$ 0.00 (0.56 \%)  \\
 & $3\ell$ &             7.52 $\pm$ 0.36 (0.87 \%)       & 1.10 $\pm$ 0.05 (1.18 \%) & 0.19 $\pm$ 0.01 (1.37 \%)  \\
 & $4\ell$ &             2.67 $\pm$ 0.21 (0.31 \%)       & 0.35 $\pm$ 0.03 (0.37 \%) & 0.06 $\pm$ 0.00 (0.47 \%)  \\
\noalign{\smallskip}\hline\hline\noalign{\smallskip}
%%%%%
 & $\llSC$ SR &  0.45 $\pm$ 0.09 (0.05 \%)       & 0.02 $\pm$ 0.01 (0.02 \%) &  0.00 $\pm$ 0.00 (0.01 \%) \\
\noalign{\smallskip}\hline\hline\noalign{\smallskip}
%%%%%
 & $3\ell$ SR  &  2.95 $\pm$ 0.23 (0.34 \%)       & 0.47 $\pm$ 0.03 (0.50 \%)  & 0.06 $\pm$ 0.00 (0.44 \%) \\
\noalign{\smallskip}\hline\hline\noalign{\smallskip}
%%%%%
 & $4\ell$ SR &  0.91 $\pm$ 0.13 (0.11 \%)       & 0.13 $\pm$ 0.02 (0.14 \%) & 0.02 $\pm$ 0.00 (0.13 \%) \\
\noalign{\smallskip}\hline\hline\noalign{\smallskip}
%%%%%
%%%%%
\end{tabular*}}
\end{table}

The number of signal events obtained for the $A^{0}H^{0}$ associated production mode with $H^0 \to ZZ$, $A^0 \to H^\pm W^{\mp *}$, and $H^\pm \to W^{\mp}Z$ decays is presented in \cref{tab:Cutlfow_PreSel_H0A0_ZZ_WHptoWZ}. 
The event count when no requirements are applied (entry ``All"), indicates that for an $\sqrt s$ of 13~TeV and an integrated luminosity of 140~fb$^{-1}$, the statics are reasonable only for the lowest mass point. 
For higher mass points, the statistics are not sufficient, which is somewhat expected given the low production cross-section $\times$ BRs values. 
Results are also presented after the lepton selection and at the pre-selection stage.
Given the high SM background expected for all $0\ell$, $1\ell$ and $2\ell$ selections,
it seems highly unlikely that a feasible analysis could be designed and performed with any of these channels using the LHC data.
The results obtained for the lowest mass point at the pre-selection level suggest that the most sensitive channel could be the \lll{} one. 
However, since this particular signal model predicts three $Z$ and two $W^\pm$ bosons, a channel defined with 4 or 5 type L leptons---of which one or two meet the type T criteria, to account for the lepton selection at trigger level~\cite{ATLAS:2019dpa,ATLAS:2020gty}---might be worth considering. 
Such high lepton multiplicity final states are dominated by very few SM processes, like $ZZZ$, or $ZZW^{\mp }$. 
Nonetheless, challenges are expected in estimating the fake/non-prompt lepton background, due to the very low number of anticipated data events.
The results in the signal regions optimized for this study are also displayed. They corroborate the observations and conclusions drawn from the previous selections.
Overall, with the increase in center-of-mass energy and the expected higher luminosity at the LHC Run-3 and HL-LHC, pursuing a search for $A^{0}H^{0}$ with $H^0 \to ZZ$, $A^0 \to H^\pm W^{\mp *}$, and $H^\pm \to W^{\mp }Z$ decays may be worthwhile.

%%%%%
%%%%%
%%%%%
\begin{table}[!ht]
\centering
\def\arraystretch{1.5}
    \caption{
        The signal yields, and the acceptance, obtained for the $A^{0}H^{0}$ associated production processes with $H^0 \to ZZ$, $A^0 \to H^\pm W^{\mp *}$, $H^\pm \to H^{\pm\pm}W^{\mp *}$ and $H^{\pm\pm} \to W^{\pm }W^{\pm }$ decays. 
         Three steps are considered: no-selection, lepton selection and event pre-selection.
         Only the MC statistical uncertainty is shown, and the considered integrated luminosity is 140~fb$^{-1}$, $\sqrt s = 13$~TeV.
    }
\label{tab:Cutlfow_PreSel_H0A0_ZZ_WHptoWHpptoWW}
\setlength{\tabcolsep}{0.0pc}
\resizebox{0.8\textwidth}{!}{
\begin{tabular*}{\textwidth}{@{\extracolsep{\fill}}llccc}
            \noalign{\smallskip}\hline\noalign{\smallskip}
                &  & $H^0 / A^0$ 243~GeV & $H^0 / A^0$ 325~GeV & $H^0 / A^0$ 437~GeV \\
                & Selection & N events ($A$) & N events ($A$) & N events ($A$) \\
                \noalign{\smallskip}\hline\noalign{\smallskip}
%%%%%
  & All & 1522.66 $\pm$ 6.82 (100.00 \%)          & 73.25 $\pm$ 0.33  & 26.85 $\pm$ 0.12  \\
\noalign{\smallskip}\hline\hline\noalign{\smallskip}
%%%%%
\multirow[c]{6}{*}[0in]{\rotatebox{90}{$\ell$ selection}}
 & $0\ell$ &             890.49 $\pm$ 5.22 (58.48 \%)    & 41.43 $\pm$ 0.25 (56.56 \%) & 14.20 $\pm$ 0.09 (52.87 \%)  \\
 & $1\ell$ &             328.79 $\pm$ 3.17 (21.59 \%)    & 16.81 $\pm$ 0.16 (22.95 \%) & 6.40 $\pm$ 0.06 (23.83 \%)   \\
 & $2\ell$ &             108.42 $\pm$ 1.82 (7.12 \%)     & 5.79 $\pm$ 0.09 (7.91 \%)   & 2.47 $\pm$ 0.04 (9.21 \%)    \\
 & $2\ell^{\mathrm{SC}}$&36.82 $\pm$ 1.06 (2.42 \%)      & 1.98 $\pm$ 0.05 (2.70 \%)   & 0.84 $\pm$ 0.02 (3.15 \%)    \\
 & $3\ell$ &             26.02 $\pm$ 0.89 (1.71 \%)      & 1.43 $\pm$ 0.05 (1.96 \%)   & 0.66 $\pm$ 0.02 (2.44 \%)    \\
 & $4\ell$ &             7.26 $\pm$ 0.47 (0.48 \%)       & 0.42 $\pm$ 0.02 (0.57 \%)   & 0.19 $\pm$ 0.01 (0.72 \%)    \\
\noalign{\smallskip}\hline\hline\noalign{\smallskip}
%%%%%
\multirow[c]{3}{*}[0in]{\rotatebox{90}{pre-selection}}
 & $2\ell^{\mathrm{SC}}$&20.23 $\pm$ 0.79 (1.33 \%)      & 1.13 $\pm$ 0.04 (1.54 \%) & 0.53 $\pm$ 0.02 (1.97 \%) \\
 & $3\ell$ &             14.40 $\pm$ 0.66 (0.95 \%)      & 0.85 $\pm$ 0.04 (1.16 \%) & 0.41 $\pm$ 0.01 (1.51 \%) \\
 & $4\ell$ &             4.88 $\pm$ 0.39 (0.32 \%)       & 0.29 $\pm$ 0.02 (0.40 \%) & 0.14 $\pm$ 0.01 (0.50 \%) \\
\noalign{\smallskip}\hline\hline\noalign{\smallskip}
%%%%%
 & $\llSC$ SR  &  7.35 $\pm$ 0.47 (0.48 \%)       & 0.33 $\pm$ 0.02 (0.45 \%)  & 0.09 $\pm$ 0.01 (0.35 \%)   \\
\noalign{\smallskip}\hline\hline\noalign{\smallskip}
%%%%%
 & $3\ell$ SR  & 6.13 $\pm$ 0.43 (0.40 \%)       & 0.35 $\pm$ 0.02 (0.48 \%) & 0.14 $\pm$ 0.01 (0.52 \%)  \\
\noalign{\smallskip}\hline\hline\noalign{\smallskip}
%%%%%
 & $4\ell$ SR & 1.37 $\pm$ 0.20 (0.09 \%)       & 0.07 $\pm$ 0.01 (0.09 \%) & 0.03 $\pm$ 0.00 (0.12 \%) \\
\noalign{\smallskip}\hline\hline\noalign{\smallskip}
%%%%%
%%%%%
\end{tabular*}}
\end{table}

\cref{tab:Cutlfow_PreSel_H0A0_ZZ_WHptoWHpptoWW} presents the number of signal events for the $A^{0}H^{0}$ associated production mode with $H^0 \to ZZ$, $A^0 \to H^\pm W^{\mp *}$, $H^\pm \to H^{\pm\pm}W^{\mp *}$, and $H^{\pm\pm} \to W^{\mp }W^{\mp }$ decays.
This extended decay chain offers intriguing possibilities for experimental final states suitable for analysis.
As for the other production modes, event yields are displayed at various selection stages.
Similar to the previously discussed decay chain, the statistics at the LHC Run-2 are sufficient primarily for the lowest mass point.
Upon examining these results, both the \llSC{} and \lll{} channels appear promising.
The \llll{} channel exhibits low statistics, likely due to its definition involving only type T leptons, which are subject to stringent isolation and identification criteria.
Investigating a channel with five type L leptons could also be interesting, and 
a six lepton channel would certainly be fascinating to examine. 
Given that some of these leptons may be very soft, reducing the lepton \pt{} for some of the sub-leading leptons might significantly increase the statistics.
With the expanded search capabilities afforded by machine learning algorithms, such searches could become feasible, as they allow for the mitigation of fake/non-prompt lepton backgrounds with minimal signal loss.

The anticipated increase in center-of-mass energy and the considerably higher luminosity at the HL-LHC make the pursuit of searches for $A^0H^0$ associated processes particularly promising. These conditions are expected to enhance the potential for discovery and provide a richer dataset for analysis. Furthermore, the advancements in detection technology and data analysis methods, including machine learning algorithms, will significantly aid in distinguishing signal from background sources.

%%%%%%%%%%%%%%%%%%%%%%%%%%%%%%%%%%%%%%%%%%%%%%%%
%%%%%%%%%%%%%%%%%%%%%%%%%%%%%%%%%%%%%%%%%%%%%%%%
%%%%%%%%%%%%%%%%%%%%%%%%%%%%%%%%%%%%%%%%%%%%%%%%
\FloatBarrier
\section{Further discussions}
\label{sec:further_discussions}

As shown in \cref{sec:decayModes,sec:allFinalStates}, several decay modes can occur with significant relative contributions. Here we discuss qualitatively their possible impact on the assessment of the experimental search potential when specific decay modes are considered, such as the ones investigated in the previous section. \vspace{0.5cm}

\noindent
\underline{\Hpp \sl pair production sector:} In \cref{sec:HppHpp}, the search analysis for \Hpp pair production through multi-lepton final states assumed 100\% BR for $H^{\pm \pm} \rightarrow W^\pm W^\pm$.  Although the \Hp mass does not enter here, the validity of the conclusions ensuing from this analysis presupposes implicitly that $m_{H^\pm}$ is sufficiently close to or above  $m_{H^{\pm \pm}}$, in order to justify this BR assumption, cf. \cref{fig:BR_Hpp}. In contrast, if \Hp is much lighter than \Hpp then the decay channel $H^{\pm \pm} \rightarrow H^\pm f \bar{f}'$ becomes sizeable, and could even largely dominate, leading to quite different final states as stressed in \cref{sec:Hp}. 
This happens more easily for relatively light \Hpp bosons, for instance for the two first benchmark signal points given in \cref{tab:charged_ap_pp_nominals}, if the nominal \Hp masses are lowered by say 10 to 15~GeV. The decay $H^{\pm \pm} \rightarrow H^\pm f \bar{f}'$ is then followed by the decay of $H^\pm$ dominantly to an off-shell $W^\pm$ in association with an on-shell $H^0$ or $A^0$, with 50\% BR each, and further cascade decays of the latter. This brings in another important uncertainty in identifying the dominant final states, due to the strong sensitivity to \sina parameter as shown in \cref{sec:H0,sec:A0}: $H^0$ dominantly decays to $WW$ or $ZZ$ for some ranges of \sina, and the considered benchmark points $m_{H^{\pm\pm}}=220$ and $300$~GeV (\cref{fig:BR_H1,fig:BR_H2}) lead to intermediate states with high W or Z multiplicity---$4W^* + 4W$, or $4Z$, or $2W +2Z$---thus to higher lepton multiplicity for which the analysis of \cref{sec:HppHpp} with $\geq 4\ell$ selection criteria could apply. 
However, keeping in mind that four $W^\pm$'s are off-shell and assuming the narrow width approximation (for the on-shell objects) reliable throughout the decay chain, the signal yield would be around a factor four smaller as compared to the $H^{\pm \pm} \rightarrow W^\pm W^\pm$ case.
The reason is that the yield comes only from the two decay chains containing an $H^0$ boson each, $H^{\pm\pm} \to H^\pm \to H^0$, and not from the decay chain containing the $A^0$ boson, since the latter decays to $WW$ or $ZZ$ are forbidden at the tree-level. The decrease in the yield could be compensated by higher $4\ell$ statistics when pairs of
$Z$'s are present, although this suffers from the irreducible uncertainty on \sina.

For some other ranges of \sina, $H^0$ decays dominantly to $h^0h^0$, a pair of SM-like scalars, as shown in \cref{fig:BR_H2,fig:BR_H3}. Here, the final state mainly consists of $8b$ jets in addition to the decay products of the four off-shell $W^\pm$ bosons.
The analysis selection presented in \cref{sec:HppHpp} is not adequate for this topology, as the $b$-jets are explicitly removed from the event (see \cref{Tab:preselection}).
A dedicated optimization is needed to cover such cases, one that will take advantage of the very low SM background in selections with $\geq 4b$ jets.
Despite the experimental challenges of $b$-jet tagging, an analysis with leptons and many jets is very interesting and worth performing.

If $A^0$ is present in the cascade, instead of $H^0$, its dominant decay channels in the lower part of its mass range under consideration are $b\bar{b}$ (\cref{fig:BR_A1}) or $h Z^{(*)}$ (\cref{fig:BR_A2}), independently of \sina, or
$h Z$, $t\bar{t}$, or a mixture of the two depending on \sina (\cref{fig:BR_A3}), respectively for increasing \Hpp masses. We see that, independently from \sina and the $A^0$ and \Hpp masses, all final states contain exactly $4b$ jets and at least $4\ell$. Note also that requiring events with exactly $4\ell$ would select uniquely $A^0$ decays to $b\bar{b}$ which, if dominant, would favor qualitatively low scalar masses and low \sina values. 
%for the lighter \Hpp mass points, while for heavier mass points the uncertainty on \sina translates into possible increase of the $b$-jet multiplicity from top-quark decays.
As for the previous case of $H^0$, with multiple $b$-jets in the event the analysis from \cref{sec:HppHpp} will not be directly applicable, and a dedicated signal regions optimization is needed.

Finally, since $H^0$ and $A^0$ have equal masses, one expects the various final states from $H^0$ and $A^0$ described above to be present in an inclusive analysis, resulting in a correspondingly higher yield. \vspace{0.5cm}

\noindent
\underline{\sl \Hpp\Hm associated production sector}: If the \Hp mass is 10 to 15~GeV lower than those of the benchmark signal points, the implications are essentially the same as above and will not be repeated here.
We only note that even though one of the two decay chains will have now one less off-shell $W^\pm$ in the cascade, the overall lepton and b-jet multiplicities remains comparable to that of the \Hpp pair production. The selection criteria of \cref{sec:HppHp} that accept $b$-tagged jets could then apply. Nevertheless, as mentioned earlier, it would be better to have a dedicated analysis optimization for the \Hpp\Hm associated production, also considering the presence or absence of $b$-tagged jets. Using the same strategy for both associated and pair production modes has already been found to be ineffective.

For \Hp masses as selected for the benchmark signal points, or in a range roughly $m_{H^{\pm}}{}_{-10 \rm{GeV}}^{+20 \rm{GeV}}$ around these values, but with somewhat higher values of \sina than the benchmark value of  \cref{tab:charged_ap_pp_nominals}, \Hp can decay significantly, or even dominantly, to $h^0W^\pm$ instead of $tb$ or $ZW^\pm$ that were considered for the analysis in \cref{sec:HppHp}. As far as lepton and b-jet multiplicities are concerned, the $h^0W^\pm$ and $tb$ channels could contribute similarly. Combining them inclusively would increase the signal yields well above the ones reported in
\cref{tab:Cutlfow_PreSel_HdbHsg_WWtb}, especially for the heavier \Hpp benchmark signal points,
see \cref{fig:BR_Hp2,fig:BR_Hp3}.
In addition to improving the analysis sensitivity, this could also help reduce the sensitivity to the values of \sina.

Increasing the \Hp mass by $\lesssim 30$~GeV above the benchmark signal points, the decay channel $H^\pm \to H^{\pm\pm} W^{*\mp}$ with an on-shell \Hpp, opens up and can become dominant especially for the lowest \Hpp mass point, as illustrated in \cref{fig:BR_Hp1,fig:BR_Hp2,fig:BR_Hp3}. Since in this part of the parameter space \Hpp decays 100\% to $W^\pm W^\pm$, the resulting cascade decay could yield the same lepton multiplicity as the the $ZW^\pm$ channel. 
A search for five leptons would be background-free but not necessarily feasible, as lepton identification could be problematic. Most likely, a requirement of $\geq 4$ leptons, as in \cref{sec:HppHp}, might work better.
On the other hand, a search with fewer leptons, plus several non-$b$-tagged jets, could be of interest as the differences in the signal and background kinematics can be used to define sensitive signal regions. Nevertheless, if a channel with only two leptons of the same electric charge, or with three leptons, jets, and missing transverse momentum is considered---to avoid the loss in production cross-section times BRs due to the low $W/Z\to\ell$ BRs---a similar analysis strategy could again be applied to both decay modes. 
Note that, in this case as well, taking the two channels inclusively would increase the signal yields reported in \cref{tab:Cutlfow_SRs_HdbHsg_WWWZ}.\vspace{0.5cm}

\noindent
\underline{\sl $H^0A^0$ associated production sector}: The analysis discussed in \cref{sec:H0A0} relies on mass points satisfying the hierarchy
$m_{H^{\pm\pm}} \leq m_{H^{\pm}} \leq m_{H^0} (\simeq m_{A^0})$. This leads to interesting cascade decays depending on the decay mode of the intermediate \Hp boson. However, lowering the $H^0 (A^0)$ mass by a few GeV while maintaining the same mass hierarchy and keeping the $H^0 \to ZZ$ channel dominant would result in a direct decay of $A^0$ to $h^0 Z^{(*)}$ (and for the highest mass point to $t\bar t$ as well). This would lead to shorter decay chains and a reduction of the lepton multiplicity by at least 2 for comparable cross-sections, assuming the dominant $h^0$ decay to two b-quarks.
As already hinted, $b$-tagged jets will also be present in the event, and the analysis from \cref{sec:H0A0} will not apply as it is. 

A more important modification can come from the sensitivity to \sina. 
The $H^0$ mainly decaying to $ZZ$ in the analysis of \cref{sec:H0A0} occurs for very small \sina values, as chosen for the benchmark points (\cref{app:NominalP_Details}). 
An increase of this unknown parameter by, say, 50\% drastically changes the configuration, especially for the higher \Hpp mass points, as seen in \cref{fig:BR_H2,fig:BR_H3}.
The dominant decays become $H^0 \to h^0h^0 \to 4b$
and $A^0\to h^0 Z \to 2b Z$, reducing the lepton multiplicity by 6 compared to the analysis in 
\cref{sec:H0A0} for the same benchmark points (\cref{tab:neutral_ap_nominals}), and essentially the same, or even higher, cross-section.
This remains true even beyond the chosen benchmark points for virtually the whole range of $m_{H^0},m_{A^0}$ masses. 
To account for the new final state signatures, a complete optimization of the \cref{sec:H0A0} analysis is needed. A search with same-sign or multiple leptons could be performed, though this would come at the cost of a decrease in the production cross-section times BR. In this case, signal regions with $3b$-tagged jets will ensure a small SM background.
Alternatively, one could consider a final state with two leptons of different charge and at least $3$ to $6$ $b$-tagged jets---the $3b$ requirement effectively removing a significant fraction of the SM background (such as $t\bar{t}$). Signal regions defined in bins of the number of $b$-tagged jets will ensure full coverage of the phase space.

%%%%%%%%%%%%%%%%%%%%%%%%%%%%%%%%%%%%%%%%%%%%%%%%
%%%%%%%%%%%%%%%%%%%%%%%%%%%%%%%%%%%%%%%%%%%%%38%%%
%%%%%%%%%%%%%%%%%%%%%%%%%%%%%%%%%%%%%%%%%%%%%%%%
\FloatBarrier
\section{Conclusion \label{sec:conclusion}}

In this paper, we considered the phenomenology at the LHC of a scalar $\rm SU(2)_L$ complex triplet with hypercharge $2$, typically present in the Type-II Seesaw Model. If lepton-number-violating decays of the physical states are suppressed, the present LHC mass exclusion limits for the charged and doubly-charged states produced in pairs or in association are so far mild, around $300$~GeV, while experimental searches for the neutral states are yet to come.
Under this assumption, we carried out a detailed survey of all decay channels of all the scalars of the model, decaying either directly to SM particles or through cascades.
In particular, we pinpointed an important sensitivity of the decay branching ratios to the mixing angle between the CP-even scalars, in regions where this angle is too small to be independently probed by the SM-Higgs experimental studies. This sensitivity entails the coexistence of different decay channels with relative contributions that can be quite comparable~---~once kinematically open~---~independently of their available phase-space. 
This gives rise to a theoretical uncertainty that calls for a comprehensive strategy for experimental searches, rather than one based on specific channels for given mass ranges. Although illustrated in this paper for a fixed triplet VEV, the genericity of the effects is clearly demonstrated. The study is complementary to the existing literature where the relevance of the various channels is often presented as a function of the triplet VEV and phase space, through which the important effect of the mixing angle is not manifest.

We also carried out a detailed prospective study for the LHC, assuming an ATLAS-like detector, relying on a few benchmark scenarios for the production and decays of (doubly-)charged scalars following existing ATLAS search analyses, and for the associated neutral scalar production, which is presented for the first time. Results were obtained for center-of-mass energies of 13~TeV (LHC Run-2), 13.6~TeV (LHC Run-3), or 14~TeV (predicted for HL-LHC), and for integrated luminosities of 140~fb$^{-1}$ (achieved at the end of LHC Run-2), 300~fb$^{-1}$ (probably at the end of LHC Run-3), or 3000~fb$^{-1}$ (possibly at HL-LHC). The obtained projections show which mass points could be excluded or potentially discovered if the model is real. Other prospective future studies should be considered in light of wider search strategies, including scenarios where lepton-number and flavor-violating decays become comparable to those studied in this paper. In addition, studies for future accelerator machines~--~like ILC and FCC~--~would also be interesting to perform, as these make a case for a TeV-range \Hpp{} mass for the leptonic decays and a half-TeV range for di-boson decay modes.
 % \cite{Yang-Yang:2022prospect,CMS:2022prospect}

%%%%%%%%%%%%%%%%%%%%%%%%%%%%%%%%%%%%%%%%%%%%%%%%
%%%%%%%%%%%%%%%%%%%%%%%%%%%%%%%%%%%%%%%%%%%%%%%%
%%%%%%%%%%%%%%%%%%%%%%%%%%%%%%%%%%%%%%%%%%%%%%%%
\acknowledgments 
We would like to thank Lorenzo Basso for his crucial input, as well as Cristinel Diaconu, Venugopal Ellajosyula and Yanwen Liu for contributing to the present work at an early stage.
We also benefited from insightful discussions with  Calin Alexa, Julien Maurer, Elisabeth Petit, Dorel Pietreanu,  Marina Rotaru and Valentina Tudorache.
%We thank ZZ for a critical reading of the manuscript, and WW from
%IFIN-HH for support.
%This work has been carried out thanks to the support of the OCEVU Labex (ANR-11-LABX-0060), the  A$\star$MIDEX project (ANR-11-IDEX-0001-02) funded by the ``Investissements d'Avenir" French government program managed by the ANR, 
This work received support from the French government under the France 2030 investment plan, as part of the Excellence Initiative of Aix Marseille University - A*MIDEX (AMX-19-IET-008 - IPhU), and  support from IFIN-HH under the Contract ATLAS CERN-RO with the Romanian MCID / IFA. GM has received partial support from the European Union’s Horizon 2020 research and innovation programme under the Marie Skłodowska-Curie grant agreements No 860881-HIDDeN and No
101086085–ASYMMETRY.

%%%%%%%%%%%%%%%%%%%%%%%%%%%%%%%%%%%%%%%%%%%%%%%%
%%%%%%%%%%%%%%%%%%%%%%%%%%%%%%%%%%%%%%%%%%%%%%%%
%%%%%%%%%%%%%%%%%%%%%%%%%%%%%%%%%%%%%%%%%%%%%%%%
%\newpage
%\FloatBarrier
\appendix
%%%%%%%%%%%%%%%%%%%%%%%%%%%%%%%%%%%%%%%%%%%%%%%%
%%%%%%%%%%%%%%%%%%%%%%%%%%%%%%%%%%%%%%%%%%%%%%%%
\section{Parameterization strategy}
\label{app:paramstrategy}

We summarize here, without entering into the detailed derivation, the main steps of the procedure advocated in \cref{sec:parameters}, where the set of parameters $m_{h^0}$, $m_{H^\pm}$, $m_{H^{\pm \pm}}$, $\sin\alpha$, $\lambda_2$, $\lambda_3,v_d$ and $v_t$ is taken as input.
Starting from the expression of $m_{h^0}$ given by \cref{eq:mh0}
and \sina as given by \cref{eq:sinalpha}, one can solve for $A$ and $B$
leading to a double solution.
However, the mathematical consistency requirement $m_{h^0}^2 < m_{H^0}^2$, cf.  \cref{eq:mh0}, 
implies the consistency condition
\begin{equation}
    C > m_{h^0}^2, \label{eq:Ccondition}
\end{equation}
and eliminates  a spurious solution. One is left with the unique solutions
\begin{align}
&A=m_{h^0}^2 + (C - m_{h^0}^2) \tan^2\alpha, \label{eq:A}\\
&B= (m_{h^0}^2-C) \tan\alpha. \label{eq:B}
\end{align}
 In practice we choose by convention $\cos\alpha >0$, i.e. $\epsilon_{\alpha}=+$ in \cref{eq:sinalpha}, so that without loss of generality
 $\tan\alpha$ and the input \sina have the same sign.

The parameters $\mu$ and $\lambda_4$ are straightforwardly obtained 
from \cref{eq:mHpm,eq:mHpmpm}. 
\begin{align}
 &\mu = \sqrt{2} v_t\left(2 \, \frac{m_{H^+}^2}{v_d^2 + 2 v_t^2} - \frac{m_{H^{++}}^2 + \lambda_3 v_t^2}{v_d^2} \right), \label{eq:mu}\\
 &\lambda_4 = 4 \, \frac{m_{H^+}^2}{v_d^2 + 2 v_t^2} - 4 \, \frac{m_{H^{++}}^2 + \lambda_3 v_t^2}{v_d^2} \label{eq:lambda4}  
\end{align}
It is now clear that taking as input $m_{h^0}$, $m_{H^\pm}$, $m_{H^{\pm \pm}}$, $\sin\alpha$, $\lambda_2$, $\lambda_3$, $v_d$ and $v_t$, one determines $\mu$ and $\lambda_4$ from the above two equations. $C$ is then fully determined, cf. \cref{eq:ABC},  which allows to check the consistency condition  \cref{eq:Ccondition}. Finally, $\lambda$ and $\lambda_1$ that enter linearly $A$ and $B$, \cref{eq:ABC}, are uniquely determined 
from \cref{eq:A,eq:B}. The knowledge of all these parameters fixes then the masses of $A^0$ and $H^0$ through \cref{eq:mA0,eq:mh0}.

Although in the present phenomenological study we stick only to the configurations where \cref{eq:Ccondition} is satisfied, we outline hereafter for completeness how to treat the opposite case. 

If \cref{eq:Ccondition} is violated, then the input set of values can still be made consistent provided that the input value for $m_{h^0}$ is interpreted as the heavier
rather than the lighter CP-even state. Let us relabel this input $m_{H^0}$ and write
\begin{align}
    C < m_{H^0}^2. \label{eq:Cconditionother}
\end{align}

In practice this would mean the presence of a CP-even (and a CP-odd) state lighter than 125~GeV~\cite{Arhrib:2014nya}. In this case, \cref{eq:A,eq:mu,eq:lambda4} preserve their forms (apart from the relabelling $m_{h^0} \to m_{H^0}$), while \cref{eq:B} becomes
\begin{align}
 B= (C-m_{H^0}^2) \cot\alpha.   
\end{align}
The rest of the procedure works as in the previous case.

%%%%%%%%%%%%%%%%%%%%%%%%%%%%%%%%%%%%%%%%%%%%%%%%
%%%%%%%%%%%%%%%%%%%%%%%%%%%%%%%%%%%%%%%%%%%%%%%%
\clearpage % needed, to not have the title only on the previous page
\section{Scalar sector BSM Higgs widths}
\label{app:widths}

We list hereafter the tree-level analytical expressions of the widths for the two-body  decays  of the neutral and (doubly-)charged scalars, but {\sl only the  on-shell} configurations. These serve mainly as guides when discussing the sensitivities to the model parameters and as validation of the UFO files output. The (approximate) analytical expressions when one of the gauge bosons decays off-shell, have been derived in several places \cite{PhysRevD.22.722,PhysRevD.30.248,Cahn:1990xc}, \cite{Aoki:2011pz,Djouadi:1995gv}.
We do not reproduce them here as we relied on the fully numerical evaluation with \texttt{Madgraph}. An updated list with complete expressions can be found in \cite{Ashanujjaman:2021txz}. In the following, $e$ stands for the electric charge and $s_W$ ($c_W$) for $\sin\theta_W$ ($\cos\theta_W$); all other quantities have been defined previously. For a comparison among the various widths, one can when needed re-express  $v_d$ approximately in terms of $M_W$ using \cref{eq:mW} and the relation $e=g \sin\theta_W$.

\begin{align}
&\Gamma_{H^{\pm\pm} \to W^\pm W^\pm}= \frac{v_t^2 e^4}{64 \pi s_{\rm W}^4}\sqrt{m_{H^{++}}^2 - 4 M_W^2} \, \frac{8 M_W^4 + (m_{H^{++}}^2 - 2 M_W^2)^2}{m_{H^{++}}^2 M_W^4},\label{eq:H++W+W+}\\
&\Gamma_{H^{\pm\pm} \to W^\pm H^\pm}=  \cos^2\beta' \frac{e^2}{16 \pi s_{\rm W}^2} \frac{\left((m_{H^{++}}^2 - (m_{H^+} - M_W)^2) (m_{H^{++}}^2 - (m_{H^+} + M_W)^2)\right)^{\frac32}}{m_{H^{++}}^3 M_W^2},\label{eq:H++W+H+}\\
&\Gamma_{H^{\pm\pm} \to H^\pm H^\pm}= \left(4 \mu \sin^2\beta' +  \lambda_4 v_d \sin2\beta' - 
   2 \sqrt{2} \lambda_3 v_t \cos^2\beta' \right)^2 \frac{\sqrt{m_{H^{++}}^2 - 4 m_{H^+}^2}}{128 \pi m_{H^{++}}^2 }, \label{eq:H++H+H+}
\end{align}
\begin{align}
&\Gamma_{H^\pm \to Z W^\pm}= 
%\frac{e^4\left(\sin\beta' s_{\rm W}^2 v_d - \sqrt{2} \cos\beta' (1 + s_{\rm W}^2) v_t\right)^2} {256 \pi c_{\rm W}^2 s_{\rm W}^4 M_W^2 M_Z^2 m_{H^+}^3} \times \nonumber \\
\frac{e^4 {\cos\beta'}^2 v_t^2}{128 \pi c_{\rm W}^2 s_{\rm W}^4 M_W^2 M_Z^2 m_{H^+}^3} \times \nonumber \\
& \sqrt{\left(m_{H^+}^2 \!-\! (M_W \!-\! M_Z)^2\right) \left(m_{H^+}^2 \!-\! (M_W \!+\! M_Z)^2\right)} \left(8 M_W^2 M_Z^2 \!+\! (m_{H^+}^2 \!-\! M_W^2 \!-\! M_Z^2)^2\right),\label{eq:H+ZW+}\\
&\Gamma_{H^\pm \to \gamma W^\pm}=   \frac{3 e^4 (m_{H^+}^2 - M_W^2)}{64\pi s_{\rm W}^2 m_{H^+}^3}(\sin\beta' v_d -  \sqrt{2} \cos\beta' v_t)^2 =0,\label{eq:H+photonW+}\\
&\Gamma_{H^\pm \to t b}=   \frac{3 \sin^2\beta'}{8 \pi m_{H^+}^3  v_d^2}  [V_{{}_{\rm CKM}}]_{33}^2 \sqrt{(m_{H^+}^2 \!-\! ( m_t\!-\!m_b )^2) (m_{H^+}^2 \!-\! (m_t \!+\! m_b)^2)} \times \nonumber \\
&~~~~~~~~~~~~~(m_{H^+}^2 (m_b^2 \!+\! m_t^2)\!-\!(m_t^2\!-\!m_b^2 )^2  ), \label{eq:H+tb} \\
&\Gamma_{H^\pm \to A^0 W^\pm}= \frac{e^2(\sin\beta \sin\beta'\!+\!\sqrt{2} \cos\beta \cos\beta')^2}{64 \pi s_{\rm W}^2 m_{H^+}^3 M_W^2} \left(( m_{H^+}^2 \!-\! (m_{A^0} \!-\! M_W)^2) (m_{H^+}^2 \!-\! (m_{A^0} \! +\! M_W)^2)\right)^\frac32, \label{eq:H+A0W+} \\
&\Gamma_{H^\pm \to H^0 W^\pm}= \frac{e^2( \sin\alpha \sin\beta'\!+\!\sqrt{2} \cos\alpha \cos\beta')^2}{64 \pi s_{\rm W}^2 m_{H^+}^3 M_W^2} \left(( m_{H^+}^2 \!-\! (m_{H^0} \!-\! M_W)^2) (m_{H^+}^2 \!-\! (m_{H^0} \!+\! M_W)^2)\right)^\frac32,\label{eq:H+H0W+} \\
&\Gamma_{H^\pm \to h^0 W^\pm}= \frac{e^2(\cos\alpha \sin\beta'\!-\!\sqrt{2} \sin\alpha \cos\beta')^2}{64 \pi s_{\rm W}^2 m_{H^+}^3 M_W^2} (( m_{H^+}^2 \!-\! (m_{h^0} \!-\! M_W)^2) (m_{H^+}^2 \!-\! (m_{h^0} \!+\! M_W)^2))^\frac32,\label{eq:H+h0W+}
\end{align}
\begin{align}
%&\Gamma_{H^0 \to b \bar b}=  \frac{3 \sin^2\!\alpha \, m_b^2 (m_{H^0}^2-4 m_b^2)^\frac32}{8 \pi m_{H^0}^2  v_d^2} ,\label{eq:H0bb}\\
%&\Gamma_{H^0 \to t \bar t}=  \frac{3 \sin^2\!\alpha \, m_t^2 (m_{H^0}^2-4 m_t^2)^\frac32}{8 \pi m_{H^0}^2  v_d^2},\label{eq:H0tt}\\
&\Gamma_{H^0 \to W^+ W^-}=   \frac{e^4(\sin\!\alpha \, v_d - 
   2 \cos\!\alpha \, v_t)^2}{256 \pi s_{\rm W}^4 m_{H^0}^2 M_W^4} \sqrt{
 m_{H^0}^2 - 4 M_W^2} (8 M_W^4 + (m_{H^{0}}^2 - 2 M_W^2)^2),\label{eq:H0W+W-} \\
&\Gamma_{H^0 \to Z Z}=   \frac{e^4(\sin\!\alpha \, v_d - 
   4 \cos\!\alpha \, v_t)^2}{256 \pi s_{\rm W}^4 m_{H^0}^2 M_Z^4} \sqrt{
 m_{H^0}^2 - 4 M_Z^2} (8 M_Z^4 + (m_{H^{0}}^2 - 2 M_Z^2)^2),\label{eq:H0ZZ}\\
%&\Gamma_{H^0 \to H^\pm W^\mp}= \frac{e^2 ( \sin\alpha \sin\beta' + \sqrt{2} \cos\alpha \cos\beta' )^2}{64 \pi s_{\rm W}^2 m_{H^0}^3 M_W^2} ((m_{H^0}^2 - (m_{H^\pm} - M_W)^2) (m_{H^0}^2 - (m_{H^\pm} + M_W)^2))^\frac32, \label{eq:H0H+W-}\\
&\Gamma_{H^0 \to h^0 h^0}=  \frac{\sqrt{
  m_{H^0}^2-4 m_{h^0}^2}}{32 \pi m_{H^0}^2 }  \left(\sqrt{2} \mu - 
  (\lambda_1 + \lambda_4) v_t \right) \left(\sqrt{2} \mu + 
   3 \lambda \sin\!\alpha \,v_d - (\lambda_1 + \lambda_4) (4 \sin\!\alpha \, v_d + v_t)\right) \nonumber \\
&~~~~~~~~~~~~~~~~~~~   + {\cal O}(\sin^2\alpha), \label{eq:H0h0h0} \\
&~~~~~~~~~~~~~= \frac{\sqrt{
  m_{H^0}^2-4 m_{h^0}^2}}{32 \pi m_{H^0}^2 v_d^2}\left(2 m_{h^0}^2 + 3 m_{H^{\pm \pm}}^2 - 6 m_{H^\pm}^2\right)^2 \sin^2\alpha + {\cal O}(\sin^3\alpha) , \label{eq:H0h0h0prime}\\
&\Gamma_{H^0 \to H^\pm W^\mp} = \left[\Gamma_{H^\pm \to H^0 W^\pm}\right]_{| m_{H^0} \leftrightarrow  m_{H^+}},\label{eq:H0H+W-}\\
&\Gamma_{H^0 \to q \bar q}=  \frac{3 \sin^2\!\alpha \, m_q^2 (m_{H^0}^2-4 m_q^2)^\frac32}{8 \pi m_{H^0}^2  v_d^2} ,\label{eq:H0qq}
\end{align}
%\gilbertcom{expressions below can be simplified much further}
\begin{align} 
%&\Gamma_{A^{0} \to H^0 Z} = \frac{e^2}{64 c_{\rm W}^2 m_{A^0}^3 M_Z^2 \pi s_{\rm W}^2} \left[ (m_{H^0} - m_{A^0} - M_Z) (m_{H^0} + m_{A^0} - M_Z) (m_{H^0} - m_{A^0} + M_Z ) \right]^{\frac{3}{2}} \times \nonumber \\ 
%&~~~~~~~~~~~~~~~~~~~ (m_{H^0} + m_{A^0} + M_Z)^{\frac{3}{2}}
% \times (2 \cos\alpha \cos\beta + \sin\alpha \sin \beta)^2, \label{eq:A0H0Z} \\
%&\Gamma_{A^0 \to H^+ W^-} = \frac{e^2}{64 M_W^2 \pi s_{\rm W}^2 |m_A|^3} [(m_{A^0} - m_{H^+} - M_W) (m_{A^0} + m_{H^+} - M_W) (m_{A^0} - m_{H^+} + M_W)]^\frac{3}{2} \times \nonumber \\
%&~~~~~~~~~~~~~~~~~~~ (m_{A^0} + m_{H^+} + M_W)^\frac{3}{2} (2 \cos^2\beta \cos^2\beta' + 2 \sqrt{2} \cos\beta \cos\beta' \sin\beta \sin\beta' + \sin^2\beta \sin^2\beta'), \label{eq:A0H+W} \\
&\Gamma_{A^0 \to h^0 Z} = \frac{e^2 (\cos\alpha \sin\beta-2 \cos\beta \sin\alpha)^2}{64 c_{\rm W}^2 M_Z^2 \pi s_{\rm W}^2 m_{A^0}^3} ((m_{A^0}^2 - (m_{h^0} - M_Z)^2) (m_{A^0}^2 - (m_{h^0} + M_Z)^2))^\frac32 , \label{eq:A0hZ} \\
&\Gamma_{A^0 \to H^\pm W^\mp} = \left[\Gamma_{H^\pm \to A^0 W^\pm}\right]_{| m_{A^0} \leftrightarrow  m_{H^+}}, \label{eq:A0H+W}\\
& \Gamma_{A^0 \to q \bar q} = \frac{3  \sin^2\!\beta m_q^2 \sqrt{m_{A^0}^2 - 4  m_q^2}}{8 \pi  v_d^2}.\label{eq:A0qqbar}
\end{align}
Note a disagreement between \cref{eq:H++W+W+} and the expression of $\Gamma_{H^{\pm\pm} \to W^\pm W^\pm}$ given in \cite{Ashanujjaman:2021txz} where, we believe, a factor `$-3$' should read `$-4$', as is the case for $\Gamma_{H^0 \to W^+ W^-}$ and $\Gamma_{H^0 \to Z Z}$ given by \cite{Ashanujjaman:2021txz} with which \Cref{eq:H0W+W-,eq:H0ZZ} agree.
%\gilbertcom{ $y_t$, $y_b$ should be replaced in the two last equations and $M_A$ should be $m_t$, check!? DONE and simplified}

%%%%%%%%%%%%%%%%%%%%%%%%%%%%%%%%%%%%%%%%%%%%%%%%
%%%%%%%%%%%%%%%%%%%%%%%%%%%%%%%%%%%%%%%%%%%%%%%%
\FloatBarrier
\clearpage
\section{Benchmark signal points chosen for the analysis}
\label{app:NominalP_Details}

\begin{table}[h!]
\def\arraystretch{1.5}
\caption{
    Table presenting additional information for selected benchmark signal points for the associated production of doubly- and singly-charged Higgs bosons,
    as well as for the pair production of doubly-charged Higgs bosons. The value of $\operatorname{sin}\alpha$ parameter is set to  $8 \times 10^{-4}$, and the Standard Model Higgs boson mass is set to 125~GeV, respectively.
}
\label{tab:charged_ap_pp_nominals}
\scalebox{0.72}{
\begin{tabular}{ c  c| c c c c c c c c  }
    \hline \hline
    \multicolumn{10}{c}{\Hpp\Hm associated production: $H^{\pm \pm} \rightarrow W^\pm W^\pm$ , $H^\mp \rightarrow W^{\mp}Z$} \\
    \hline 
    $m_{H^{\pm \pm}}$ & $m_{H^\pm}$ & $m_{H^0}$   & $m_{A^0}$ & $\mu$ & $\lambda$ & $\lambda_1$ & $\lambda_2$ & $\lambda_3$ & $\lambda_4$  \\
    \hline \hline
        220 & 215.987 & 211.898 & 211.898 & 0.10493 & 0.516393 &  0.647518 & 0.776123348651 & -0.712629386820 &  -0.115646 \\
        300 & 296.005 & 291.956 & 291.956 & 0.199195 & 0.516394 &  0.710579 & 0.403454510208 & 0.163018451623 & -0.157377 \\
        400 & 393.934 & 387.773 & 387.773 & 0.351398 & 0.516395 & 0.905983 & 0.225335474785 & 0.421030521633 & -0.318342 \\
    \hline 
    \multicolumn{10}{c}{\Hpp\Hm associated production: $H^{\pm \pm} \rightarrow W^{\pm}W^{\pm}$ , $H^{\mp} \rightarrow tb$ } \\
    \hline
        \multicolumn{10}{c}{\multirow{1}{*}{same values as above}} \\
    \hline
    \multicolumn{10}{c}{\Hpp\Hmm pair production: $H^{\pm\pm} \rightarrow W^{\pm}W^{\pm}$} \\
    \hline
    \multicolumn{10}{c}{\multirow{1}{*}{same values as above}}\\
    \hline\hline
\end{tabular}
}
\end{table}

\begin{table}[h!]
\def\arraystretch{1.5}
\caption{
    Table presenting additional information for selected benchmark signal points for the associated production of neutral Higgs bosons $A^0 H^0$. The value of $\operatorname{sin}\alpha$ parameter is set to $8 \times 10^{-4}$, and the Standard Model Higgs boson mass is set to 125~GeV, respectively.
}
\label{tab:neutral_ap_nominals}
\scalebox{0.72}{
\begin{tabular}{ c c| c c c c c c c c  }
    \hline\hline
    \multicolumn{10}{c}{$H^0A^0$ associated production: $H^0 \rightarrow  Z Z$, $A^0 \rightarrow H^\pm W^{\pm *}, H^\pm \rightarrow W^{\pm} Z$} \\
    \hline 
    $m_{H^{\pm \pm}}$ & $m_{H^\pm}$ & $m_{H^0}$   & $m_{A^0}$ & $\mu$ & $\lambda$ & $\lambda_1$ & $\lambda_2$ & $\lambda_3$ & $\lambda_4$  \\
    \hline \hline
        220 & 231.788  &  243.005& 243.005 &  0.137998  &  0.516393&   0.187342&   -0.843376311614 & 0.474753444756  &   0.352014\\
        300 &  312.761 &  325.022&  325.022 &  0.24687 & 0.516394  &   0.0471361&  -0.337686002995 &  -0.505602322679 & 0.516854  \\
        400 &  418.933 &  437.046 & 437.046 &   0.446374 & 0.516396 & -0.415692&  0.918875311665 &  -0.391835433198&   1.02482\\
    \hline
    \multicolumn{10}{c}{$H^0A^0$ associated production: $H^0 \rightarrow Z Z $, $A^0 \rightarrow H^\pm W^{\pm*}, H^\pm \rightarrow H^{\pm \pm} W^{\mp*}, H^{\pm \pm} \rightarrow W^{\pm }W^{\pm}$  } \\
    \hline
    \multicolumn{10}{c}{same values as above} \\
    \hline \hline
\end{tabular}
}
\end{table}

%%%%%%%%%%%%%%%%%%%%%%%%%%%%%%%%%%%%%%%%%%%%%%%%
%%%%%%%%%%%%%%%%%%%%%%%%%%%%%%%%%%%%%%%%%%%%%%%%
\clearpage
\section{Additional details from the ATLAS \texorpdfstring{$H^{\pm\pm}/H^\pm$}{} analysis
}
\label{app:Rel21Paper_AdditionalDetails}

The variables used in the Ref.~\cite{ATLAS:2021pairbosons} analysis are reminded below:
\begin{itemize}
    \item The missing transverse momentum in the event, with magnitude $\met$.
    \item The magnitude of the momentum in the plane transverse to the beam axis, \pt. 
    \item The invariant mass of the same-flavor opposite-charge leptons, $m_{\ell\ell}^{oc}$.
	\item The invariant mass of all jets in the event, $m_\mathrm{jets}$. When the event has more than four jets, only the four leading jets are considered for the computation.
    \item The invariant mass of all selected leptons in the event, $m_{x\ell}$. Here, x takes values of 2, 3 or 4, and corresponds to the \llSC, \lll{} and \llll{} cases.
    \item Inclusive effective mass, \meff{}, defined by summing the \pt{} of all leptons, jets and the $\met{}$ present in the event 
    \item The angular distance in $\eta$ and $\phi$, between two same-charge leptons in the event, $\Delta R_{\ell^{\pm}\ell^{\pm}}$. In the \lll{} lepton channel, two variables can be computed, $\Delta R^\text{max}_{\ell^{\pm}\ell^{\pm}}$ and $\Delta R^\text{min}_{\ell^{\pm}\ell^{\pm}}$, and corresponds to the maximum and minimum values of $\Delta R_{\ell^{\pm}\ell^{\pm}}$.
    \item The azimuthal distance between the two same charge leptons and and $\met$, $\Delta \phi_{\ell\ell,\met}$.
    \item The smallest angular distance in $\eta$ and $\phi$ between any lepton and its closed jet, $\Delta R_{\ell\mathrm{jet}}$.
    \item The transverse momentum of the highest-\pt{} lepton, $p_\text{T}^{\ell_1}$. It is used only in the \lll{} lepton channel.
    \item The transverse momentum of the highest-\pt{} jet, $p_\text{T}^{\text{leading jet}}$.
    \item The variable $S$, that accounts for the event topology in the transverse plane, and defined using the spread of the $\phi$ angles of the leptons, $\met$, and jets. In the~\cite{ATLAS:2021pairbosons} analysis, this variable was found to have a negligible impact on the considered signal. Thus, for the studies done in this phenomenological paper the requirement on $S$ were was dropped.
\end{itemize}

\begin{table}[!ht]
\begin{center}
	\def\arraystretch{1.5}
	\caption{
     Table showing the event selection criteria in the signal regions, from Ref.~\cite{ATLAS:2021pairbosons}.
     These requirements are applied on top of the preselection shown in Table~\ref{Tab:preselection}.
     The $m_{H^{\pm\pm}}=200$~GeV signal regions are used for $m_{H^{\pm\pm}}=220$~GeV signal mass point.
	 The bold-slashed criteria show the differences with respect to Ref.~\cite{ATLAS:2021pairbosons}. 
	}
	\label{Tab:SRselection}
 	% \scriptsize
 	\resizebox{0.9\textwidth}{!}
    % \resizebox{0.3\paperheight}{!}
	{
		\begin{tabular}{l||c|c|c}
		\hline\hline
			Charged Higgs  & \multirow{2}{*}{$m_{H^{\pm\pm}} = 200 $~GeV} & \multirow{2}{*}{$m_{H^{\pm\pm}} = 300$~GeV} & \multirow{2}{*}{$m_{H^{\pm\pm}} = 400$~GeV}\\
			boson mass     & & & \\   

			\hline
			Selection criteria &  \multicolumn{3}{c}{\llSC channel}  \\
			\hline			
			
			$m_\mathrm{jets}$ [GeV]                 & [100, 450] &  [100, 500] &  [300, 700]   \\
		    \textbf{\cancel{$S$}}                       & \textbf{\cancel{$\boldsymbol{<}$0.3}}    & \textbf{\cancel{ $\boldsymbol{<}$0.6}}     & \textbf{\cancel{$\boldsymbol{<}$0.6}}  \\
			$\Delta R_{\ell^{\pm}\ell^{\pm}}$        & $<$1.9     &  $<$2.1     &  $<$2.2  \\
			$\Delta\phi_{\ell\ell,\met}$             & $<$0.7     &  $<$0.9     &  $<$1.0 \\
			$m_{x\ell}$ [{GeV}]                      & [40, 150]  &  [90, 240]  &  [130, 340]  \\
	    	$\met{}$ [{GeV}]                         & $>$100     &  $>$130     &  $>$170 \\

			\hline
		    Selection criteria & \multicolumn{3}{c}{\lll channel}  \\
		    \hline
			
			$\Delta R_{\ell^{\pm}\ell^{\pm}}$        &  [0.2, 1.7]  &  [0.0, 2.1]  &  [0.2, 2.5] \\
			$m_{x\ell}$ [{GeV}]                      &  $>$160      &  $>$190     &  $>$240 \\
			$\met{}$ [{GeV}]                         &  $>$30       &  $>$55       &  $>$80 \\
			$\Delta R_{\ell\mathrm{jet}}$            &  [0.1, 1.5]  &  [0.1, 2.0]  &  [0.1,2.3] \\
			$p_\text{T}^{\text{leading jet}}$ [{GeV}]& $>$40       &  $>$70       &  $>$100 \\

			\hline
		    Selection criteria & \multicolumn{3}{c}{\llll channel}  \\
		    \hline
			
			$m_{x\ell}$ [{GeV}]                           &  $>$230~~      &  $>$270~~      &  $>$360~~ \\
			$\met{}$  [{GeV}]                             &  $>$60       &  $>$60       &  $>$60   \\ 
			$p_\text{T}^{\ell_1}$ [{GeV}]                 &  $>$65       &  $>$80       &  $>$110~~  \\
			$\Delta R^\text{min}_{\ell^{\pm}\ell^{\pm}}$  &  [0.2, 1.2]  &  [0.2, 2.0]  &  [0.5, 2.4]  \\
			$\Delta R^\text{max}_{\ell^{\pm}\ell^{\pm}}$  &  [{0.3}, \textbf{\cancel{2.0}}]  & [{0.5}, \textbf{\cancel{2.6}}]  &   [{0.6}, \textbf{\cancel{3.1}}] \\
		\hline\hline
		\end{tabular}
	}
\end{center}
\end{table}

%%%%%%%%%%%%%%%%%%%%%%%%%%%%%%%%%%%%%%%%%%%%%%%% 
%%%%%%%%%%%%%%%%%%%%%%%%%%%%%%%%%%%%%%%%%%%%%%%%
\clearpage
\section{Potential signal regions for the \texorpdfstring{$H^{\mp} \rightarrow t b$}{} decay mode}
 
 \begin{table}[!ht]
\begin{center}
	\def\arraystretch{1.5}
	\caption{
     Table showing the event selection criteria for potential signal regions proposed for the \Hpp\Hm associated production sector with $H^{\mp} \rightarrow t b$ decays. 
     They are applied on top of the preselection discussed in the text.
	}
	\label{Tab:MySRselection_HdbHsg_HsgTOtb}
 	% \scriptsize
 	\resizebox{0.9\textwidth}{!}
    % \resizebox{0.3\paperheight}{!}
	{
		\begin{tabular}{l||c|c|c}
		\hline\hline
			Charged Higgs  & \multirow{2}{*}{$m_{H^{\pm\pm}} = 220 $~GeV} & \multirow{2}{*}{$m_{H^{\pm\pm}} = 300$~GeV} & \multirow{2}{*}{$m_{H^{\pm\pm}} = 400$~GeV}\\
			boson mass     & & & \\   

			\hline
			Selection criteria &  \multicolumn{3}{c}{\llSC channel}  \\
			\hline			
			$N_{jets}$ & $\geq 4$ & $\geq 2$ & $\geq 2$ \\
			\meff{} [GeV] & $>500$  & $>500$ & $>500$\\
			$m_\mathrm{jets}$ [GeV]  & [150, 300] &  [250, 400] &  [300, 500]   \\
		    $\Delta R_{\ell^{\pm}\ell^{\pm}}$        & $<$2.5     &  $<$2.5     &  $<$2.5  \\
			$\Delta\phi_{\ell\ell,\met}$             & $<$1.0     &  $<$1.0     &  $<$1.0 \\
			$m_{x\ell}$ [{GeV}] & [40, 150]  &  [90, 240]  &  [130, 340]  \\
	    	$\met{}$ [{GeV}] & $>$100     &  $>$130     &  $>$170 \\
	    	
			\hline
		    Selection criteria & \multicolumn{3}{c}{\lll channel}  \\
		    \hline
            \meff{} [GeV] & $>350$  & $>400$ & $>450$\\
			$\Delta R_{\ell^{\pm}\ell^{\pm}}$        &  [0.2, 2.5]  &  [0.0, 3.0]  &  [0.2, 3.5] \\
			$m_{x\ell}$ [{GeV}] &  $>$90      &  $>$90     &  $>$90 \\
			$\met{}$ [{GeV}] &  $>$50 &  $>$30 &  $>$30 \\
			$\Delta R_{\ell\mathrm{jet}}$ &  [0.4, 1.0]  &  [0.4, 1.0]  &  [0.4, 1.0] \\
			$p_\text{T}^{\text{leading jet}}$ [{GeV}]& $>$40       &  $>$40       &  $>$40 \\
		\hline\hline
		\end{tabular}
	}
\end{center}
\end{table}

%%%%%%%%%%%%%%%%%%%%%%%%%%%%%%%%%%%%%%%%%%%%%%%% 
%%%%%%%%%%%%%%%%%%%%%%%%%%%%%%%%%%%%%%%%%%%%%%%%
\clearpage
\section{Potential signal regions for the \texorpdfstring{$A^0H^0$}{} pair production mode, without \texorpdfstring{$H^{\pm\pm}$}{} in the decay chain}
 
 \begin{table}[hb!]
\begin{center}
	\def\arraystretch{1.5}
	\caption{
     Table showing the event selection criteria for potential signal regions proposed for the $H^0A^0$ associated production mode, with $H^0 \to ZZ$, $A^0 \to H^\pm W^{\mp }$ and $H^\pm \to W^{\mp }Z$ decays. 
     They are applied on top of the preselection discussed in the text, and independent of the $H^0$ boson mass.
	}
	\label{Tab:MySRselection_A0H0_withoutHppInTheDecayCh}
 	% \scriptsize
    \resizebox{0.4\textwidth}{!}
    % \resizebox{0.3\paperheight}{!}
	{
		\begin{tabular}{l||c}
		\hline\hline
			Selection criteria &  \multicolumn{1}{c}{\llSC channel}  \\
			\hline	
			\meff{} [GeV] & [350, 550] \\
			$m_\mathrm{jets}$ [GeV]  & $>150$  \\
			$\Delta\phi_{\ell\ell,\met}$ & $<$2.0 \\
			$m_{x\ell}$ [{GeV}] & [40, 180] \\
	    	$\met{}$ [{GeV}] & $<$80 \\       

			\hline
		    Selection criteria & \multicolumn{1}{c}{\lll channel}  \\
		    \hline
            \meff{} [GeV] & $>400$ \\
			$\Delta R_{\ell^{\pm}\ell^{\pm}}$ & $<3$ \\
			$m_{x\ell}$ [{GeV}] &  $>$100  \\
			$\met{}$ [{GeV}] &  $<$100  \\
			$\Delta R_{\ell\mathrm{jet}}$ &  [0.5, 3.0] \\
			$p_\text{T}^{\text{leading jet}}$ [{GeV}]& $>$50  \\

			\hline
		    Selection criteria & \multicolumn{1}{c}{\llll channel}  \\
		    \hline
            \meff{} [GeV] & $>400$ \\
			$m_{x\ell}$ [{GeV}] &  $>$200  \\
			$\met{}$ [{GeV}] &  $<$100  \\
			$p_\text{T}^{\text{leading } \ell}$ [{GeV}]& $>$100  \\
			$\Delta R^\text{max}_{\ell^{\pm}\ell^{\pm}}$  & $>2.5$ \\
		\hline\hline
		\end{tabular}
	}
\end{center}
\end{table}

%%%%%%%%%%%%%%%%%%%%%%%%%%%%%%%%%%%%%%%%%%%%%%%% 
%%%%%%%%%%%%%%%%%%%%%%%%%%%%%%%%%%%%%%%%%%%%%%%%
\clearpage 
\section{Potential signal regions for the \texorpdfstring{$A^0H^0$}{} pair production mode, with \texorpdfstring{$H^{\pm\pm}$}{} in the decay chain}
 
 \begin{table}[hb!]
\begin{center}
	\def\arraystretch{1.5}
	\caption{
     Table showing the event selection criteria for potential signal regions proposed for the $H^0A^0$ associated production mode, with $H^0 \to ZZ$, $A^0 \to H^\pm W^{\mp }$,  $H^\pm \to H^{\pm \pm} W^{\mp }$ and $H^{\pm \pm} \to W^{\mp }W^{\mp }$ decays. 
     They are applied on top of the preselection discussed in the text, and independent of the $H^0$ boson mass.
	}
	\label{Tab:MySRselection_A0H0_withHppInTheDecayCh}
 	% \scriptsize
    \resizebox{0.4\textwidth}{!}
    % \resizebox{0.3\paperheight}{!}
	{
		\begin{tabular}{l||c}
		\hline\hline
			Selection criteria &  \multicolumn{1}{c}{\llSC channel}  \\
			\hline	
			\meff{} [GeV] & $>350$ \\
			$m_\mathrm{jets}$ [GeV]  & $>100$  \\
			$\Delta\phi_{\ell\ell,\met}$ & $<$1.5 \\
			$m_{x\ell}$ [{GeV}] & [40, 180] \\
	    	$\met{}$ [{GeV}] & $<$200 \\  

			\hline
		    Selection criteria & \multicolumn{1}{c}{\lll channel}  \\
		    \hline
            \meff{} [GeV] & $>400$ \\
			$m_{x\ell}$ [{GeV}] &  $>$100  \\
			$\met{}$ [{GeV}] &  $<$100  \\
			$\Delta R_{\ell\mathrm{jet}}$ &  [0.5, 3.0] \\
			$p_\text{T}^{\text{leading jet}}$ [{GeV}]& $>$50  \\

			\hline
		    Selection criteria & \multicolumn{1}{c}{\llll channel}  \\
		    \hline
            \meff{} [GeV] & $>400$ \\
			$m_{x\ell}$ [{GeV}] &  $>$200  \\
			$\met{}$ [{GeV}] &  $<$100  \\
			$p_\text{T}^{\text{leading } \ell}$ [{GeV}]& $>$60  \\
			$\Delta R^\text{max}_{\ell^{\pm}\ell^{\pm}}$  & $>2.5$ \\
		\hline\hline
		\end{tabular}
	}
\end{center}
\end{table}

%%%%%%%%%%%%%%%%%%%%%%%%%%%%%%%%%%%%%%%%%%%%%%%%
%%%%%%%%%%%%%%%%%%%%%%%%%%%%%%%%%%%%%%%%%%%%%%%%
%%%%%%%%%%%%%%%%%%%%%%%%%%%%%%%%%%%%%%%%%%%%%%%%
%\bibliographystyle{h-physrev5}
\bibliographystyle{utphys}
\bibliography{references-triplet}

\providecommand{\href}[2]{#2}\begingroup\raggedright\begin{thebibliography}{100}

\bibitem{Aad:2012tfa}
{ATLAS Collaboration}, ``{Observation of a new particle in the search for the
  Standard Model Higgs boson with the ATLAS detector at the LHC},''
  \href{http://dx.doi.org/10.1016/j.physletb.2012.08.020}{{\em Phys.Lett.}
  {\bfseries B716} (2012) 1--29},
\href{http://arxiv.org/abs/1207.7214}{{\ttfamily arXiv:1207.7214 [hep-ex]}}.
%%CITATION = ARXIV:1207.7214;%%.

\bibitem{Chatrchyan:2012ufa}
{CMS Collaboration}, ``{Observation of a new boson at a mass of 125 GeV with
  the CMS experiment at the LHC},''
  \href{http://dx.doi.org/10.1016/j.physletb.2012.08.021}{{\em Phys.Lett.}
  {\bfseries B716} (2012) 30--61},
\href{http://arxiv.org/abs/1207.7235}{{\ttfamily arXiv:1207.7235 [hep-ex]}}.
%%CITATION = ARXIV:1207.7235;%%.

\bibitem{Konetschny:1977bn}
W.~Konetschny and W.~Kummer, ``{Nonconservation of Total Lepton Number with
  Scalar Bosons},''
\href{http://dx.doi.org/10.1016/0370-2693(77)90407-5}{{\em Phys. Lett.}
  {\bfseries B70} (1977) 433}.
%%CITATION = PHLTA,B70,433;%%.

\bibitem{Cheng:1980qt}
T.~P. Cheng and L.-F. Li, ``{Neutrino Masses, Mixings and Oscillations in SU(2)
  x U(1) Models of Electroweak Interactions},''
\href{http://dx.doi.org/10.1103/PhysRevD.22.2860}{{\em Phys. Rev.} {\bfseries
  D22} (1980) 2860}.
%%CITATION = PHRVA,D22,2860;%%.

\bibitem{Lazarides:1980nt}
G.~Lazarides, Q.~Shafi, and C.~Wetterich, ``{Proton Lifetime and Fermion Masses
  in an SO(10) Model},''
\href{http://dx.doi.org/10.1016/0550-3213(81)90354-0}{{\em Nucl. Phys.}
  {\bfseries B181} (1981) 287}.
%%CITATION = NUPHA,B181,287;%%.

\bibitem{Schechter:1980gr}
J.~Schechter and J.~W.~F. Valle, ``{Neutrino Masses in SU(2) x U(1)
  Theories},''
\href{http://dx.doi.org/10.1103/PhysRevD.22.2227}{{\em Phys. Rev.} {\bfseries
  D22} (1980) 2227}.
%%CITATION = PHRVA,D22,2227;%%.

\bibitem{Mohapatra:1980yp}
R.~N. Mohapatra and G.~Senjanovic, ``{Neutrino Masses and Mixings in Gauge
  Models with Spontaneous Parity Violation},''
\href{http://dx.doi.org/10.1103/PhysRevD.23.165}{{\em Phys. Rev.} {\bfseries
  D23} (1981) 165}.
%%CITATION = PHRVA,D23,165;%%.

\bibitem{deGouvea:2006gz}
A.~de~Gouvea, J.~Jenkins, and N.~Vasudevan, ``{Neutrino Phenomenology of Very
  Low-Energy Seesaws},''
  \href{http://dx.doi.org/10.1103/PhysRevD.75.013003}{{\em Phys. Rev. D}
  {\bfseries 75} (2007) 013003},
  \href{http://arxiv.org/abs/hep-ph/0608147}{{\ttfamily arXiv:hep-ph/0608147}}.

\bibitem{Perez:2008ha}
P.~Fileviez~Perez, T.~Han, G.-y. Huang, T.~Li, and K.~Wang, ``{Neutrino Masses
  and the LHC: Testing Type II Seesaw},''
  \href{http://dx.doi.org/10.1103/PhysRevD.78.015018}{{\em Phys. Rev.}
  {\bfseries D78} (2008) 015018},
\href{http://arxiv.org/abs/0805.3536}{{\ttfamily arXiv:0805.3536 [hep-ph]}}.
%%CITATION = 0805.3536;%%.

\bibitem{Huitu:1996su}
K.~Huitu, J.~Maalampi, A.~Pietila, and M.~Raidal, ``{Doubly charged Higgs at
  LHC},'' \href{http://dx.doi.org/10.1016/S0550-3213(97)87466-4}{{\em Nucl.
  Phys.} {\bfseries B487} (1997) 27--42},
\href{http://arxiv.org/abs/hep-ph/9606311}{{\ttfamily arXiv:hep-ph/9606311}}.
%%CITATION = HEP-PH/9606311;%%.

\bibitem{Chakrabarti:1998qy}
S.~Chakrabarti, D.~Choudhury, R.~M. Godbole, and B.~Mukhopadhyaya, ``{Observing
  doubly charged Higgs bosons in photon-photon collisions},''
  \href{http://dx.doi.org/10.1016/S0370-2693(98)00743-6}{{\em Phys. Lett. B}
  {\bfseries 434} (1998) 347--353},
  \href{http://arxiv.org/abs/hep-ph/9804297}{{\ttfamily arXiv:hep-ph/9804297}}.

\bibitem{Chun:2003ej}
E.~J. Chun, K.~Y. Lee, and S.~C. Park, ``{Testing Higgs triplet model and
  neutrino mass patterns},''
  \href{http://dx.doi.org/10.1016/S0370-2693(03)00770-6}{{\em Phys. Lett.}
  {\bfseries B566} (2003) 142--151},
\href{http://arxiv.org/abs/hep-ph/0304069}{{\ttfamily arXiv:hep-ph/0304069}}.
%%CITATION = HEP-PH/0304069;%%.

\bibitem{Muhlleitner:2003me}
M.~Muhlleitner and M.~Spira, ``{A note on doubly-charged Higgs pair production
  at hadron colliders},''
  \href{http://dx.doi.org/10.1103/PhysRevD.68.117701}{{\em Phys. Rev.}
  {\bfseries D68} (2003) 117701},
\href{http://arxiv.org/abs/hep-ph/0305288}{{\ttfamily arXiv:hep-ph/0305288}}.
%%CITATION = HEP-PH/0305288;%%.

\bibitem{Akeroyd:2005gt}
A.~G. Akeroyd and M.~Aoki, ``{Single and pair production of doubly charged
  Higgs bosons at hadron colliders},''
  \href{http://dx.doi.org/10.1103/PhysRevD.72.035011}{{\em Phys. Rev.}
  {\bfseries D72} (2005) 035011},
\href{http://arxiv.org/abs/hep-ph/0506176}{{\ttfamily arXiv:hep-ph/0506176}}.
%%CITATION = HEP-PH/0506176;%%.

\bibitem{Dey:2008jm}
P.~Dey, A.~Kundu, and B.~Mukhopadhyaya, ``{Some consequences of a Higgs
  triplet},'' \href{http://dx.doi.org/10.1088/0954-3899/36/2/025002}{{\em J.
  Phys.} {\bfseries G36} (2009) 025002},
\href{http://arxiv.org/abs/0802.2510}{{\ttfamily arXiv:0802.2510 [hep-ph]}}.
%%CITATION = 0802.2510;%%.

\bibitem{FileviezPerez:2008jbu}
P.~Fileviez~Perez, T.~Han, G.-y. Huang, T.~Li, and K.~Wang, ``{Neutrino Masses
  and the CERN LHC: Testing Type II Seesaw},''
  \href{http://dx.doi.org/10.1103/PhysRevD.78.015018}{{\em Phys. Rev. D}
  {\bfseries 78} (2008) 015018},
  \href{http://arxiv.org/abs/0805.3536}{{\ttfamily arXiv:0805.3536 [hep-ph]}}.

\bibitem{delAguila:2008cj}
F.~del Aguila and J.~A. Aguilar-Saavedra, ``{Distinguishing seesaw models at
  LHC with multi-lepton signals},''
  \href{http://dx.doi.org/10.1016/j.nuclphysb.2008.12.029}{{\em Nucl. Phys.}
  {\bfseries B813} (2009) 22--90},
\href{http://arxiv.org/abs/0808.2468}{{\ttfamily arXiv:0808.2468 [hep-ph]}}.
%%CITATION = 0808.2468;%%.

\bibitem{Akeroyd:2009hb}
A.~G. Akeroyd and C.-W. Chiang, ``{Doubly charged Higgs bosons and three-lepton
  signatures in the Higgs Triplet Model},''
  \href{http://dx.doi.org/10.1103/PhysRevD.80.113010}{{\em Phys. Rev.}
  {\bfseries D80} (2009) 113010},
\href{http://arxiv.org/abs/0909.4419}{{\ttfamily arXiv:0909.4419 [hep-ph]}}.
%%CITATION = 0909.4419;%%.

\bibitem{Akeroyd:2010je}
A.~G. Akeroyd and C.-W. Chiang, ``{Phenomenology of Large Mixing for the
  CP-even Neutral Scalars of the Higgs Triplet Model},''
  \href{http://dx.doi.org/10.1103/PhysRevD.81.115007}{{\em Phys. Rev.}
  {\bfseries D81} (2010) 115007},
\href{http://arxiv.org/abs/1003.3724}{{\ttfamily arXiv:1003.3724 [hep-ph]}}.
%%CITATION = 1003.3724;%%.

\bibitem{Akeroyd:2011zza}
A.~G. Akeroyd and H.~Sugiyama, ``{Production of doubly charged scalars from the
  decay of singly charged scalars in the Higgs Triplet Model},''
  \href{http://dx.doi.org/10.1103/PhysRevD.84.035010}{{\em Phys. Rev. D}
  {\bfseries 84} (2011) 035010},
  \href{http://arxiv.org/abs/1105.2209}{{\ttfamily arXiv:1105.2209 [hep-ph]}}.

\bibitem{Arhrib:2011uy}
A.~Arhrib, R.~Benbrik, M.~Chabab, G.~Moultaka, M.~Peyranere, J.~Ramadan, and
  L.~Rahili, ``{The Higgs Potential in the Type II Seesaw Model},''
  \href{http://dx.doi.org/10.1103/PhysRevD.84.095005}{{\em Phys.Rev.}
  {\bfseries D84} (2011) 095005},
  \href{http://arxiv.org/abs/1105.1925}{{\ttfamily arXiv:1105.1925 [hep-ph]}}.

\bibitem{Melfo:2011nx}
A.~Melfo, M.~Nemevsek, F.~Nesti, G.~Senjanovic, and Y.~Zhang, ``{Type II Seesaw
  at LHC: The Roadmap},''
  \href{http://dx.doi.org/10.1103/PhysRevD.85.055018}{{\em Phys. Rev.}
  {\bfseries D85} (2012) 055018},
\href{http://arxiv.org/abs/1108.4416}{{\ttfamily arXiv:1108.4416 [hep-ph]}}.
%%CITATION = ARXIV:1108.4416;%%.

\bibitem{Aoki:2011pz}
M.~Aoki, S.~Kanemura, and K.~Yagyu, ``{Testing the Higgs triplet model with the
  mass difference at the LHC},''
  \href{http://dx.doi.org/10.1103/PhysRevD.85.055007}{{\em Phys.Rev.}
  {\bfseries D85} (2012) 055007},
\href{http://arxiv.org/abs/1110.4625}{{\ttfamily arXiv:1110.4625 [hep-ph]}}.
%%CITATION = ARXIV:1110.4625;%%.

\bibitem{Arhrib:2011vc}
A.~Arhrib, R.~Benbrik, M.~Chabab, G.~Moultaka, and L.~Rahili, ``{Higgs boson
  decay into 2 photons in the type~II Seesaw Model},''
  \href{http://dx.doi.org/10.1007/JHEP04(2012)136}{{\em JHEP} {\bfseries 1204}
  (2012) 136},
\href{http://arxiv.org/abs/1112.5453}{{\ttfamily arXiv:1112.5453 [hep-ph]}}.
%%CITATION = ARXIV:1112.5453;%%.

\bibitem{Akeroyd:2012nd}
A.~G. Akeroyd, S.~Moretti, and H.~Sugiyama, ``{Five-lepton and six-lepton
  signatures from production of neutral triplet scalars in the Higgs Triplet
  Model},'' \href{http://dx.doi.org/10.1103/PhysRevD.85.055026}{{\em Phys. Rev.
  D} {\bfseries 85} (2012) 055026},
  \href{http://arxiv.org/abs/1201.5047}{{\ttfamily arXiv:1201.5047 [hep-ph]}}.

\bibitem{Chiang:2012dk}
C.-W. Chiang, T.~Nomura, and K.~Tsumura, ``{Search for doubly charged Higgs
  bosons using the same-sign diboson mode at the LHC},''
  \href{http://dx.doi.org/10.1103/PhysRevD.85.095023}{{\em Phys. Rev. D}
  {\bfseries 85} (2012) 095023},
  \href{http://arxiv.org/abs/1202.2014}{{\ttfamily arXiv:1202.2014 [hep-ph]}}.

\bibitem{Chun:2012zu}
E.~J. Chun and P.~Sharma, ``{Same-Sign Tetra-Leptons from Type II Seesaw},''
  \href{http://dx.doi.org/10.1007/JHEP08(2012)162}{{\em JHEP} {\bfseries 08}
  (2012) 162}, \href{http://arxiv.org/abs/1206.6278}{{\ttfamily arXiv:1206.6278
  [hep-ph]}}.

\bibitem{Akeroyd:2012ms}
A.~Akeroyd and S.~Moretti, ``{Enhancement of H to gamma gamma from doubly
  charged scalars in the Higgs Triplet Model},''
  \href{http://dx.doi.org/10.1103/PhysRevD.86.035015}{{\em Phys.Rev.}
  {\bfseries D86} (2012) 035015},
\href{http://arxiv.org/abs/1206.0535}{{\ttfamily arXiv:1206.0535 [hep-ph]}}.
%%CITATION = ARXIV:1206.0535;%%.

\bibitem{Chun:2012jw}
E.~J. Chun, H.~M. Lee, and P.~Sharma, ``{Vacuum Stability, Perturbativity, EWPD
  and Higgs-to-diphoton rate in Type II Seesaw Models},''
  \href{http://dx.doi.org/10.1007/JHEP11(2012)106}{{\em JHEP} {\bfseries 11}
  (2012) 106}, \href{http://arxiv.org/abs/1209.1303}{{\ttfamily arXiv:1209.1303
  [hep-ph]}}.

\bibitem{BhupalDev:2013xol}
P.~S. Bhupal~Dev, D.~K. Ghosh, N.~Okada, and I.~Saha, ``{125 GeV Higgs Boson
  and the Type-II Seesaw Model},''
  \href{http://dx.doi.org/10.1007/JHEP03(2013)150}{{\em JHEP} {\bfseries 03}
  (2013) 150}, \href{http://arxiv.org/abs/1301.3453}{{\ttfamily arXiv:1301.3453
  [hep-ph]}}. [Erratum: JHEP 05, 049 (2013)].

\bibitem{Englert:2013wga}
C.~Englert, E.~Re, and M.~Spannowsky, ``{Pinning down Higgs triplets at the
  LHC},'' \href{http://dx.doi.org/10.1103/PhysRevD.88.035024}{{\em Phys. Rev.}
  {\bfseries D88} (2013) 035024},
\href{http://arxiv.org/abs/1306.6228}{{\ttfamily arXiv:1306.6228 [hep-ph]}}.
%%CITATION = ARXIV:1306.6228;%%.

\bibitem{Kanemura:2013vxa}
S.~Kanemura, K.~Yagyu, and H.~Yokoya, ``{First constraint on the mass of
  doubly-charged Higgs bosons in the same-sign diboson decay scenario at the
  LHC},'' \href{http://dx.doi.org/10.1016/j.physletb.2013.08.054}{{\em
  Phys.Lett.} {\bfseries B726} (2013) 316--319},
\href{http://arxiv.org/abs/1305.2383}{{\ttfamily arXiv:1305.2383 [hep-ph]}}.
%%CITATION = ARXIV:1305.2383;%%.

\bibitem{Chun:2013vma}
E.~J. Chun and P.~Sharma, ``{Search for a doubly-charged boson in four lepton
  final states in type II seesaw},''
  \href{http://dx.doi.org/10.1016/j.physletb.2013.11.056}{{\em Phys. Lett. B}
  {\bfseries 728} (2014) 256--261},
  \href{http://arxiv.org/abs/1309.6888}{{\ttfamily arXiv:1309.6888 [hep-ph]}}.

\bibitem{Kang:2014lwn}
Z.~Kang, J.~Li, T.~Li, Y.~Liu, and G.-Z. Ning, ``{Light Doubly Charged Higgs
  Boson via the $WW^*$ Channel at LHC},''
  \href{http://dx.doi.org/10.1140/epjc/s10052-015-3774-1}{{\em Eur. Phys. J. C}
  {\bfseries 75} no.~12, (2015) 574},
  \href{http://arxiv.org/abs/1404.5207}{{\ttfamily arXiv:1404.5207 [hep-ph]}}.

\bibitem{kang:2014jia}
Z.~Kang, J.~Li, T.~Li, Y.~Liu, and G.-Z. Ning, ``{Light Doubly Charged Higgs
  Boson via the $WW^*$ Channel at LHC},''
  \href{http://dx.doi.org/10.1140/epjc/s10052-015-3774-1}{{\em Eur. Phys. J.}
  {\bfseries C75} no.~12, (2015) 574},
\href{http://arxiv.org/abs/1404.5207}{{\ttfamily arXiv:1404.5207 [hep-ph]}}.
%%CITATION = ARXIV:1404.5207;%%.

\bibitem{Kanemura:2014goa}
S.~Kanemura, M.~Kikuchi, K.~Yagyu, and H.~Yokoya, ``{Bounds on the mass of
  doubly-charged Higgs bosons in the same-sign diboson decay scenario},''
  \href{http://dx.doi.org/10.1103/PhysRevD.90.115018}{{\em Phys. Rev. D}
  {\bfseries 90} no.~11, (2014) 115018},
  \href{http://arxiv.org/abs/1407.6547}{{\ttfamily arXiv:1407.6547 [hep-ph]}}.

\bibitem{Arhrib:2014nya}
A.~Arhrib, R.~Benbrik, G.~Moultaka, and L.~Rahili, ``{Type II Seesaw Higgsology
  and LEP/LHC constraints},'' \href{http://arxiv.org/abs/1411.5645}{{\ttfamily
  arXiv:1411.5645 [hep-ph]}}.

\bibitem{Han:2015hba}
Z.-L. Han, R.~Ding, and Y.~Liao, ``{LHC Phenomenology of Type II Seesaw:
  Nondegenerate Case},''
  \href{http://dx.doi.org/10.1103/PhysRevD.91.093006}{{\em Phys. Rev. D}
  {\bfseries 91} (2015) 093006},
  \href{http://arxiv.org/abs/1502.05242}{{\ttfamily arXiv:1502.05242
  [hep-ph]}}.

\bibitem{Han:2015sca}
Z.-L. Han, R.~Ding, and Y.~Liao, ``{LHC phenomenology of the type II seesaw
  mechanism: Observability of neutral scalars in the nondegenerate case},''
  \href{http://dx.doi.org/10.1103/PhysRevD.92.033014}{{\em Phys. Rev. D}
  {\bfseries 92} no.~3, (2015) 033014},
  \href{http://arxiv.org/abs/1506.08996}{{\ttfamily arXiv:1506.08996
  [hep-ph]}}.

\bibitem{Das:2016bir}
D.~Das and A.~Santamaria, ``{Updated scalar sector constraints in the Higgs
  triplet model},'' \href{http://dx.doi.org/10.1103/PhysRevD.94.015015}{{\em
  Phys. Rev. D} {\bfseries 94} no.~1, (2016) 015015},
  \href{http://arxiv.org/abs/1604.08099}{{\ttfamily arXiv:1604.08099
  [hep-ph]}}.

\bibitem{Mitra:2016wpr}
M.~Mitra, S.~Niyogi, and M.~Spannowsky, ``{Type-II Seesaw Model and Multilepton
  Signatures at Hadron Colliders},''
  \href{http://dx.doi.org/10.1103/PhysRevD.95.035042}{{\em Phys. Rev.}
  {\bfseries D95} no.~3, (2017) 035042},
\href{http://arxiv.org/abs/1611.09594}{{\ttfamily arXiv:1611.09594 [hep-ph]}}.
%%CITATION = ARXIV:1611.09594;%%.

\bibitem{Babu:2016rcr}
K.~S. Babu and S.~Jana, ``{Probing Doubly Charged Higgs Bosons at the LHC
  through Photon Initiated Processes},''
  \href{http://dx.doi.org/10.1103/PhysRevD.95.055020}{{\em Phys. Rev. D}
  {\bfseries 95} no.~5, (2017) 055020},
  \href{http://arxiv.org/abs/1612.09224}{{\ttfamily arXiv:1612.09224
  [hep-ph]}}.

\bibitem{BhupalDev:2018tox}
P.~S. Bhupal~Dev and Y.~Zhang, ``{Displaced vertex signatures of doubly charged
  scalars in the type-II seesaw and its left-right extensions},''
  \href{http://dx.doi.org/10.1007/JHEP10(2018)199}{{\em JHEP} {\bfseries 10}
  (2018) 199}, \href{http://arxiv.org/abs/1808.00943}{{\ttfamily
  arXiv:1808.00943 [hep-ph]}}.

\bibitem{Du:2018eaw}
Y.~Du, A.~Dunbrack, M.~J. Ramsey-Musolf, and J.-H. Yu, ``{Type-II Seesaw Scalar
  Triplet Model at a 100 TeV $pp$ Collider: Discovery and Higgs Portal Coupling
  Determination},'' \href{http://dx.doi.org/10.1007/JHEP01(2019)101}{{\em JHEP}
  {\bfseries 01} (2019) 101}, \href{http://arxiv.org/abs/1810.09450}{{\ttfamily
  arXiv:1810.09450 [hep-ph]}}.

\bibitem{Antusch:2018svb}
S.~Antusch, O.~Fischer, A.~Hammad, and C.~Scherb, ``{Low scale type II seesaw:
  Present constraints and prospects for displaced vertex searches},''
  \href{http://dx.doi.org/10.1007/JHEP02(2019)157}{{\em JHEP} {\bfseries 02}
  (2019) 157}, \href{http://arxiv.org/abs/1811.03476}{{\ttfamily
  arXiv:1811.03476 [hep-ph]}}.

\bibitem{Li:2018jns}
T.~Li, ``{Type II Seesaw and tau lepton at the HL-LHC, HE-LHC and FCC-hh},''
  \href{http://dx.doi.org/10.1007/JHEP09(2018)079}{{\em JHEP} {\bfseries 09}
  (2018) 079}, \href{http://arxiv.org/abs/1802.00945}{{\ttfamily
  arXiv:1802.00945 [hep-ph]}}.

\bibitem{Ferreira:2019qpf}
M.~M. Ferreira, T.~B. de~Melo, S.~Kovalenko, P.~R.~D. Pinheiro, and F.~S.
  Queiroz, ``{Lepton Flavor Violation and Collider Searches in a Type I + II
  Seesaw Model},'' \href{http://dx.doi.org/10.1140/epjc/s10052-019-7422-z}{{\em
  Eur. Phys. J. C} {\bfseries 79} no.~11, (2019) 955},
  \href{http://arxiv.org/abs/1903.07634}{{\ttfamily arXiv:1903.07634
  [hep-ph]}}.

\bibitem{Anisha:2021fzf}
Anisha, U.~Banerjee, J.~Chakrabortty, C.~Englert, and M.~Spannowsky,
  ``{Extended Higgs boson sectors, effective field theory, and Higgs boson
  phenomenology},'' \href{http://dx.doi.org/10.1103/PhysRevD.103.096009}{{\em
  Phys. Rev. D} {\bfseries 103} no.~9, (2021) 096009},
  \href{http://arxiv.org/abs/2103.01810}{{\ttfamily arXiv:2103.01810
  [hep-ph]}}.

\bibitem{Banerjee:2024jwn}
U.~Banerjee, C.~Englert, and W.~Naskar, ``{Resurrecting the LHC discovery
  potential in the extended type-II seesaw model},''
  \href{http://dx.doi.org/10.1103/PhysRevD.110.055010}{{\em Phys. Rev. D}
  {\bfseries 110} no.~5, (2024) 055010},
  \href{http://arxiv.org/abs/2403.17455}{{\ttfamily arXiv:2403.17455
  [hep-ph]}}.

\bibitem{Bolton:2024thn}
P.~D. Bolton, J.~Kriewald, M.~Nemev\v{s}ek, F.~Nesti, and J.~C. Vasquez, ``{On
  Lepton Number Violation in the Type II Seesaw},''
  \href{http://arxiv.org/abs/2408.00833}{{\ttfamily arXiv:2408.00833
  [hep-ph]}}.

\bibitem{Ashanujjaman:2023tlj}
S.~Ashanujjaman and S.~P. Maharathy, ``{Probing compressed mass spectra in the
  type-II seesaw model at the LHC},''
  \href{http://dx.doi.org/10.1103/PhysRevD.107.115026}{{\em Phys. Rev. D}
  {\bfseries 107} no.~11, (2023) 115026},
  \href{http://arxiv.org/abs/2305.06889}{{\ttfamily arXiv:2305.06889
  [hep-ph]}}.

\bibitem{Giarnetti:2023dcr}
A.~Giarnetti, J.~Herrero-Garcia, S.~Marciano, D.~Meloni, and D.~Vatsyayan,
  ``{Neutrino masses from new Weinberg-like operators: phenomenology of TeV
  scalar multiplets},'' \href{http://dx.doi.org/10.1007/JHEP05(2024)055}{{\em
  JHEP} {\bfseries 05} (2024) 055},
  \href{http://arxiv.org/abs/2312.13356}{{\ttfamily arXiv:2312.13356
  [hep-ph]}}.

\bibitem{Mandal:2022zmy}
S.~Mandal, O.~G. Miranda, G.~Sanchez~Garcia, J.~W.~F. Valle, and X.-J. Xu,
  ``{Toward deconstructing the simplest seesaw mechanism},''
  \href{http://dx.doi.org/10.1103/PhysRevD.105.095020}{{\em Phys. Rev. D}
  {\bfseries 105} no.~9, (2022) 095020},
  \href{http://arxiv.org/abs/2203.06362}{{\ttfamily arXiv:2203.06362
  [hep-ph]}}.

\bibitem{Chiang:2021lsx}
C.-W. Chiang, S.~Jana, and D.~Sengupta, ``{Investigating new physics models
  with signature of same-sign diboson+$+{E\!\!\!\!/}_{T}$},''
  \href{http://dx.doi.org/10.1103/PhysRevD.105.055014}{{\em Phys. Rev. D}
  {\bfseries 105} no.~5, (2022) 055014},
  \href{http://arxiv.org/abs/2106.03888}{{\ttfamily arXiv:2106.03888
  [hep-ph]}}.

\bibitem{Primulando:2019evb}
R.~Primulando, J.~Julio, and P.~Uttayarat, ``{Scalar phenomenology in type-II
  seesaw model},'' \href{http://dx.doi.org/10.1007/JHEP08(2019)024}{{\em JHEP}
  {\bfseries 08} (2019) 024}, \href{http://arxiv.org/abs/1903.02493}{{\ttfamily
  arXiv:1903.02493 [hep-ph]}}.

\bibitem{Ashanujjaman:2021txz}
S.~Ashanujjaman and K.~Ghosh, ``{Revisiting type-II see-saw: present limits and
  future prospects at LHC},''
  \href{http://dx.doi.org/10.1007/JHEP03(2022)195}{{\em JHEP} {\bfseries 03}
  (2022) 195}, \href{http://arxiv.org/abs/2108.10952}{{\ttfamily
  arXiv:2108.10952 [hep-ph]}}.

\bibitem{Georgi:1985nv}
H.~Georgi and M.~Machacek, ``{Doubly charged Higgs bosons},''
\href{http://dx.doi.org/10.1016/0550-3213(85)90325-6}{{\em Nucl. Phys.}
  {\bfseries B262} (1985) 463}.
%%CITATION = NUPHA,B262,463;%%.

\bibitem{Chanowitz:1985ug}
M.~S. Chanowitz and M.~Golden, ``{Higgs Boson Triplets With M ($W$) = M ($Z$)
  $\cos \theta \omega$},''
\href{http://dx.doi.org/10.1016/0370-2693(85)90700-2}{{\em Phys. Lett.}
  {\bfseries B165} (1985) 105}.
%%CITATION = PHLTA,B165,105;%%.

\bibitem{HERA:2006}
{HERA Collaboration}, ``{Search for doubly-charged Higgs boson production at
  HERA},'' \href{http://dx.doi.org/10.1016/j.physletb.2006.05.061}{{\em Phys
  Letters B} {\bfseries 638} (2006) 432},
\href{http://arxiv.org/abs/hep-ex/0604027}{{\ttfamily arXiv:hep-ex/0604027
  [hep-ex]}}.
%%CITATION = ARXIV:0604027;%%.

\bibitem{CDF:2011}
{CDF Collaboration}, ``{Search for new physics in high pT like-sign dilepton
  events at CDF II},''
  \href{http://dx.doi.org/10.1103/PhysRevLett.107.181801}{{\em Phys. Rev. Lett}
  {\bfseries 107} (2011) 181801},
\href{http://arxiv.org/abs/1108.0101}{{\ttfamily arXiv:1108.0101 [hep-ex]}}.
%%CITATION = ARXIV:1108.0101;%%.

\bibitem{CMS:2015}
{CMS Collaboration}, ``{Study of Vector Boson Scattering and Search for New
  Physics in Events with Two Same-Sign Leptons and Two Jets},''
  \href{http://dx.doi.org/10.1103/PhysRevLett.114.051801}{{\em Phys. Rev.
  Lett.} {\bfseries 115} (2015) 051801},
\href{http://arxiv.org/abs/1410.6315}{{\ttfamily arXiv:1410.6315 [hep-ex]}}.
%%CITATION = ARXIV:1410.6315;%%.

\bibitem{CMS:2018}
{CMS Collaboration}, ``{Observation of Electroweak Production of Same-Sign W
  Boson Pairs in the Two Jet and Two Same-Sign Lepton Final State in
  Proton-Proton Collisions at $\sqrt{s}$=13 TeV},''
  \href{http://dx.doi.org/10.1103/PhysRevLett.120.081801}{{\em Phys. Rev.
  Lett.} {\bfseries 120} (2018) 081801},
\href{http://arxiv.org/abs/1709.05822}{{\ttfamily arXiv:1709.05822 [hep-ex]}}.
%%CITATION = ARXIV:1709.05822;%%.

\bibitem{CMS:2021wlt}
{CMS Collaboration}, ``{Search for charged Higgs bosons produced in vector
  boson fusion processes and decaying into vector boson pairs in
  proton\textendash{}proton collisions at $\sqrt{s} = 13\,{\text {TeV}} $},''
  \href{http://dx.doi.org/10.1140/epjc/s10052-021-09472-3}{{\em Eur. Phys. J.
  C} {\bfseries 81} no.~8, (2021) 723},
  \href{http://arxiv.org/abs/2104.04762}{{\ttfamily arXiv:2104.04762
  [hep-ex]}}.

\bibitem{ATLAS:2023dbw}
{ATLAS Collaboration}, ``{Measurement and interpretation of same-sign $W$ boson
  pair production in association with two jets in $pp$ collisions at $\sqrt{s}
  = 13$ TeV with the ATLAS detector}.'' \textsc{ATLAS-CONF-2023-023}, 2023.
\newblock \url{https://cds.cern.ch/record/2859330}.

\bibitem{OPAL}
{OPAL Collaboration}, ``{Search for Doubly Charged Higgs Bosons with the OPAL
  detector at LEP},''
  \href{http://dx.doi.org/10.1016/S0370-2693(01)01474-5}{{\em Phys. Lett. B}
  {\bfseries 526} (2002) 221},
\href{http://arxiv.org/abs/hep-ex/0111059}{{\ttfamily arXiv:hep-ex/0111059}}.
%%CITATION = HEP-EX/0111059;%%.

\bibitem{ATLAS:2023pairmultilep}
{ATLAS Collaboration}, ``{Search for doubly charged Higgs boson production in
  multi-lepton final states using 139 fb$^{-1}$ of proton–proton collisions
  at $\sqrt{s} = 13$ TeV with the ATLAS detector},''
  \href{http://dx.doi.org/10.1140/epjc/s10052-023-11578-9}{{\em The European
  Physical Journal C} {\bfseries 83} (2023) 605},
\href{http://arxiv.org/abs/2211.07505}{{\ttfamily arXiv:2211.07505 [hep-ex]}}.
%%CITATION = ARXIV:2211.07505;%%.

\bibitem{Zee:1985id}
A.~Zee, ``{Quantum Numbers of Majorana Neutrino Masses},''
  \href{http://dx.doi.org/10.1016/0550-3213(86)90475-X}{{\em Nucl. Phys. B}
  {\bfseries 264} (1986) 99--110}.

\bibitem{Babu:1988ki}
K.~S. Babu, ``{Model of 'Calculable' Majorana Neutrino Masses},''
  \href{http://dx.doi.org/10.1016/0370-2693(88)91584-5}{{\em Phys. Lett. B}
  {\bfseries 203} (1988) 132--136}.

\bibitem{Nebot:2007bc}
M.~Nebot, J.~F. Oliver, D.~Palao, and A.~Santamaria, ``{Prospects for the
  Zee-Babu Model at the CERN LHC and low energy experiments},''
  \href{http://dx.doi.org/10.1103/PhysRevD.77.093013}{{\em Phys. Rev. D}
  {\bfseries 77} (2008) 093013},
  \href{http://arxiv.org/abs/0711.0483}{{\ttfamily arXiv:0711.0483 [hep-ph]}}.

\bibitem{Pati:1974yy}
J.~C. Pati and A.~Salam, ``{Lepton Number as the Fourth Color},''
  \href{http://dx.doi.org/10.1103/PhysRevD.10.275}{{\em Phys. Rev. D}
  {\bfseries 10} (1974) 275--289}. [Erratum: Phys.Rev.D 11, 703--703 (1975)].

\bibitem{Mohapatra:1974hk}
R.~N. Mohapatra and J.~C. Pati, ``{Left-Right Gauge Symmetry and an
  Isoconjugate Model of CP Violation},''
  \href{http://dx.doi.org/10.1103/PhysRevD.11.566}{{\em Phys. Rev. D}
  {\bfseries 11} (1975) 566--571}.

\bibitem{Senjanovic:1975rk}
G.~Senjanovic and R.~N. Mohapatra, ``{Exact Left-Right Symmetry and Spontaneous
  Violation of Parity},''
  \href{http://dx.doi.org/10.1103/PhysRevD.12.1502}{{\em Phys. Rev. D}
  {\bfseries 12} (1975) 1502}.

\bibitem{BhupalDev:2016nfr}
P.~S. Bhupal~Dev, R.~N. Mohapatra, and Y.~Zhang, ``{Displaced photon signal
  from a possible light scalar in minimal left-right seesaw model},''
  \href{http://dx.doi.org/10.1103/PhysRevD.95.115001}{{\em Phys. Rev. D}
  {\bfseries 95} no.~11, (2017) 115001},
  \href{http://arxiv.org/abs/1612.09587}{{\ttfamily arXiv:1612.09587
  [hep-ph]}}.

\bibitem{Borah:2016hqn}
D.~Borah and A.~Dasgupta, ``{Observable Lepton Number Violation with
  Predominantly Dirac Nature of Active Neutrinos},''
  \href{http://dx.doi.org/10.1007/JHEP01(2017)072}{{\em JHEP} {\bfseries 01}
  (2017) 072}, \href{http://arxiv.org/abs/1609.04236}{{\ttfamily
  arXiv:1609.04236 [hep-ph]}}.

\bibitem{CMS:2012}
{CMS Collaboration}, ``{A search for a doubly-charged Higgs boson in pp
  collisions at $\sqrt{s}$=7 TeV},''
  \href{http://dx.doi.org/10.1140/epjc/s10052-012-2189-5}{{\em The European
  Physical Journal C} {\bfseries 72} (2012) 2189},
\href{http://arxiv.org/abs/1207.2666}{{\ttfamily arXiv:1207.2666 [hep-ex]}}.
%%CITATION = ARXIV:1207.2666;%%.

\bibitem{ATLAS:2018multilepton}
{ATLAS Collaboration}, ``{Search for doubly charged Higgs boson production in
  multi-lepton final states with the ATLAS detector using proton–proton
  collisions at $\sqrt{s}$=13TeV},''
  \href{http://dx.doi.org/10.1140/EPJC/S10052-018-5661-Z}{{\em The European
  Physical Journal C} {\bfseries 78} (2018) 199},
\href{http://arxiv.org/abs/1710.09748}{{\ttfamily arXiv:1710.09748 [hep-ex]}}.
%%CITATION = ARXIV:1710.09748;%%.

\bibitem{ATLAS:2019pair}
{ATLAS Collaboration}, ``{Search for doubly charged scalar bosons decaying into
  same-sign W boson pairs with the ATLAS detector},''
  \href{http://dx.doi.org/10.1140/epjc/s10052-018-6500-y}{{\em The European
  Physical Journal C} {\bfseries 79} (2019) 58},
\href{http://arxiv.org/abs/1808.01899}{{\ttfamily arXiv:1808.01899 [hep-ex]}}.
%%CITATION = ARXIV:1808.01899;%%.

\bibitem{ATLAS:2021pairbosons}
{ATLAS Collaboration}, ``{Search for doubly and singly charged Higgs bosons
  decaying into vector bosons in multi-lepton final states with the ATLAS
  detector using proton-proton collisions at $\sqrt{s}$ = 13 TeV},''
  \href{http://dx.doi.org/10.1007/JHEP06(2021)146}{{\em Journal of High Energy
  Physics} {\bfseries 06} (2021) 146},
\href{http://arxiv.org/abs/2101.11961}{{\ttfamily arXiv:2101.11961 [hep-ex]}}.
%%CITATION = ARXIV:2101.11961;%%.

\bibitem{ATLAS:2014otc}
{ATLAS Collaboration}, ``{Search for charged Higgs bosons decaying via $H^{\pm}
  \rightarrow \tau^{\pm}\nu$ in fully hadronic final states using $pp$
  collision data at $\sqrt{s} = 8$ TeV with the ATLAS detector},''
  \href{http://dx.doi.org/10.1007/JHEP03(2015)088}{{\em JHEP} {\bfseries 03}
  (2015) 088}, \href{http://arxiv.org/abs/1412.6663}{{\ttfamily arXiv:1412.6663
  [hep-ex]}}.

\bibitem{CMS:2015lsf}
{CMS Collaboration}, ``{Search for a charged Higgs boson in pp collisions at $
  \sqrt{s}=8 $ TeV},'' \href{http://dx.doi.org/10.1007/JHEP11(2015)018}{{\em
  JHEP} {\bfseries 11} (2015) 018},
  \href{http://arxiv.org/abs/1508.07774}{{\ttfamily arXiv:1508.07774
  [hep-ex]}}.

\bibitem{ATLAS:2018gfm}
{ATLAS Collaboration}, ``{Search for charged Higgs bosons decaying via $H^{\pm}
  \to \tau^{\pm}\nu_{\tau}$ in the $\tau$+jets and $\tau$+lepton final states
  with 36 fb$^{-1}$ of $pp$ collision data recorded at $\sqrt{s} = 13$ TeV with
  the ATLAS experiment},''
  \href{http://dx.doi.org/10.1007/JHEP09(2018)139}{{\em JHEP} {\bfseries 09}
  (2018) 139}, \href{http://arxiv.org/abs/1807.07915}{{\ttfamily
  arXiv:1807.07915 [hep-ex]}}.

\bibitem{CMS:2019bfg}
{CMS Collaboration}, ``{Search for charged Higgs bosons in the H$^{\pm}$ $\to$
  $\tau^{\pm}\nu_\tau$ decay channel in proton-proton collisions at $\sqrt{s}
  =$ 13 TeV},'' \href{http://dx.doi.org/10.1007/JHEP07(2019)142}{{\em JHEP}
  {\bfseries 07} (2019) 142}, \href{http://arxiv.org/abs/1903.04560}{{\ttfamily
  arXiv:1903.04560 [hep-ex]}}.

\bibitem{CMS:2015yvc}
{CMS Collaboration}, ``{Search for a light charged Higgs boson decaying to $
  \mathrm{c}\overline{\mathrm{s}} $ in pp collisions at $ \sqrt{s}=8 $ TeV},''
  \href{http://dx.doi.org/10.1007/JHEP12(2015)178}{{\em JHEP} {\bfseries 12}
  (2015) 178}, \href{http://arxiv.org/abs/1510.04252}{{\ttfamily
  arXiv:1510.04252 [hep-ex]}}.

\bibitem{ATLAS:2013uxj}
{ATLAS Collaboration}, ``{Search for a light charged Higgs boson in the decay
  channel $H^+ \to c\bar{s}$ in $t\bar{t}$ events using pp collisions at
  $\sqrt{s}$ = 7 TeV with the ATLAS detector},''
  \href{http://dx.doi.org/10.1140/epjc/s10052-013-2465-z}{{\em Eur. Phys. J. C}
  {\bfseries 73} no.~6, (2013) 2465},
  \href{http://arxiv.org/abs/1302.3694}{{\ttfamily arXiv:1302.3694 [hep-ex]}}.

\bibitem{CMS:2018dzl}
{CMS Collaboration}, ``{Search for a charged Higgs boson decaying to charm and
  bottom quarks in proton-proton collisions at $ \sqrt{s}=8 $ TeV},''
  \href{http://dx.doi.org/10.1007/JHEP11(2018)115}{{\em JHEP} {\bfseries 11}
  (2018) 115}, \href{http://arxiv.org/abs/1808.06575}{{\ttfamily
  arXiv:1808.06575 [hep-ex]}}.

\bibitem{ATLAS:2023bzb}
{ATLAS Collaboration}, ``{Search for a light charged Higgs boson in $t
  \rightarrow H^{\pm}b$ decays, with $H^{\pm} \rightarrow cb$, in the
  lepton+jets final state in proton-proton collisions at $\sqrt{s}=13$ TeV with
  the ATLAS detector},'' \href{http://dx.doi.org/10.1007/JHEP09(2023)004}{{\em
  JHEP} {\bfseries 09} (2023) 004},
  \href{http://arxiv.org/abs/2302.11739}{{\ttfamily arXiv:2302.11739
  [hep-ex]}}.

\bibitem{CMS:2019idx}
{CMS Collaboration}, ``{Search for a light charged Higgs boson decaying to a W
  boson and a CP-odd Higgs boson in final states with e$\mu\mu$ or $\mu\mu\mu$
  in proton-proton collisions at $\sqrt{s} =$ 13 TeV},''
  \href{http://dx.doi.org/10.1103/PhysRevLett.123.131802}{{\em Phys. Rev.
  Lett.} {\bfseries 123} no.~13, (2019) 131802},
  \href{http://arxiv.org/abs/1905.07453}{{\ttfamily arXiv:1905.07453
  [hep-ex]}}.

\bibitem{ATLAS:2021xhq}
{ATLAS Collaboration}, ``{Search for $H^{\pm} \rightarrow W^{\pm}A \rightarrow
  W^{\pm}\mu\mu$ in $pp \rightarrow t\overline{t}$ events using an $e\mu\mu$
  signature with the ATLAS detector at $\sqrt{s}=13$ TeV}.''
  \textsc{ATLAS-CONF-2021-047}, 2021.
\newblock \url{https://cds.cern.ch/record/2780092}.

\bibitem{ATLAS:2015nkq}
{ATLAS Collaboration}, ``{Search for charged Higgs bosons in the $H^{\pm}
  \rightarrow tb$ decay channel in $pp$ collisions at $\sqrt{s}=8 $ TeV using
  the ATLAS detector},'' \href{http://dx.doi.org/10.1007/JHEP03(2016)127}{{\em
  JHEP} {\bfseries 03} (2016) 127},
  \href{http://arxiv.org/abs/1512.03704}{{\ttfamily arXiv:1512.03704
  [hep-ex]}}.

\bibitem{ATLAS:2018ntn}
{ATLAS Collaboration}, ``{Search for charged Higgs bosons decaying into top and
  bottom quarks at $\sqrt{s}$ = 13 TeV with the ATLAS detector},''
  \href{http://dx.doi.org/10.1007/JHEP11(2018)085}{{\em JHEP} {\bfseries 11}
  (2018) 085}, \href{http://arxiv.org/abs/1808.03599}{{\ttfamily
  arXiv:1808.03599 [hep-ex]}}.

\bibitem{CMS:2019rlz}
{CMS Collaboration}, ``{Search for a charged Higgs boson decaying into top and
  bottom quarks in events with electrons or muons in proton-proton collisions
  at $ \sqrt{\mathrm{s}} $ = 13 TeV},''
  \href{http://dx.doi.org/10.1007/JHEP01(2020)096}{{\em JHEP} {\bfseries 01}
  (2020) 096}, \href{http://arxiv.org/abs/1908.09206}{{\ttfamily
  arXiv:1908.09206 [hep-ex]}}.

\bibitem{CMS:2020imj}
{CMS Collaboration}, ``{Search for charged Higgs bosons decaying into a top and
  a bottom quark in the all-jet final state of pp collisions at $ \sqrt{s} $ =
  13 TeV},'' \href{http://dx.doi.org/10.1007/JHEP07(2020)126}{{\em JHEP}
  {\bfseries 07} (2020) 126}, \href{http://arxiv.org/abs/2001.07763}{{\ttfamily
  arXiv:2001.07763 [hep-ex]}}.

\bibitem{ATLAS:2016avi}
{ATLAS Collaboration}, ``{Search for charged Higgs bosons produced in
  association with a top quark and decaying via $H^{\pm} \rightarrow \tau\nu$
  using $pp$ collision data recorded at $\sqrt{s} = 13$ TeV by the ATLAS
  detector},'' \href{http://dx.doi.org/10.1016/j.physletb.2016.06.017}{{\em
  Phys. Lett. B} {\bfseries 759} (2016) 555--574},
  \href{http://arxiv.org/abs/1603.09203}{{\ttfamily arXiv:1603.09203
  [hep-ex]}}.

\bibitem{ATLAS:2015edr}
{ATLAS Collaboration}, ``{Search for a Charged Higgs Boson Produced in the
  Vector-Boson Fusion Mode with Decay $H^\pm \to W^\pm Z$ using $pp$ Collisions
  at $\sqrt{s}=8$ TeV with the ATLAS Experiment},''
  \href{http://dx.doi.org/10.1103/PhysRevLett.114.231801}{{\em Phys. Rev.
  Lett.} {\bfseries 114} no.~23, (2015) 231801},
  \href{http://arxiv.org/abs/1503.04233}{{\ttfamily arXiv:1503.04233
  [hep-ex]}}.

\bibitem{CMS:2017fgp}
{CMS Collaboration}, ``{Search for Charged Higgs Bosons Produced via Vector
  Boson Fusion and Decaying into a Pair of $W$ and $Z$ Bosons Using $pp$
  Collisions at $\sqrt{s}=13\text{ }\text{ }\mathrm{TeV}$},''
  \href{http://dx.doi.org/10.1103/PhysRevLett.119.141802}{{\em Phys. Rev.
  Lett.} {\bfseries 119} no.~14, (2017) 141802},
  \href{http://arxiv.org/abs/1705.02942}{{\ttfamily arXiv:1705.02942
  [hep-ex]}}.

\bibitem{ATLAS:2020zzb}
{ATLAS Collaboration}, ``{Search for dijet resonances in events with an
  isolated charged lepton using $\sqrt{s} = 13$ TeV proton-proton collision
  data collected by the ATLAS detector},''
  \href{http://dx.doi.org/10.1007/JHEP06(2020)151}{{\em JHEP} {\bfseries 06}
  (2020) 151}, \href{http://arxiv.org/abs/2002.11325}{{\ttfamily
  arXiv:2002.11325 [hep-ex]}}.

\bibitem{CMS:2022jqc}
{CMS Collaboration}, ``{Search for a charged Higgs boson decaying into a heavy
  neutral Higgs boson and a W boson in proton-proton collisions at $ \sqrt{s} $
  = 13 TeV},'' \href{http://dx.doi.org/10.1007/JHEP09(2023)032}{{\em JHEP}
  {\bfseries 09} (2023) 032}, \href{http://arxiv.org/abs/2207.01046}{{\ttfamily
  arXiv:2207.01046 [hep-ex]}}.

\bibitem{ALEPH:2006tnd}
{ALEPH, DELPHI, L3, OPAL, LEP Working Group for Higgs Boson Searches},
  ``{Search for neutral MSSM Higgs bosons at LEP},''
  \href{http://dx.doi.org/10.1140/epjc/s2006-02569-7}{{\em Eur. Phys. J. C}
  {\bfseries 47} (2006) 547--587},
  \href{http://arxiv.org/abs/hep-ex/0602042}{{\ttfamily arXiv:hep-ex/0602042}}.

\bibitem{CDF:2009esh}
{CDF Collaboration}, ``{Search for Higgs bosons predicted in two-Higgs-doublet
  models via decays to tau lepton pairs in 1.96-TeV p anti-p collisions},''
  \href{http://dx.doi.org/10.1103/PhysRevLett.103.201801}{{\em Phys. Rev.
  Lett.} {\bfseries 103} (2009) 201801},
  \href{http://arxiv.org/abs/0906.1014}{{\ttfamily arXiv:0906.1014 [hep-ex]}}.

\bibitem{D0:2010etq}
{D0 Collaboration}, ``{Search for Neutral Higgs Bosons in the Multi-$b$-Jet
  Topology in 5.2fb$^{-1}$ of $p\bar{p}$ Collisions at $\sqrt{s} = 1.96$
  TeV},'' \href{http://dx.doi.org/10.1016/j.physletb.2011.02.062}{{\em Phys.
  Lett. B} {\bfseries 698} (2011) 97--104},
  \href{http://arxiv.org/abs/1011.1931}{{\ttfamily arXiv:1011.1931 [hep-ex]}}.

\bibitem{D0:2011yor}
{D0 Collaboration}, ``{Search for Higgs bosons decaying to $\tau\tau$ pairs in
  $p\bar {p}$ collisions at $\sqrt{s} = 1.96$ TeV},''
  \href{http://dx.doi.org/10.1016/j.physletb.2011.12.050}{{\em Phys. Lett. B}
  {\bfseries 707} (2012) 323--329},
  \href{http://arxiv.org/abs/1106.4555}{{\ttfamily arXiv:1106.4555 [hep-ex]}}.

\bibitem{CDF:2011jcu}
{ CDF Collaboration}, ``{Search for Higgs Bosons Produced in Association with
  $b$-quarks},'' \href{http://dx.doi.org/10.1103/PhysRevD.85.032005}{{\em Phys.
  Rev. D} {\bfseries 85} (2012) 032005},
  \href{http://arxiv.org/abs/1106.4782}{{\ttfamily arXiv:1106.4782 [hep-ex]}}.

\bibitem{ATLAS:2019tpq}
{ATLAS Collaboration}, ``{Search for heavy neutral Higgs bosons produced in
  association with $b$-quarks and decaying into $b$-quarks at $\sqrt{s}=13$ TeV
  with the ATLAS detector},''
  \href{http://dx.doi.org/10.1103/PhysRevD.102.032004}{{\em Phys. Rev. D}
  {\bfseries 102} no.~3, (2020) 032004},
  \href{http://arxiv.org/abs/1907.02749}{{\ttfamily arXiv:1907.02749
  [hep-ex]}}.

\bibitem{ATLAS:2012ube}
{ATLAS Collaboration}, ``{Search for the neutral Higgs bosons of the Minimal
  Supersymmetric Standard Model in $pp$ collisions at $\sqrt{s}=7$ TeV with the
  ATLAS detector},'' \href{http://dx.doi.org/10.1007/JHEP02(2013)095}{{\em
  JHEP} {\bfseries 02} (2013) 095},
  \href{http://arxiv.org/abs/1211.6956}{{\ttfamily arXiv:1211.6956 [hep-ex]}}.

\bibitem{ATLAS:2019odt}
{ATLAS Collaboration}, ``{Search for scalar resonances decaying into
  $\mu^{+}\mu^{-}$ in events with and without $b$-tagged jets produced in
  proton-proton collisions at $\sqrt{s}=13$ TeV with the ATLAS detector},''
  \href{http://dx.doi.org/10.1007/JHEP07(2019)117}{{\em JHEP} {\bfseries 07}
  (2019) 117}, \href{http://arxiv.org/abs/1901.08144}{{\ttfamily
  arXiv:1901.08144 [hep-ex]}}.

\bibitem{ATLAS:2014vhc}
{ATLAS Collaboration}, ``{Search for neutral Higgs bosons of the minimal
  supersymmetric standard model in pp collisions at $\sqrt{s}$ = 8 TeV with the
  ATLAS detector},'' \href{http://dx.doi.org/10.1007/JHEP11(2014)056}{{\em
  JHEP} {\bfseries 11} (2014) 056},
  \href{http://arxiv.org/abs/1409.6064}{{\ttfamily arXiv:1409.6064 [hep-ex]}}.

\bibitem{ATLAS:2016ivh}
{ATLAS Collaboration}, ``{Search for Minimal Supersymmetric Standard Model
  Higgs bosons $H/A$ and for a $Z^{\prime}$ boson in the $\tau \tau$ final
  state produced in $pp$ collisions at $\sqrt{s}=13$ TeV with the ATLAS
  Detector},'' \href{http://dx.doi.org/10.1140/epjc/s10052-016-4400-6}{{\em
  Eur. Phys. J. C} {\bfseries 76} no.~11, (2016) 585},
  \href{http://arxiv.org/abs/1608.00890}{{\ttfamily arXiv:1608.00890
  [hep-ex]}}.

\bibitem{ATLAS:2017eiz}
{ATLAS Collaboration}, ``{Search for additional heavy neutral Higgs and gauge
  bosons in the ditau final state produced in 36 fb$^{-1}$ of pp collisions at
  $ \sqrt{s}=13 $ TeV with the ATLAS detector},''
  \href{http://dx.doi.org/10.1007/JHEP01(2018)055}{{\em JHEP} {\bfseries 01}
  (2018) 055}, \href{http://arxiv.org/abs/1709.07242}{{\ttfamily
  arXiv:1709.07242 [hep-ex]}}.

\bibitem{ATLAS:2020zms}
{ATLAS Collaboration}, ``{Search for heavy Higgs bosons decaying into two tau
  leptons with the ATLAS detector using $pp$ collisions at $\sqrt{s}=13$
  TeV},'' \href{http://dx.doi.org/10.1103/PhysRevLett.125.051801}{{\em Phys.
  Rev. Lett.} {\bfseries 125} no.~5, (2020) 051801},
  \href{http://arxiv.org/abs/2002.12223}{{\ttfamily arXiv:2002.12223
  [hep-ex]}}.

\bibitem{CMS:2013baf}
{CMS Collaboration}, ``{Search for a Higgs Boson Decaying into a b-Quark Pair
  and Produced in Association with b Quarks in Proton\textendash{}Proton
  Collisions at 7 TeV},''
  \href{http://dx.doi.org/10.1016/j.physletb.2013.04.017}{{\em Phys. Lett. B}
  {\bfseries 722} (2013) 207--232},
  \href{http://arxiv.org/abs/1302.2892}{{\ttfamily arXiv:1302.2892 [hep-ex]}}.

\bibitem{CMS:2015grx}
{CMS Collaboration}, ``{Search for neutral MSSM Higgs bosons decaying into a
  pair of bottom quarks},''
  \href{http://dx.doi.org/10.1007/JHEP11(2015)071}{{\em JHEP} {\bfseries 11}
  (2015) 071}, \href{http://arxiv.org/abs/1506.08329}{{\ttfamily
  arXiv:1506.08329 [hep-ex]}}.

\bibitem{CMS:2018hir}
{CMS Collaboration}, ``{Search for beyond the standard model Higgs bosons
  decaying into a $\mathrm{b\overline{b}}$ pair in pp collisions at $\sqrt{s}
  =$ 13 TeV},'' \href{http://dx.doi.org/10.1007/JHEP08(2018)113}{{\em JHEP}
  {\bfseries 08} (2018) 113}, \href{http://arxiv.org/abs/1805.12191}{{\ttfamily
  arXiv:1805.12191 [hep-ex]}}.

\bibitem{CMS:2015ooa}
{CMS Collaboration}, ``{Search for neutral MSSM Higgs bosons decaying to
  $\mu^{+} \mu^{-}$ in pp collisions at $ \sqrt{s} =$ 7 and 8 TeV},''
  \href{http://dx.doi.org/10.1016/j.physletb.2015.11.042}{{\em Phys. Lett. B}
  {\bfseries 752} (2016) 221--246},
  \href{http://arxiv.org/abs/1508.01437}{{\ttfamily arXiv:1508.01437
  [hep-ex]}}.

\bibitem{CMS:2019mij}
{CMS Collaboration}, ``{Search for MSSM Higgs bosons decaying to
  \ensuremath{\mu} + \ensuremath{\mu} \ensuremath{-} in proton-proton
  collisions at s=13TeV},''
  \href{http://dx.doi.org/10.1016/j.physletb.2019.134992}{{\em Phys. Lett. B}
  {\bfseries 798} (2019) 134992},
  \href{http://arxiv.org/abs/1907.03152}{{\ttfamily arXiv:1907.03152
  [hep-ex]}}.

\bibitem{CMS:2011lzj}
{CMS Collaboration}, ``{Search for Neutral MSSM Higgs Bosons Decaying to Tau
  Pairs in $pp$ Collisions at $\sqrt{s}=7$ TeV},''
  \href{http://dx.doi.org/10.1103/PhysRevLett.106.231801}{{\em Phys. Rev.
  Lett.} {\bfseries 106} (2011) 231801},
  \href{http://arxiv.org/abs/1104.1619}{{\ttfamily arXiv:1104.1619 [hep-ex]}}.

\bibitem{CMS:2012bkm}
{CMS Collaboration}, ``{Search for Neutral Higgs Bosons Decaying to Tau Pairs
  in $pp$ Collisions at $\sqrt{s}=7$ TeV},''
  \href{http://dx.doi.org/10.1016/j.physletb.2012.05.028}{{\em Phys. Lett. B}
  {\bfseries 713} (2012) 68--90},
  \href{http://arxiv.org/abs/1202.4083}{{\ttfamily arXiv:1202.4083 [hep-ex]}}.

\bibitem{CMS:2014ccx}
{CMS Collaboration}, ``{Search for neutral MSSM Higgs bosons decaying to a pair
  of tau leptons in pp collisions},''
  \href{http://dx.doi.org/10.1007/JHEP10(2014)160}{{\em JHEP} {\bfseries 10}
  (2014) 160}, \href{http://arxiv.org/abs/1408.3316}{{\ttfamily arXiv:1408.3316
  [hep-ex]}}.

\bibitem{CMS:2018rmh}
{CMS Collaboration}, ``{Search for additional neutral MSSM Higgs bosons in the
  $\tau\tau$ final state in proton-proton collisions at $\sqrt{s}=$ 13 TeV},''
  \href{http://dx.doi.org/10.1007/JHEP09(2018)007}{{\em JHEP} {\bfseries 09}
  (2018) 007}, \href{http://arxiv.org/abs/1803.06553}{{\ttfamily
  arXiv:1803.06553 [hep-ex]}}.

\bibitem{ATLAS:2023szc}
{ATLAS Collaboration}, ``{Search for a CP-odd Higgs boson decaying to a heavy
  CP-even Higgs boson and a $Z$ boson in the $\ell\ell t\bar{t}$ and
  $\nu\bar{\nu}b\bar{b}$ final states using 140 fb$^{-1}$ of data collected
  with the ATLAS detector}.'' \textsc{ATLAS-CONF-2023-034}, 2023.
\newblock \url{https://cds.cern.ch/record/2862023}.

\bibitem{CMS:2024mtn}
{CMS Collaboration}, ``{Search for heavy neutral Higgs bosons A and H in the
  t$\bar t$Z final state}.'' \textsc{CMS-PAS-B2G-23-006}, 2024.
\newblock \url{https://cds.cern.ch/record/2892681}.

\bibitem{ParticleDataGroup:2024cfk}
P.~D. Group, ``{Review of particle physics},''
  \href{http://dx.doi.org/10.1103/PhysRevD.110.030001}{{\em Phys. Rev. D}
  {\bfseries 110} no.~3, (2024) 030001}.

\bibitem{Bonilla:2015eha}
C.~Bonilla, R.~M. Fonseca, and J.~W.~F. Valle, ``{Consistency of the triplet
  seesaw model revisited},''
  \href{http://dx.doi.org/10.1103/PhysRevD.92.075028}{{\em Phys. Rev. D}
  {\bfseries 92} no.~7, (2015) 075028},
  \href{http://arxiv.org/abs/1508.02323}{{\ttfamily arXiv:1508.02323
  [hep-ph]}}.

\bibitem{Moultaka:2020dmb}
G.~Moultaka and M.~C. Peyran\`ere, ``{Vacuum stability conditions for Higgs
  potentials with $SU(2)_L$ triplets},''
  \href{http://dx.doi.org/10.1103/PhysRevD.103.115006}{{\em Phys. Rev. D}
  {\bfseries 103} no.~11, (2021) 115006},
  \href{http://arxiv.org/abs/2012.13947}{{\ttfamily arXiv:2012.13947
  [hep-ph]}}.

\bibitem{Peskin:1991sw}
M.~E. Peskin and T.~Takeuchi, ``{Estimation of oblique electroweak
  corrections},'' \href{http://dx.doi.org/10.1103/PhysRevD.46.381}{{\em Phys.
  Rev. D} {\bfseries 46} (1992) 381--409}.

\bibitem{Burgess:1993mg}
C.~P. Burgess, S.~Godfrey, H.~Konig, D.~London, and I.~Maksymyk, ``{A Global
  fit to extended oblique parameters},''
  \href{http://dx.doi.org/10.1016/0370-2693(94)91322-6}{{\em Phys. Lett. B}
  {\bfseries 326} (1994) 276--281},
  \href{http://arxiv.org/abs/hep-ph/9307337}{{\ttfamily arXiv:hep-ph/9307337}}.

\bibitem{Lavoura1994}
L.~Lavoura and L.-F. Li, ``Making the small oblique parameters large,''
  \href{http://dx.doi.org/10.1103/physrevd.49.1409}{{\em Physical Review D}
  {\bfseries 49} no.~3, (Feb., 1994) 1409–1416}.
  \url{http://dx.doi.org/10.1103/PhysRevD.49.1409}.

\bibitem{Cheng:2022hbo}
Y.~Cheng, X.-G. He, F.~Huang, J.~Sun, and Z.-P. Xing, ``{Electroweak precision
  tests for triplet scalars},''
  \href{http://dx.doi.org/10.1016/j.nuclphysb.2023.116118}{{\em Nucl. Phys. B}
  {\bfseries 989} (2023) 116118},
  \href{http://arxiv.org/abs/2208.06760}{{\ttfamily arXiv:2208.06760
  [hep-ph]}}.

\bibitem{Ashanujjaman:2022tdn}
S.~Ashanujjaman, K.~Ghosh, and K.~Huitu, ``{Type-II see-saw: searching the LHC
  elusive low-mass triplet-like Higgses at $e^-e^+$ colliders},''
  \href{http://dx.doi.org/10.1103/PhysRevD.106.075028}{{\em Phys. Rev. D}
  {\bfseries 106} no.~7, (2022) 075028},
  \href{http://arxiv.org/abs/2205.14983}{{\ttfamily arXiv:2205.14983
  [hep-ph]}}.

\bibitem{CMS:2018piu}
{CMS Collaboration}, ``{Measurements of Higgs boson properties in the diphoton
  decay channel in proton-proton collisions at $\sqrt{s} =$ 13 TeV},''
  \href{http://dx.doi.org/10.1007/JHEP11(2018)185}{{\em JHEP} {\bfseries 11}
  (2018) 185}, \href{http://arxiv.org/abs/1804.02716}{{\ttfamily
  arXiv:1804.02716 [hep-ex]}}.

\bibitem{ATLAS:2018hxb}
{ATLAS Collaboration}, ``{Measurements of Higgs boson properties in the
  diphoton decay channel with 36 fb$^{-1}$ of $pp$ collision data at $\sqrt{s}
  = 13$ TeV with the ATLAS detector},''
  \href{http://dx.doi.org/10.1103/PhysRevD.98.052005}{{\em Phys. Rev. D}
  {\bfseries 98} (2018) 052005},
  \href{http://arxiv.org/abs/1802.04146}{{\ttfamily arXiv:1802.04146
  [hep-ex]}}.

\bibitem{ATLAS:2023yqk}
{ATLAS and CMS Collaborations}, ``{Evidence for the Higgs Boson Decay to a Z
  Boson and a Photon at the LHC},''
  \href{http://dx.doi.org/10.1103/PhysRevLett.132.021803}{{\em Phys. Rev.
  Lett.} {\bfseries 132} no.~2, (2024) 021803},
  \href{http://arxiv.org/abs/2309.03501}{{\ttfamily arXiv:2309.03501
  [hep-ex]}}.

\bibitem{Smolyakov:1982}
N.~V. {Smolyakov}, ``{Furry theorem for non-abelian gauge Lagrangians},''
  \href{http://dx.doi.org/10.1007/BF01016449}{{\em Theoretical and Mathematical
  Physics} {\bfseries 50} no.~3, (Mar., 1982) 225--228}.

\bibitem{delAguila:1990yw}
F.~del Aguila and L.~Ametller, ``{On the detectability of sleptons at large
  hadron colliders},''
  \href{http://dx.doi.org/10.1016/0370-2693(91)90336-O}{{\em Phys. Lett. B}
  {\bfseries 261} (1991) 326--333}.

\bibitem{Hessler:2014ssa}
A.~G. Hessler, A.~Ibarra, E.~Molinaro, and S.~Vogl, ``{Impact of the Higgs
  boson on the production of exotic particles at the LHC},''
  \href{http://dx.doi.org/10.1103/PhysRevD.91.115004}{{\em Phys. Rev. D}
  {\bfseries 91} no.~11, (2015) 115004},
  \href{http://arxiv.org/abs/1408.0983}{{\ttfamily arXiv:1408.0983 [hep-ph]}}.

\bibitem{Fuks:2019clu}
B.~Fuks, M.~Nemev\v{s}ek, and R.~Ruiz, ``{Doubly Charged Higgs Boson Production
  at Hadron Colliders},''
  \href{http://dx.doi.org/10.1103/PhysRevD.101.075022}{{\em Phys. Rev. D}
  {\bfseries 101} no.~7, (2020) 075022},
  \href{http://arxiv.org/abs/1912.08975}{{\ttfamily arXiv:1912.08975
  [hep-ph]}}.

\bibitem{Alwall:2014hca}
J.~Alwall, R.~Frederix, S.~Frixione, V.~Hirschi, F.~Maltoni, O.~Mattelaer,
  H.~S. Shao, T.~Stelzer, P.~Torrielli, and M.~Zaro, ``{The automated
  computation of tree-level and next-to-leading order differential cross
  sections, and their matching to parton shower simulations},''
  \href{http://dx.doi.org/10.1007/JHEP07(2014)079}{{\em JHEP} {\bfseries 07}
  (2014) 079}, \href{http://arxiv.org/abs/1405.0301}{{\ttfamily arXiv:1405.0301
  [hep-ph]}}.

\bibitem{Frederix:2018nkq}
R.~Frederix, S.~Frixione, V.~Hirschi, D.~Pagani, H.~S. Shao, and M.~Zaro,
  ``{The automation of next-to-leading order electroweak calculations},''
  \href{http://dx.doi.org/10.1007/JHEP11(2021)085}{{\em JHEP} {\bfseries 07}
  (2018) 185}, \href{http://arxiv.org/abs/1804.10017}{{\ttfamily
  arXiv:1804.10017 [hep-ph]}}. [Erratum: JHEP 11, 085 (2021)].

\bibitem{Carrazza_2015}
S.~Carrazza, S.~Forte, Z.~Kassabov, J.~I. Latorre, and J.~Rojo, ``{An Unbiased
  Hessian Representation for Monte Carlo PDFs},''
  \href{http://dx.doi.org/10.1140/epjc/s10052-015-3590-7}{{\em Eur. Phys. J. C}
  {\bfseries 75} no.~8, (2015) 369},
  \href{http://arxiv.org/abs/1505.06736}{{\ttfamily arXiv:1505.06736
  [hep-ph]}}.

\bibitem{Christensen:2008py}
N.~D. Christensen and C.~Duhr, ``{FeynRules - Feynman rules made easy},''
  \href{http://dx.doi.org/10.1016/j.cpc.2009.02.018}{{\em Comput. Phys.
  Commun.} {\bfseries 180} (2009) 1614--1641},
  \href{http://arxiv.org/abs/0806.4194}{{\ttfamily arXiv:0806.4194 [hep-ph]}}.

\bibitem{Degrande:2011ua}
C.~Degrande, C.~Duhr, B.~Fuks, D.~Grellscheid, O.~Mattelaer, and T.~Reiter,
  ``{UFO - The Universal FeynRules Output},''
  \href{http://dx.doi.org/10.1016/j.cpc.2012.01.022}{{\em Comput. Phys.
  Commun.} {\bfseries 183} (2012) 1201--1214},
  \href{http://arxiv.org/abs/1108.2040}{{\ttfamily arXiv:1108.2040 [hep-ph]}}.

\bibitem{Alloul:2013bka}
A.~Alloul, N.~D. Christensen, C.~Degrande, C.~Duhr, and B.~Fuks, ``{FeynRules
  2.0 - A complete toolbox for tree-level phenomenology},''
  \href{http://dx.doi.org/10.1016/j.cpc.2014.04.012}{{\em Comput. Phys.
  Commun.} {\bfseries 185} (2014) 2250--2300},
  \href{http://arxiv.org/abs/1310.1921}{{\ttfamily arXiv:1310.1921 [hep-ph]}}.

\bibitem{PhysRevD.22.722}
T.~G. Rizzo, ``Decays of heavy higgs bosons,''
  \href{http://dx.doi.org/10.1103/PhysRevD.22.722}{{\em Phys. Rev. D}
  {\bfseries 22} (Aug, 1980) 722--726}.
  \url{https://link.aps.org/doi/10.1103/PhysRevD.22.722}.

\bibitem{PhysRevD.30.248}
W.-Y. Keung and W.~J. Marciano, ``Higgs-scalar decays:
  $h\ensuremath{\rightarrow}{W}^{\ifmmode\pm\else\textpm\fi{}}+x$,''
  \href{http://dx.doi.org/10.1103/PhysRevD.30.248}{{\em Phys. Rev. D}
  {\bfseries 30} (Jul, 1984) 248--250}.
  \url{https://link.aps.org/doi/10.1103/PhysRevD.30.248}.

\bibitem{Cahn:1990xc}
R.~N. Cahn, ``{A Higgs primer. GIF 90: 22nd Summer School on Particle Physics:
  Where is the Higgs?}.'' Lbl-29789, 1990.
\newblock \url{https://lib-extopc.kek.jp/preprints/PDF/1991/9102/9102156.pdf}.

\bibitem{Djouadi:1995gv}
A.~Djouadi, J.~Kalinowski, and P.~M. Zerwas, ``{Two- and Three-Body Decay Modes
  of SUSY Higgs Particles},''
  \href{http://dx.doi.org/10.1007/s002880050121}{{\em Z. Phys.} {\bfseries C70}
  (1996) 435--448},
\href{http://arxiv.org/abs/hep-ph/9511342}{{\ttfamily arXiv:hep-ph/9511342}}.
%%CITATION = HEP-PH/9511342;%%.

\bibitem{Bierlich:2022pfr}
C.~Bierlich {\em et~al.}, ``{A comprehensive guide to the physics and usage of
  PYTHIA 8.3},'' \href{http://dx.doi.org/10.21468/SciPostPhysCodeb.8}{{\em
  SciPost Phys. Codeb.} {\bfseries 2022} (2022) 8},
  \href{http://arxiv.org/abs/2203.11601}{{\ttfamily arXiv:2203.11601
  [hep-ph]}}.

\bibitem{Artoisenet:2012st}
P.~Artoisenet, R.~Frederix, O.~Mattelaer, and R.~Rietkerk, ``{Automatic
  spin-entangled decays of heavy resonances in Monte Carlo simulations},''
  \href{http://dx.doi.org/10.1007/JHEP03(2013)015}{{\em JHEP} {\bfseries 03}
  (2013) 015}, \href{http://arxiv.org/abs/1212.3460}{{\ttfamily arXiv:1212.3460
  [hep-ph]}}.

\bibitem{deFavereau:2013fsa}
{DELPHES 3 Collaboration}, ``{DELPHES 3, A modular framework for fast
  simulation of a generic collider experiment},''
  \href{http://dx.doi.org/10.1007/JHEP02(2014)057}{{\em JHEP} {\bfseries 02}
  (2014) 057}, \href{http://arxiv.org/abs/1307.6346}{{\ttfamily arXiv:1307.6346
  [hep-ex]}}.

\bibitem{PERF-2007-01}
{ATLAS Collaboration}, ``{The ATLAS Experiment at the CERN Large Hadron
  Collider},'' \href{http://dx.doi.org/10.1088/1748-0221/3/08/S08003}{{\em
  JINST} {\bfseries 3} (2008) S08003}.

\bibitem{Cacciari:2008gp}
M.~Cacciari, G.~P. Salam, and G.~Soyez, ``{The anti-\(k_{t}\) jet clustering
  algorithm},'' \href{http://dx.doi.org/10.1088/1126-6708/2008/04/063}{{\em
  JHEP} {\bfseries 04} (2008) 063},
\href{http://arxiv.org/abs/0802.1189}{{\ttfamily arXiv:0802.1189 [hep-ph]}}.
%%CITATION = 0802.1189;%%.

\bibitem{ATLAS:2023dxj}
{ATLAS Collaboration}, ``{Electron and photon efficiencies in LHC Run 2 with
  the ATLAS experiment},''
  \href{http://dx.doi.org/10.1007/JHEP05(2024)162}{{\em JHEP} {\bfseries 05}
  (2024) 162}, \href{http://arxiv.org/abs/2308.13362}{{\ttfamily
  arXiv:2308.13362 [hep-ex]}}.

\bibitem{ATLAS:2020auj}
{ATLAS Collaboration}, ``{Muon reconstruction and identification efficiency in
  ATLAS using the full Run 2 $pp$ collision data set at $\sqrt{s}=13$ TeV},''
  \href{http://dx.doi.org/10.1140/epjc/s10052-021-09233-2}{{\em Eur. Phys. J.
  C} {\bfseries 81} no.~7, (2021) 578},
  \href{http://arxiv.org/abs/2012.00578}{{\ttfamily arXiv:2012.00578
  [hep-ex]}}.

\bibitem{ATLAS:2022yru}
{ATLAS Collaboration}, ``{SimpleAnalysis: Truth-level Analysis Framework}.''
  \textsc{ATL-PHYS-PUB-2022-017}, 2022.
\newblock \url{https://cds.cern.ch/record/2805991}.

\bibitem{Cowan:2010js}
G.~Cowan, K.~Cranmer, E.~Gross, and O.~Vitells, ``{Asymptotic formulae for
  likelihood-based tests of new physics},''
  \href{http://dx.doi.org/10.1140/epjc/s10052-011-1554-0}{{\em Eur. Phys. J. C}
  {\bfseries 71} (2011) 1554}, \href{http://arxiv.org/abs/1007.1727}{{\ttfamily
  arXiv:1007.1727 [physics.data-an]}}. [Erratum: Eur.Phys.J.C 73, 2501 (2013)].

\bibitem{ATLAS:2022swp}
{ATLAS Collaboration}, ``{Tools for estimating fake/non-prompt lepton
  backgrounds with the ATLAS detector at the LHC},''
  \href{http://dx.doi.org/10.1088/1748-0221/18/11/T11004}{{\em JINST}
  {\bfseries 18} no.~11, (2023) T11004},
  \href{http://arxiv.org/abs/2211.16178}{{\ttfamily arXiv:2211.16178
  [hep-ex]}}.

\bibitem{ATLAS:2022qxm}
{ ATLAS Collaboration }, ``{ATLAS flavour-tagging algorithms for the LHC Run 2
  pp collision dataset},''
  \href{http://dx.doi.org/10.1140/epjc/s10052-023-11699-1}{{\em Eur. Phys. J.
  C} {\bfseries 83} no.~7, (2023) 681},
  \href{http://arxiv.org/abs/2211.16345}{{\ttfamily arXiv:2211.16345
  [physics.data-an]}}.

\bibitem{ATLAS:2019dpa}
{ATLAS Collaboration}, ``{Performance of electron and photon triggers in ATLAS
  during LHC Run 2},''
  \href{http://dx.doi.org/10.1140/epjc/s10052-019-7500-2}{{\em Eur. Phys. J. C}
  {\bfseries 80} no.~1, (2020) 47},
  \href{http://arxiv.org/abs/1909.00761}{{\ttfamily arXiv:1909.00761
  [hep-ex]}}.

\bibitem{ATLAS:2020gty}
{ATLAS Collaboration}, ``{Performance of the ATLAS muon triggers in Run 2},''
  \href{http://dx.doi.org/10.1088/1748-0221/15/09/p09015}{{\em JINST}
  {\bfseries 15} no.~09, (2020) P09015},
  \href{http://arxiv.org/abs/2004.13447}{{\ttfamily arXiv:2004.13447
  [physics.ins-det]}}.

\end{thebibliography}\endgroup

\end{document}